\documentclass[11pt,a4paper]{article}
\usepackage[a4paper,margin=2.2cm]{geometry}
\usepackage{authblk}
\usepackage{latexsym}
\usepackage{graphicx}
\newtheorem{assumption}{Assumption}
\usepackage{multirow}
\usepackage{tikz, subcaption}
\usepackage{xcolor}
\usepackage[compress]{cite}
\usepackage{hyperref}
\usepackage{url}

\usepackage{amsmath}
\usepackage{amssymb}
\usepackage[normalem]{ulem}
\begin{document}
\title{A novel phenomenological approach to \\total charm cross-section measurements at the LHC}
\author[1,3]{Yewon Yang \thanks{yewon.yang@cern.ch}}
\author[1]{Achim Geiser}
\author[2]{Sven-Olaf Moch}
\author[2]{Oleksandr Zenaiev}
\affil[1]{Deutsches Elektronen-Synchrotron DESY, Notkestr. 85, 22607 Hamburg, Germany}
\affil[2]{Universit\"at Hamburg, II. Institute for Theoretical Physics, Luruper Chaussee 149, 22761 Hamburg, Germany}
\affil[3]{now at E\"otv\"os Lor\'and University, Egyetem t\'er 1-3, H-1053 Budapest, Hungary}
\date{March 5, 2026}
\maketitle
\begin{center}
\textit{Published in Eur.Phys.J.C 86 (2026) 3, 225}
\end{center}
\vspace{1cm}
\begin{abstract}
We propose a novel, data-driven method for determining total charm cross sections in proton-proton collisions by extrapolating measured fiducial cross sections without assuming any particular fragmentation model.  
The recently observed charm fragmentation non-universality at the LHC experimentally establishes strongly increased baryon production fractions and correspondingly decreased meson production fractions compared to electron-positron collisions, with a very significant $p_T$ dependence. The novel method accounts for this non-universality and its $p_T$-dependence through a data-driven extrapolation function called ddFONLL.
Applied to $D^0$ production at 5 and 13 TeV, this approach yields total charm cross sections that fully incorporate the fragmentation non-universality and increase significantly compared to the previous measurements still based on fragmentation universality.
The results are consistent with NNLO QCD predictions and enable direct comparisons free from fragmentation assumptions.
We use this to evaluate the sensitivity of total cross-section measurements to parton distribution functions and the charm-quark mass. 
An outlook is given on the potential of further expanding the use of the ddFONLL method.
\end{abstract} %end of abstract
\sloppy % for an automatic line break in texts
\section{Introduction}

The theory of Quantum-Chromo-Dynamics (QCD) is a well established part of the Standard Model which describes many of the processes occurring in particular in proton-proton collisions at LHC. Predictions for charm production are particularly challenging since, due to the closeness of the charm mass $m_c\sim 1.5$ GeV to the fundamental QCD parameter $\Lambda_{QCD} \sim 250$ MeV, cross sections can still be calculated perturbatively, but the convergence of the perturbative series is slow since the value of the strong coupling constant $\alpha_s$ is large, resulting in large theoretical uncertainties. Charm measurements thus test QCD in the transition region of the perturbative and nonperturbative regimes. 

Measuring the total charm cross section without any cuts on phase space is particularly important since for charm the corresponding theoretical predictions are the only ones available at next-to-next-to-leading order (NNLO, terms up to $\alpha_s^4$), and do furthermore not depend on charm fragmentation. 
Currently differential cross-section calculations are known for top \cite{Czakon:2015owf,top_diffXsec_NNLO} and beauty \cite{beauty_diffXsec_NNLO,Czakon:2024tjr} up to NNLO, while for charm they are known only up to next-to-leading order (NLO, up to $\alpha_s^3$) without \cite{MNRtheory} or with \cite{fonll1,fonll2} next-to-leading logarithmic (NLL) contributions.
In the case of the total cross section, calculations are known up to NNLO~\cite{Czakon:2013goa} for all three heavy quarks
and have been implemented in public codes~\cite{hathor,Czakon:2011xx}. 

For the total charm cross-section measurement, differential cross sections measured in limited kinematic ranges and for a restricted set of hadronic final states need to be extrapolated to the total cross section, under certain theoretical or phenomenological assumptions.
Several differential cross-section measurements have been performed and published on different inclusive charm-hadron final states in $pp$ collisions by the LHC experiments so far: ALICE at $\sqrt{s} =$ 2.76, 5, 7, 13 TeV \cite{ALICE_Dmeson_2p76TeV, ALICE_DmesonRatios_5TeV, ALICE_LcToD0_5TeV, ALICE_Dmesons_5TeV, ALICE_LcToD0_5TeV_update, ALICE_Dmeson_7TeV, ALICE_Dmeson_7TeV_2, ALICE_cTotXsec_7TeV, ALICE_cHad_13TeV, ALICE_cFragFrac_13TeV}, LHCb at $\sqrt{s} =$ 5, 7, 13 TeV \cite{LHCb_Dmesons_5TeV, LHCb_Dmesons_7TeV, LHCb_Dmesons_13TeV}, ATLAS at $\sqrt{s} =$ 7, 13 TeV \cite{ATLAS_Dmeson_7TeV, ATLAS:2024vne} and CMS at $\sqrt{s} =$ 5, 13 TeV \cite{CMS_D0_5TeV, CMS_Lc_5TeV, CMS_Lc_5TeV_update, CMS_Dmesons_13TeV}. Some of these measurements were extrapolated to the full kinematic phase space in order to extract the total charm cross section \cite{ALICE_Dmeson_2p76TeV, ALICE_Dmeson_7TeV_2, ALICE_cTotXsec_7TeV, nnloCharm1}, with the assumption of charm-fragmentation universality, i.e., that the fragmentation is independent of either collision systems and kinematics. The corresponding total cross-section values are strongly theory and model dependent.

Recent results from the LHC experiments
\cite{ALICE_cFragFrac_5TeV,ALICE:2020wfu,ALICE_nonUni1,LHCb_Dmesons_7TeV,CMS_Lc_5TeV,CMS_Lc_5TeV_update}
%\cite{ALICE:2021dhb,ALICE:2020wfu,Aaij:2013mga,CMS:2019uws,CMS:2023frs}
have reported violations of the charm fragmentation universality assumption in $pp$ collisions by up to an order of magnitude.
This shatters the typical inputs to the extrapolation, of which a specific example is the direct usage of charm fragmentation fractions, i.e. the frequency with which a charm quark fragments into specific charm-hadron final states, averaged over the full phase space. 

A pre-release version \cite{eps2023_pos,moriond2024_Achim} of the almost entirely data-driven
approach advocated in this work to mitigate this fragmentation non-universality issue, conceptually
applicable for any LHC $pp$ center-of-mass energy,
has so far been applied to $pp$ data at 5 and 13 TeV.
It has also already been used for a preliminary CMS measurement \cite{bph_c7TeV} at 7 TeV.
This approach is explained in detail in the following.
An alternative approach aiming at mitigating the same issue through the
assumption of a particular fragmentation model,
specifically tuned to 5 TeV $pp$ collisions used as a reference for heavy ion
collisions, has also recently been advocated \cite{totXsec_5TeV_LHC}.  

\begin{figure}
 \begin{center}
  \includegraphics[width=0.5\textwidth]{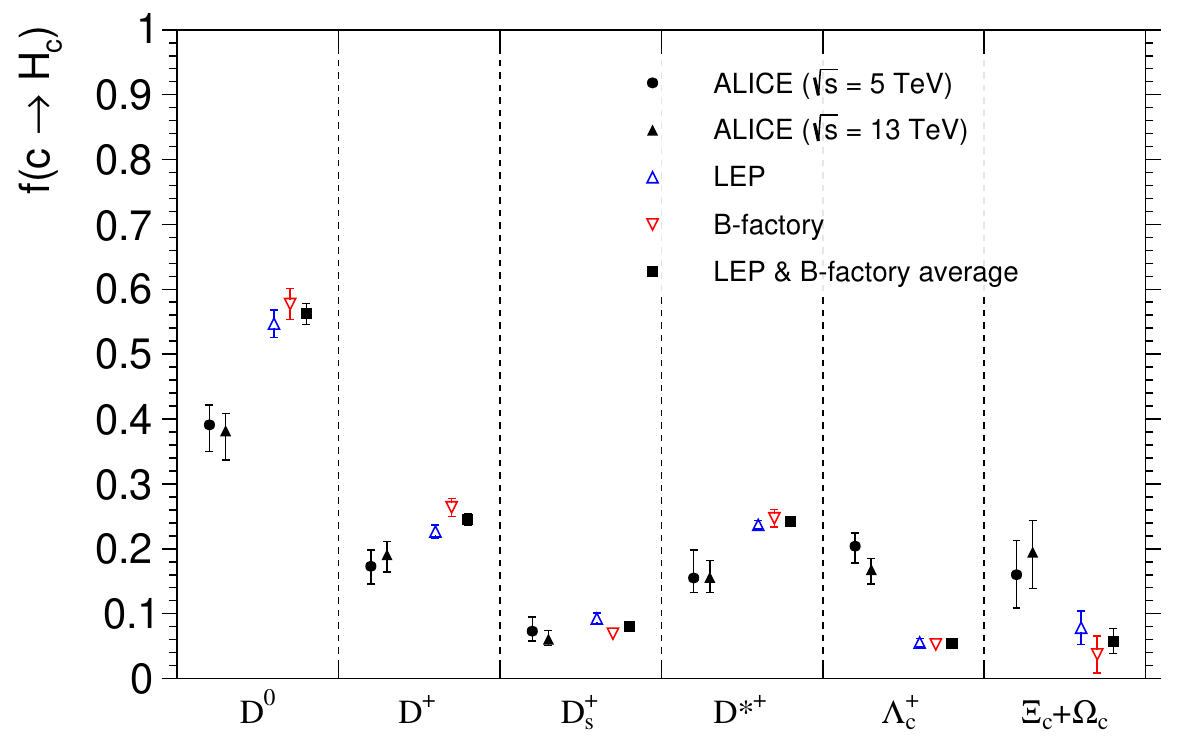}
 \end{center}
 \caption{Fragmentation fractions measured from the ALICE \cite{ALICE_cFragFrac_5TeV,ALICE_cFragFrac_13TeV} and $e^+e^-$ experiments \cite{fragfrac_comb}. The label $\Xi_c + \Omega_c$ indicates $\Xi_c^0 + \Xi_c^+ + \Omega_c^0$.
} \label{fig:fragfrac_ee_ep}
\end{figure}

The charm fragmentation fractions have been measured mostly from $e^+e^-$ or $ep$ collisions. No significant discrepancy has been reported between the measurements in $e^+e^-$ and $ep$ collisions (see e.g. \cite{fragfrac_ep}). 
Thus, although not theoretically required, in practice it has been assumed that the fragmentation is independent of the collision system, including $pp$ collisions. 
Recent reports from LHC experiments, specifically from ALICE \cite{ALICE_cFragFrac_5TeV, ALICE_cFragFrac_13TeV}, however, show large differences in the fragmentation fractions between $e^+e^-$/$ep$ and $pp$ collisions. A summary of these measurements has been compiled in Fig. \ref{fig:fragfrac_ee_ep}, with more details given in the next section.
Especially the overall $\Lambda_c^+$ fragmentation fraction shows a big discrepancy of $\sim 5\sigma$ between $pp$ and $e^+e^-/ep$ collisions, while the overall meson fractions in $pp$ collisions are smaller compared to the other collisions. 
This is strongly related to a clear dependence on transverse momentum ($p_T$) of the cross-section ratio $\Lambda_c^+$ \cite{ALICE_cFragFrac_13TeV, CMS_Lc_5TeV_update} (and $\Xi_c^0$ \cite{ALICE_cFragFrac_13TeV, ALICE_Xc_5TeV}) to $D^0$ observed in $pp$ collisions in the lower $p_T$ region, while it is asymptotically approaching the $e^+e^-$/$ep$ data at high $p_T$. 
The averaged effect of this over the full kinematic range leads to an increase of the total baryon fraction from about 10\% to about 30-40\%, and a corresponding decrease of the meson fraction from about 90\% to about 60-70\%.
On the other hand, none of the currently available theoretical fragmentation models are able to describe all the relevant data \cite{ALICE_qcdreview}. 

\begin{figure*}
 \begin{center}
  \includegraphics[width=0.45\textwidth]{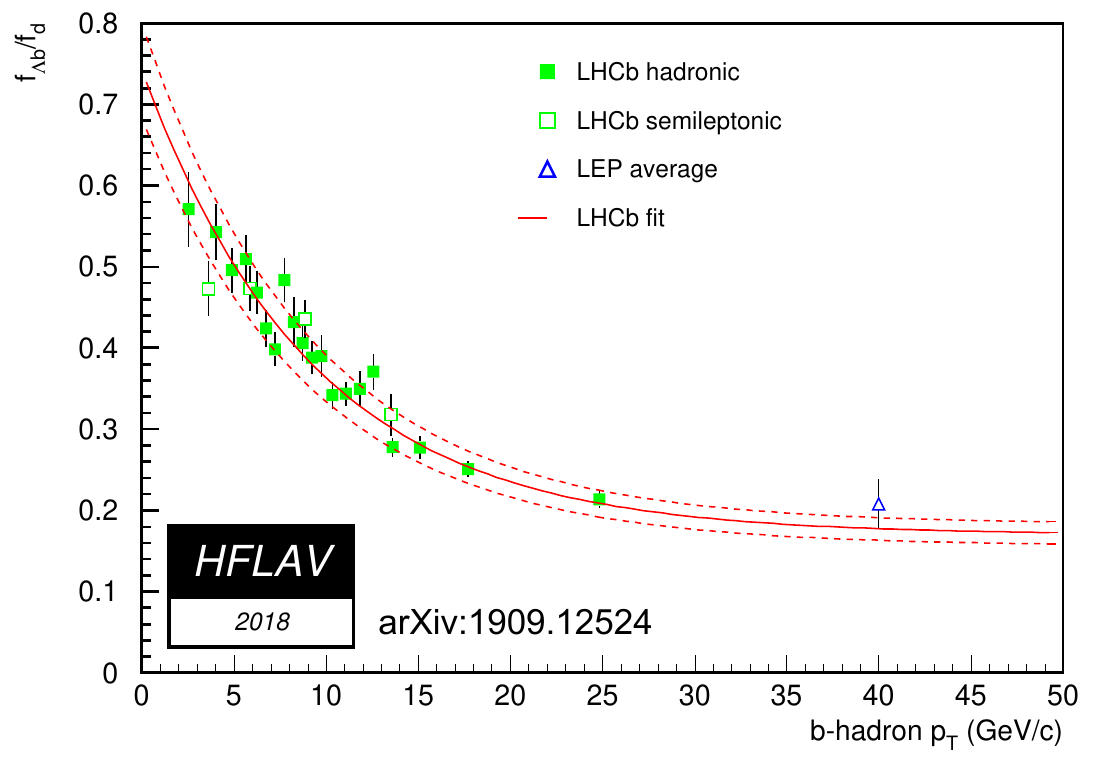}
  \includegraphics[width=0.43\textwidth]{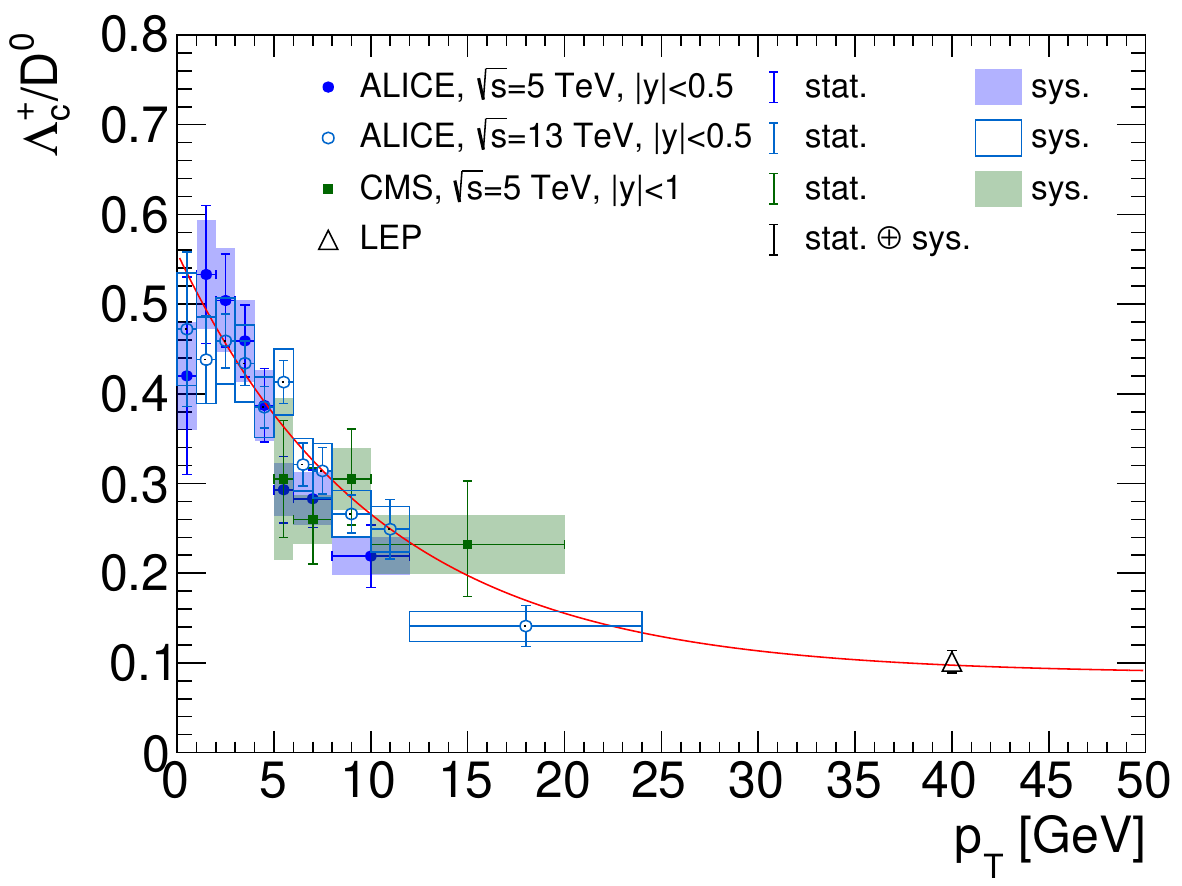}
 \end{center}
 \caption{Measurements of the ratio of $\Lambda_b^0$ to $B^0$ (left figure) and the ratio of $\Lambda_c^+$ to $D^0$ (right figure). The left figure is from \cite{LbToB0}, where the exponential fit to LHCb data gives consistent results to LEP at high $p_T$. In the right figure, the ALICE and CMS points are collected from \cite{ALICE_LcToD0_5TeV_update, ALICE_cFragFrac_13TeV} and \cite{CMS_Lc_5TeV}, respectively, and the LEP point is derived based on the numbers given in \cite{fragfrac_comb}. Apart from the LEP value, these points are placed at the center of each bin, i.e. not displaced to bin barycenters, for which additional information would be required. For details of the fit in the right figure see text. This fit is given only for illustration purposes to show qualitative consistency with what is observed from beauty production, using the same exponential functional form (with somewhat different parameters, see text).} \label{fig:ratio_MsToBy_LHCandLEP}
\end{figure*}

Similar phenomena were observed also from beauty-hadron production.
The production fractions have been measured for beauty, e.g. by some of the LEP, $B$-factory, and LHC experiments \cite{LbToB0}. Especially, it is observed that the $\Lambda_b^0/B^0$ measurements from the LHC, which show a clear $p_T$ dependence, are asymptotically consistent with the LEP measurement at high $p_T$ (the left panel of Figure \ref{fig:ratio_MsToBy_LHCandLEP}). 
In this figure a fit is provided by the HFLAV group \cite{LbToB0} for the LHCb \cite{LHCb:2011leg,LHCb:2014ofc} data using an exponential function, and the fit results agree well with the LEP average value \cite{LbToB0} positioned at an approximate $p_T$ as it occurs in $Z$ decays. 

In the right panel of Figure \ref{fig:ratio_MsToBy_LHCandLEP}, we collected $\Lambda_c^+$-to-$D^0$ ratio measurements from ALICE \cite{ALICE_LcToD0_5TeV_update, ALICE_cFragFrac_13TeV}, CMS \cite{CMS_Lc_5TeV} and LEP \cite{fragfrac_comb}.
% Lambda_b to B0
The LEP point is again added at the approximate $p_T$ in $Z$ decays.
A fit (the red curve) was performed to the ALICE data at $\sqrt{s} =$ 13 TeV including the LEP point, using an exponential function like the one used in the left figure.
The fit result is given here only for illustration purposes without uncertainties, and is not used further anywhere else in this report; rather, the measurements will be used directly.
The resulting parametrization is
$\Lambda_c/D^0 \sim 0.083 + \exp\big[-0.748 -0.095 \times p_T/{\rm GeV}\big]$,
to be compared to \cite{LbToB0} 
$\Lambda_b/B^0 \sim 0.151 + \exp\big[-0.57 -0.095 \times p_T/{\rm GeV}\big]$
for the beauty case. Note that the coefficient for the $p_T$ dependence turns out to be
similar in both cases.
In the charm case, the $\Lambda_c/D^0$ ratio in Fig. 2 changes from about 0.1 at high $p_T$ (consistent with $e^+e^-$) to about 0.5 at low $p_T$, i.e. a change by a factor 5.

%%Although not included in the fit, this also shows good agreement with the ALICE and CMS data at $\sqrt{s} =$ 5 TeV shown in Fig.~\ref{fig:ratio_MsToBy_LHCandLEP}.
Although not included in the fit, this parametrization also shows good agreement with the ALICE and CMS data at $\sqrt{s} =$ 5 TeV shown in Fig.~\ref{fig:ratio_MsToBy_LHCandLEP}.
Thus the heavy-flavour baryon-to-meson production ratios in $pp$ collisions agree well with an asymptotically flat $p_T$ dependence and with the LEP value at high $p_T$, both for charm and beauty production.

In order to properly treat this non-universal charm fragmentation, in Section \ref{sec:cHadProdFrac} we introduce $p_T$-dependent binned functions $\tilde f(p_T)$, the so-called \emph{$p_T$-dependent production fractions}, instead of using the fragmentation fractions $f^{uni}$ (defined to be independent of kinematics). These binned functions are derived based on the measured $\Lambda_c^+$-to-$D^0$ ratio as a function of $p_T$, such that we do not need to assume any particular non-universal fragmentation model.

These $p_T$-dependent production fractions are then applied to extrapolate the fiducial hadronic cross sections from LHC experiments.
For this, in Section \ref{sec:ddFONLL} we modify the FONLL \cite{fonll1, fonll2} theory calculations and introduce what we call the \emph{data-driven FONLL (ddFONLL)} approach, for which the only change in the theory parametrization is the empirical replacement of $f^{uni}$ by the measured $\tilde f(p_T)$. Furthermore, we phenomenologically fit all other parameters occurring in the calculation to available fiducial data, with a fully data-based uncertainty treatment. This is then obviously no longer a theory prediction, but a purely phenomenological parametrization starting from theory with all parameters adjusted to data.

Section \ref{sect:validation} is devoted to some validation checks of this scheme, and to the addition of some information needed for its application to non--groundstates like $D^*$ mesons. 
All the code and procedures are available in a public repository \cite{ddfonll}, such that
they can be used by third parties for other similar applications.

Eventually, total charm-pair cross-section measurements are presented at various center-of-mass energies ($\sqrt{s}$) at the LHC in Section \ref{sec:totXsec}. 
Providing these measurements allows for a comparison to NNLO theory. Furthermore, the total charm cross section can be predicted in perturbative QCD totally free from fragmentation inputs. Therefore, the measurements can be used to constrain QCD parameters like the charm-quark mass and the low-$x$ part of parton density functions (PDFs).
%(some PDF sets show still very large uncertainties at the low $x$, i.e., the large $\sqrt{s}$). 
The first qualitative examples of applying the results to constrain these QCD parameters
from total charm cross sections are presented in Section~\ref{sec:QCDsens}.

The results are then summarized in Section~\ref{sect:con}, together with an outlook on future applications.

\section{Charm hadron production fractions in \texorpdfstring{$pp$}{pp} collisions} \label{sec:cHadProdFrac}

In Figure \ref{fig:fragfrac_ee_ep}, the fragmentation fractions $f$ from $e^+e^-$ data were already compared to the ALICE measurements at $\sqrt{s} =$ 5 TeV \cite{ALICE_cFragFrac_5TeV} and 13 TeV \cite{ALICE_cFragFrac_13TeV}. For the latter, $f$ conceptually corresponds to the average of the function $\tilde f(p_T)$ over the full total cross-section phase space, weighted by the measured hadron-$p_T$ spectrum of the respective hadron differential cross section. This is possible since some of the ALICE measurements extend down to zero in $p_T$.
For the $e^+e^-$ measurements, consistent with each other (Fig. \ref{fig:fragfrac_ee_ep}) and with $ep$ measurements\footnote{For $ep$ measurements, a $p_T$ dependence could also be expected for the hadron-like resolved photon contribution to the charm cross section. However, this contribution is negligible for electroproduction (DIS), and only about 10\% for photoproduction \cite{hq_HERA}. Although presumably within measurement uncertainties, here we avoid further discussion by not including the $ep$ and $\gamma p$ measurements.} \cite{fragfrac_comb}, $f$ for the LEP and $B$-factory average corresponds to what we will refer to as $f^{uni}$. 
We now give some more details.

The $f(c\rightarrow \Xi_c + \Omega_c)$ (here $\Xi_c + \Omega_c$ indicates $\Xi_c^0 + \Xi_c^+ + \Omega_c^0$) of the 5 TeV ALICE data is twice the measured $f(c\rightarrow \Xi_c^0)$ to account for the additional $\Xi_c^+$ contribution, where the $\Omega_c^0$ contribution is already accounted for in the uncertainties as described in \cite{ALICE_cFragFrac_5TeV}.
Recently the 5 TeV ALICE results were updated in \cite{ALICE_cFragFrac_13TeV}, but the differences are not significant in the present context. Thus, here the numbers are still based on \cite{ALICE_cFragFrac_5TeV} (i.e., the numbers are the same as the ones used for the results presented in \cite{eps2023_pos}). 
The $f(c\rightarrow \Xi_c + \Omega_c)$ of the 13 TeV ALICE data\footnote{The fragmentation fractions of the 13 TeV ALICE data \cite{ALICE_cFragFrac_13TeV} were measured by also including the $J/\psi$ contribution. However, this contribution is less than 1\%, which is neglected in this work.} is given by the sum of the measured $f(c\rightarrow \Xi_c^0)$ and $f(c\rightarrow \Xi_c^+)$, where the $\Omega_c^0$ contribution is accounted for in the uncertainties as described in \cite{ALICE_cFragFrac_13TeV}. For simplicity, the $f(c\rightarrow \Xi_c + \Omega_c)$ uncertainties of the 5 and 13 TeV ALICE data were calculated by assuming that the measurement uncertainties are fully uncorrelated.

\begin{figure}
 \begin{center}
  \includegraphics[width=0.5\textwidth]{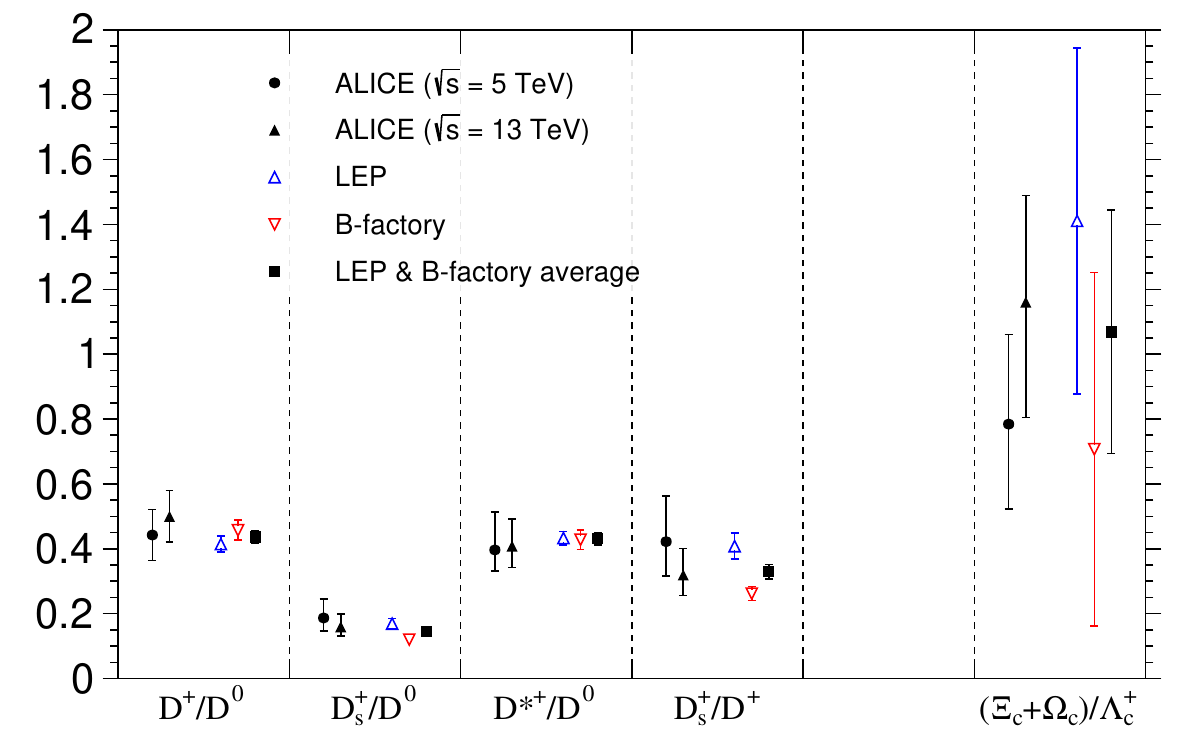}
 \end{center}
 \caption{Comparison of meson-to-$D^0$, $D_s^+$-to-$D^+$ and baryon-to-$\Lambda_c^+$ ratios of fragmentation fractions between $e^+e^-$ and $pp$ collisions. The label $\Xi_c + \Omega_c$ indicates $\Xi_c^0 + \Xi_c^+ + \Omega_c^0$. The uncertainties were derived under the simplifying assumption that the measurement uncertainties are fully uncorrelated.} \label{fig:ms_by_ratios}
\end{figure}

Based on the fragmentation fractions shown in Figure \ref{fig:fragfrac_ee_ep}, the ratios of meson to $D^0$ and baryon to $\Lambda_c^+$ are derived and shown in Figure \ref{fig:ms_by_ratios}. 
The uncertainties were calculated again assuming that all the initial measurement uncertainties are fully uncorrelated.

Interestingly and very importantly, these ratios show consistency with the assumption that the charm meson-to-meson and baryon-to-baryon ratios remain essentially universal for all collision systems.
%(Assumption \ref{ast:nonUni1}),
Direct LHC measurements \cite{ALICE_cFragFrac_13TeV} indeed confirm that there is no significant $p_T$ dependence of these ratios within uncertainties.
An uncertainty for a possible non-negligible $p_T$-dependence of $D_s/D$ will be derived
and applied later.

The direct evidence for the rapidity (in)dependence of these ratios is less
clear. There are however LHCb measurements for both the meson-to-meson and
baryon-to-meson ratios for beauty production
that indicate independence of rapidity \cite{LHCb_Bfrac_13TeV}.
The assumption that this is also true for charm will be validated later
by the very good fit to fiducial charm data based on this assumption,
over a wide rapidity range. 

Since the LHC measurements show consistency with LEP data at high $p_T$,
which in turn show consistency with all other $e^+e^-$ data even at low $p_T$,
the LEP and $B$-factory average is taken to be the asymptotic value $f^{uni}$
at high $p_T$,
with a $p_T$ dependent correction introduced for the lower $p_T$ region in
$pp$ collisions.

As a starting point, motivated by the considerations above, our approach follows the most simple assumption, that the meson-to-meson and baryon-to-baryon ratios are still universal, i.e., independent of either the collision system or kinematics.
%since no measurements show significant evidence contradicts to this assumption (see more later).
Then, under this data-motivated assumption, the $p_T$-dependence of the baryon-to-meson ratio in $pp$ collisions at LHC can be defined by a general form.
%It is also assumed that the charm fragmentation is independent of (pseudo-)rapidity (refer to e.g. \cite{LHCb_Bfrac_13TeV}).
In other words, we follow the assumptions:
\begin{assumption} \label{ast:nonUni1}
 meson-to-meson and baryon-to-baryon ratios are universal, i.e., independent of kinematics and the collision system,
\end{assumption}
and
\begin{assumption} \label{ast:nonUni2}
 baryon-to-meson ratios are dependent on transverse momentum, while independent of (pseudo-)rapidity.
\end{assumption}
Both of these assumptions will get uncertainty treatments that will allow
deviations from them within measured uncertainties.

In this report, $p_T$-dependent production fractions ($\tilde{f}(p_T)$) are defined for $pp$ collisions as cross-section fractions of each hadron state relative to the sum of all the weakly-decaying ground states ($w.d.$) as a function of $p_T$:
\begin{equation} \label{eq:def_ftilde}
 \tilde{f}_{H_c}(p_T) = \frac{d\sigma_{H_c}}{\Sigma_{w.d.} d\sigma_{H_c}},
\end{equation}
where $d\sigma$ is the $p_T$-differential cross-section and $p_T$ is the transverse momentum of each hadron. The weakly-decaying ground states\footnote{Here multi-charm states are neglected.} are taken to be the mesons ($MS$) $D^0$, $D^+$, $D_s^+$, and the baryons ($BY$) $\Lambda_c^+$, $\Xi_c^0$, $\Xi_c^+$ and $\Omega_c^0$. 

To derive $\tilde{f}(p_T)$, so-called $p_T$-dependent factors ($F(p_T)$) are applied to the fragmentation fractions of $e^+e^-$ collisions ($f^{uni}$):
\begin{equation} \label{eq:fTilde_D0}
 \tilde{f}_{D^0}(p_T) = f_{D^0}^{uni}F_{MS}(p_T),
\end{equation}
\begin{equation}
 \tilde{f}_{D^+}(p_T) = f_{D^+}^{uni}F_{MS}(p_T),
\end{equation}
\begin{equation}
 \tilde{f}_{D_s^+}(p_T) = f_{D_s^+}^{uni}F_{MS}(p_T),
\end{equation}
\begin{equation} \label{eq:fTilde_Lc}
 \tilde{f}_{\Lambda_c^+}(p_T) = f_{\Lambda_c^+}^{uni}F_{BY}(p_T),
\end{equation}
\begin{equation}
 \tilde{f}_{\Xi_c^0}(p_T) = f_{\Xi_c^0}^{uni}F_{BY}(p_T),
\end{equation}
\begin{equation}
 \tilde{f}_{\Xi_c^+}(p_T) = f_{\Xi_c^+}^{uni}F_{BY}(p_T),
\end{equation}
and
\begin{equation}
 \tilde{f}_{\Omega_c^0}(p_T) = f_{\Omega_c^0}^{uni}F_{BY}(p_T),
\end{equation}
where the same factors $F_{MS}(p_T)$ and $F_{BY}(p_T)$ are applied to each meson and baryon state, respectively, making use of the assumptions stated above. By definition, the sum of the production fractions for all the weakly-decaying ground states is unity:
\begin{equation} \label{eq:fragFracSum}
 f_{MS}^{uni} F_{MS}(p_T) + f_{BY}^{uni} F_{BY}(p_T) = 1
\end{equation}
where $f_{MS}^{uni}$ is the sum of all meson fractions:
\begin{equation} \label{eq:funi_ms}
 f_{MS}^{uni} = f_{D^0}^{uni} + f_{D^+}^{uni} + f_{D_s^+}^{uni},
\end{equation}
and $f_{BY}^{uni}$ is the sum of all baryon fractions:
\begin{equation} \label{eq:funi_by}
 f_{BY}^{uni} = f_{\Lambda_c^+}^{uni} + f_{\Xi_c^0}^{uni} + f_{\Xi_c^+}^{uni} + f_{\Omega_c^0}^{uni}.
\end{equation}
Then the relation between meson and baryon modifiers can be given by
\begin{equation}
 F_{BY}(p_T) = \frac{1 - f_{MS}^{uni} F_{MS}(p_T)}{f_{BY}^{uni}}.
\end{equation}

To determine $F_{MS}(p_T)$ and $F_{BY}(p_T)$, the most precise measurements of the ratio of baryon to meson are to be taken. Currently, those are the measurements of the ratio of $\Lambda_c^+$ to $D^0$. Denoting the $p_T$-dependent cross-section ratio of $\Lambda_c^+$ to $D^0$ in Fig. \ref{fig:ratio_MsToBy_LHCandLEP} by $R(p_T)$, a relation between $R(p_T)$ and $\tilde{f}(p_T)$ can be given as
\begin{align}
 R(p_T) \equiv \frac{\tilde{f}_{\Lambda_c^+}(p_T)}{\tilde{f}_{D^0}(p_T)} &= \frac{f_{\Lambda_c^+}^{uni} F_{BY}(p_T)}{f_{D^0}^{uni} F_{MS}(p_T)} \\ \nonumber
 &= C\bigg(\frac{1}{f_{MS}^{uni} F_{MS}(p_T)} - 1 \bigg),
\end{align}
where the constant term $C$ is defined by:
\begin{equation}
 C \equiv \frac{f_{\Lambda_c}^{uni}}{f_{D^0}^{uni}}\frac{f_{MS}^{uni}}{f_{BY}^{uni}}.
\end{equation}
As a result, $F_{MS}(p_T)$ and $F_{BY}(p_T)$ can be determined by the fragmentation fractions measured in $e^+e^-$ collisions and $R(p_T)$:
\begin{equation} \label{eq:Fms}
 F_{MS}(p_T) = \frac{1}{f_{MS}^{uni}}\frac{C}{R(p_T) + C}
\end{equation}
and
\begin{equation} \label{eq:Fby}
 F_{BY}(p_T) = \frac{1}{f_{BY}^{uni}}\bigg(1 - \frac{C}{R(p_T) + C}\bigg).
\end{equation}

Therefore, the $p_T$-dependent production fractions can be described e.g., for $D^0$ and $\Lambda_c^+$ as
\begin{equation} \label{eq:pTD0frac}
 \tilde{f}_{D^0}(p_T) = \frac{f_{D^0}^{uni}}{f_{MS}^{uni}} \times \frac{C}{R(p_T) + C}
\end{equation}
and
\begin{equation} \label{eq:pTLcfrac}
 \tilde{f}_{\Lambda_c^+}(p_T) = \frac{f_{\Lambda_c^+}^{uni}}{f_{BY}^{uni}} \times \bigg(1 - \frac{C}{R(p_T) + C}\bigg)
\end{equation}
by inserting Eq.(\ref{eq:Fms}) into Eq.(\ref{eq:fTilde_D0}) and Eq.(\ref{eq:Fby}) into Eq.(\ref{eq:fTilde_Lc}), respectively.

For this report, the $f^{uni}$s were extracted from $e^+e^-$ data.
Then the fragmentation fraction sum of all the other states not yet measured ($\Xi_c^0$, $\Xi_c^+$ and $\Omega_c^0$) in $e^+e^-$ collisions is assumed to be
%\begin{linenomath}
\begin{align} \label{eq:fragFracXO_ee}
 &f(c\rightarrow \Xi_c^0) + f(c\rightarrow \Xi_c^+) + f(c\rightarrow \Omega_c^0) \equiv 1 \\ \nonumber
 &- [f(c\rightarrow D^0) + f(c\rightarrow D^+) + f(c\rightarrow D_s^+) + f(c\rightarrow \Lambda_c^+)]
\end{align}
%\end{linenomat}
so that the sum of all the known weakly-decaying ground states is unity. 
The fragmentation fractions measured in $e^+e^-$ collisions were
taken\footnote{As motivated and described in \cite{Yewonthesis}, the fragmentation fractions based on the precisely known $e^+e^-$ charm cross sections were used to normalize the baryon fractions.} from \cite{fragfrac_comb}. 
The collected $e^+e^-$ fragmentation fractions for $D^0$, $D^+$, $D_s^+$ and $\Lambda_c^+$ (including $D^{*+}$) can be found in Table~\ref{tb:fragfrac_ee}.
\begin{table}
\begin{center}
\caption{Fragmentation fractions from $e^+e^-$ data extracted from \cite{fragfrac_comb}.}
\renewcommand{\arraystretch}{1.5}
\resizebox{0.5\textwidth}{!}{
\begin{tabular}{|c|c|c|c|}
\hline
$f(c\rightarrow H_c)$ & LEP & B-factory & $e^+e^-$ averaged \\
\hline
$D^0$ & $0.547\pm0.022$ & $0.577\pm0.024$ & $0.562\pm0.016$ \\
\hline
$D^+$ & $0.227\pm0.010$ & $0.264\pm0.014$ & $0.245\pm0.009$ \\
\hline
$D_s^+$ & $0.093\pm0.008$ & $0.069\pm0.005$ & $0.081\pm0.005$ \\
\hline
$D^{*+}$ & $0.237\pm0.006$ & $0.247\pm0.014$ & $0.242\pm0.008$ \\
\hline
$\Lambda_c^+$ & $0.056\pm0.007$ & $0.053\pm0.003$ & $0.054\pm0.004$ \\
\hline
$\Xi_c^0 + \Xi_c^+ + \Omega_c^0$ & $0.078\pm0.026$ & $0.037\pm0.028$ & $0.058\pm0.019$ \\
\hline
\end{tabular}
}
%The fractions of $\Xi_c^0 + \Xi_c^+ + \Omega_c^0$ were derived so that the sum of fragmentation fractions of all the weakly-decaying ground states is unity by assuming that measurement uncertainties are fully uncorrelated.}
\label{tb:fragfrac_ee}
\end{center}
\end{table}
In deriving the fractions of the others ($\Xi_c^0$, $\Xi_c^+$ and $\Omega_c^0$) using Eq.(\ref{eq:fragFracXO_ee}), the uncertainty was calculated under the assumption that all the measured fraction uncertainties are fully uncorrelated.

The ratios of $\Lambda_c^+$ to $D^0$ to be used for $R(p_T)$ were collected from the measurements as a function of $p_T$ at $\sqrt{s} = 5$~TeV from ALICE \cite{ALICE_LcToD0_5TeV_update} and CMS \cite{CMS_Lc_5TeV}, and at $\sqrt{s} =$ 13 TeV from ALICE \cite{ALICE_cFragFrac_13TeV}, as shown in Fig.~\ref{fig:ratio_MsToBy_LHCandLEP}. As an asymptotic value at high $p_T$, the $e^+e^-$ averaged numbers shown in Table~\ref{tb:fragfrac_ee} were used.

In the case of 5~TeV, since the ALICE measurements are more precise at lower $p_T$, by default the ALICE points were used if applicable, otherwise the CMS points were taken. The $R(p_T)$ values are explicitly written in Table~\ref{tab:r_alice5TeV}. The values in the range $0 < p_T < 8$~GeV were taken from the ALICE measurements as they are and the value of $8 < p_T < 10$~GeV was taken from the ALICE measurement of $8 < p_T < 12$~GeV. For $p_T > 10$~GeV, the CMS measurement in the range $10 < p_T < 20$~GeV was taken for $10 < p_T < 20$~GeV and the averaged $e^+e^-$ value was used as the $p_T > 20$~GeV point. For the extrapolation which will be introduced in the next section, the two values were combined to give an overflow bin $p_T > 10$~GeV by applying weights determined based on the FONLL predictions. 
Tentative uncertainties for these weights are negligible compared to the measurement uncertainties in the end. The statistical and systematic uncertainties of the ALICE and CMS measurements were summed in quadrature.
\begin{table}
 {\small
 \begin{center}
  \caption{$R(p_T)$ at $\sqrt{s} =$ 5 TeV. The third column of $p_T > 10$ GeV was derived by the sum of the two values in the second column applying weights determined based on the FONLL predictions.} \label{tab:r_alice5TeV}
  \renewcommand{\arraystretch}{1.5}
  \resizebox{0.5\textwidth}{!}{
  \begin{tabular}{|c|c|c|}
   \hline
   [GeV] & \multicolumn{2}{c|}{$R(p_T)$} \\
   \hline
   $0 < p_T < 1$ & \multicolumn{2}{c|}{$0.420 + 0.125 - 0.125$} \\
   \hline
   $1 < p_T < 2$ & \multicolumn{2}{c|}{$0.533 + 0.098 - 0.098$} \\
   \hline
   $2 < p_T < 3$ & \multicolumn{2}{c|}{$0.504 + 0.078 - 0.077$} \\
   \hline
   $3 < p_T < 4$ & \multicolumn{2}{c|}{$0.459 + 0.061 - 0.061$} \\
   \hline
   $4 < p_T < 5$ & \multicolumn{2}{c|}{$0.387 + 0.057 - 0.057$} \\
   \hline
   $5 < p_T < 6$ & \multicolumn{2}{c|}{$0.293 + 0.048 - 0.047$} \\
   \hline
   $6 < p_T < 8$ & \multicolumn{2}{c|}{$0.283 + 0.044 - 0.043$} \\
   \hline
   $8 < p_T < 10$ & \multicolumn{2}{c|}{$0.219 + 0.041 - 0.041$} \\
   \hline
   $10 < p_T < 20$ & $0.232 + 0.078 - 0.067$ & \multirow{2}{*}{$0.223 + 0.074 - 0.063$} \\
   \cline{1-2}
   $p_T > 20$ & $0.096 + 0.007 - 0.007$ & \\
   \hline
  \end{tabular}
  }
 \end{center}
 } %small
\end{table}

Similarly, $R(p_T)$ at $\sqrt{s} =$ 13 TeV is shown in Table~\ref{tab:r_alice13TeV}. 
\begin{table}
 {\small
 \begin{center}
  \caption{$R(p_T)$ at $\sqrt{s} =$ 13 TeV. The third column of $p_T > 10$ GeV was derived by the sum of the three values in the second column applying weights determined based on the FONLL predictions.} \label{tab:r_alice13TeV}
  \renewcommand{\arraystretch}{1.5}
  \resizebox{0.5\textwidth}{!}{
  \begin{tabular}{|c|c|c|}
   \hline
   [GeV] & \multicolumn{2}{c|}{$R(p_T)$} \\
   \hline
   $0 < p_T < 1$ & \multicolumn{2}{c|}{$0.472 + 0.106 - 0.106$} \\
   \hline
   $1 < p_T < 2$ & \multicolumn{2}{c|}{$0.438 + 0.068 - 0.069$} \\
   \hline
   $2 < p_T < 3$ & \multicolumn{2}{c|}{$0.459 + 0.056 - 0.056$} \\
   \hline
   $3 < p_T < 4$ & \multicolumn{2}{c|}{$0.434 + 0.050 - 0.050$} \\
   \hline
   $4 < p_T < 5$ & \multicolumn{2}{c|}{$0.385 + 0.041 - 0.041$} \\
   \hline
   $5 < p_T < 6$ & \multicolumn{2}{c|}{$0.413 + 0.044 - 0.044$} \\
   \hline
   $6 < p_T < 7$ & \multicolumn{2}{c|}{$0.321 + 0.038 - 0.038$} \\
   \hline
   $7 < p_T < 8$ & \multicolumn{2}{c|}{$0.314 + 0.040 - 0.040$} \\
   \hline
   $8 < p_T < 10$ & \multicolumn{2}{c|}{$0.266 + 0.033 - 0.033$} \\
   \hline
   $10 < p_T < 12$ & $0.249 + 0.042 - 0.042$ & \multirow{3}{*}{$0.189 + 0.033 - 0.034$} \\
   \cline{1-2}
   $12 < p_T < 24$ & $0.141 + 0.028 - 0.029$ & \\
   \cline{1-2}
   $p_T > 24$ & $0.096 + 0.007 - 0.007$ & \\
   \hline
  \end{tabular}
  }
 \end{center}
 } %small
\end{table}
All the points except the overflow bin were collected directly from the ALICE measurements. The overflow bin is given by a combined point of the ALICE measurements in the range $10 < p_T < 24$~GeV and the $e^+e^-$ point defined for $p_T > 24$~GeV, applying again weights determined based on the FONLL prediction.

\begin{figure}
 \begin{center}
  \includegraphics[width=0.4\textwidth]{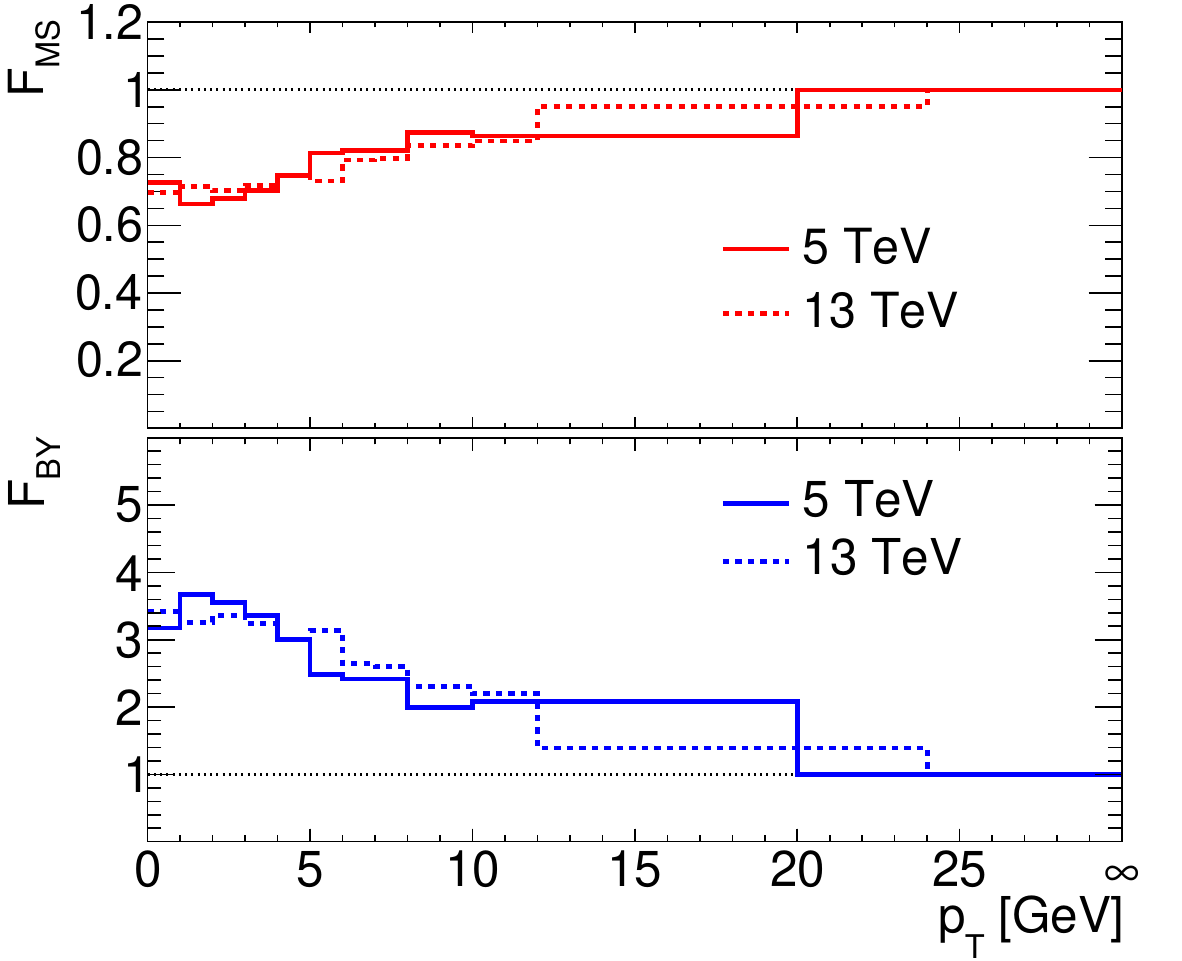}
 \end{center}
 \caption{$F_{MS}(p_T)$ (red histogram) and $F_{BY}(p_T)$ (blue histogram) at $\sqrt{s} =$ 5 (solid line) and 13 (dashed line) TeV. These are asymptotically close to 1 at high $p_T$ by definition, Eq.(\ref{eq:fragFracSum}).} \label{fig:Fms_Fby}
\end{figure}

\begin{figure}
 \begin{center}
  \includegraphics[width=0.35\textwidth]{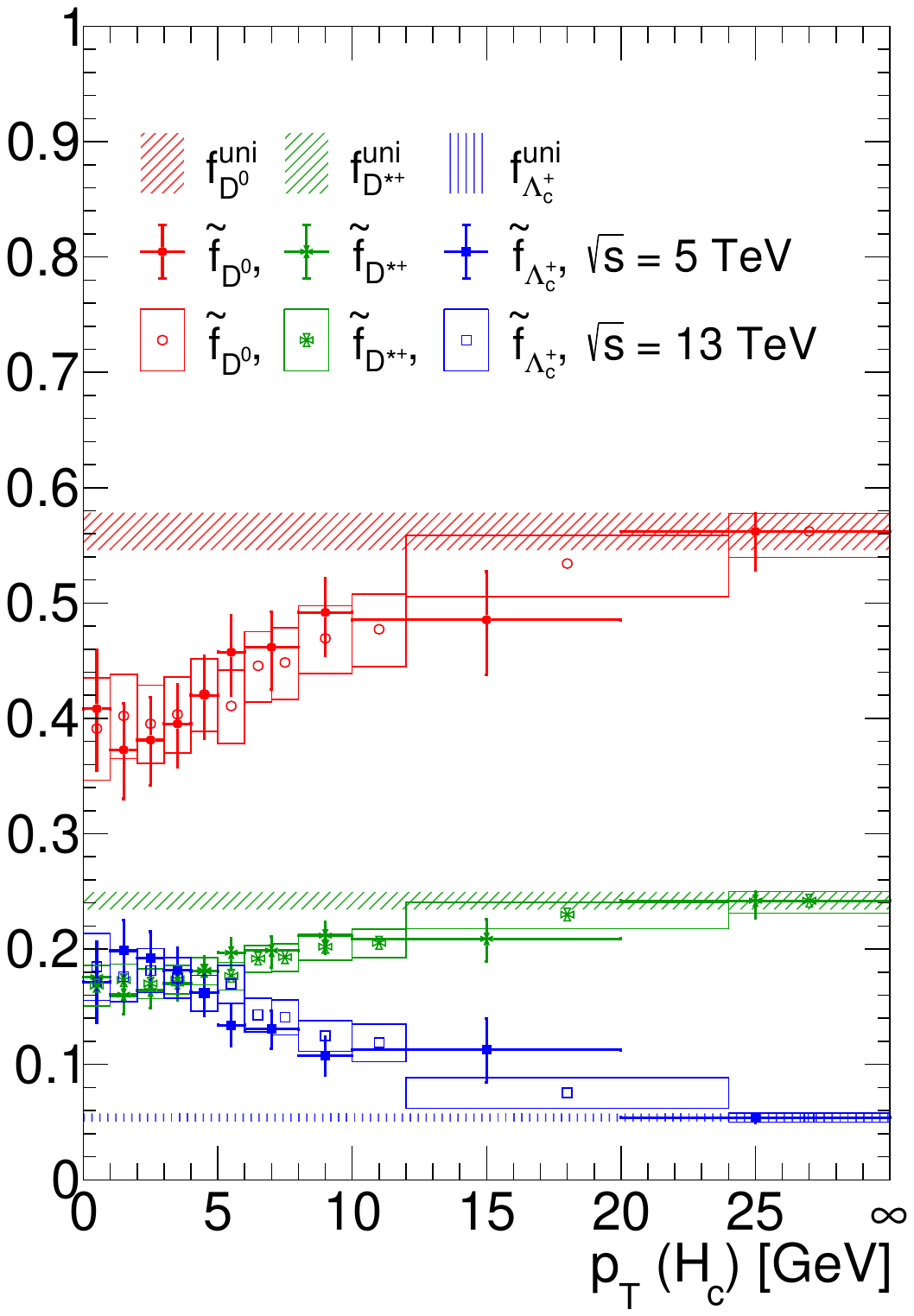}
 \end{center}
 \caption{$\tilde{f}_{D^0}(p_T)$, $\tilde{f}_{\Lambda_c^+}(p_T)$ and $\tilde{f}_{D^{*+}}(p_T)$. The red, blue and green band are the $D^0$, $\Lambda_c^+$ and $D^{*+}$ fragmentation fractions with the uncertainties, respectively, measured from $e^+e^-$ collisions.} \label{fig:fTilde}
\end{figure}

Using the numbers in Table~\ref{tab:r_alice5TeV} or Table~\ref{tab:r_alice13TeV} as $R(p_T)$ and the averaged numbers in Table~\ref{tb:fragfrac_ee} as $f^{uni}$s, $F_{MS}(p_T)$ and $F_{BY}(p_T)$ were derived with Eq.(\ref{eq:Fms}) and Eq.(\ref{eq:Fby}) and are shown in Figure \ref{fig:Fms_Fby}. Since the ratios of $\Lambda_c^+$ to $D^0$ are asymptotically identical to the ratio in $e^+e^-$ collisions at high $p_T$ by construction, the quantities $F_{MS}(p_T)$ and $F_{BY}(p_T)$ are asymptotically unity at high $p_T$ by their definition, Eq.(\ref{eq:fragFracSum}).
Lastly, $\tilde{f}_{D^0}(p_T)$ and $\tilde{f}_{\Lambda_c^+}(p_T)$ were derived by Eqs.(\ref{eq:fTilde_D0}) and (\ref{eq:fTilde_Lc}), for which results are shown in Figure \ref{fig:fTilde}.

\begin{figure*}
 \begin{subfigure}[b]{1.\linewidth}
  \centering
  \includegraphics[width=0.6\textwidth]{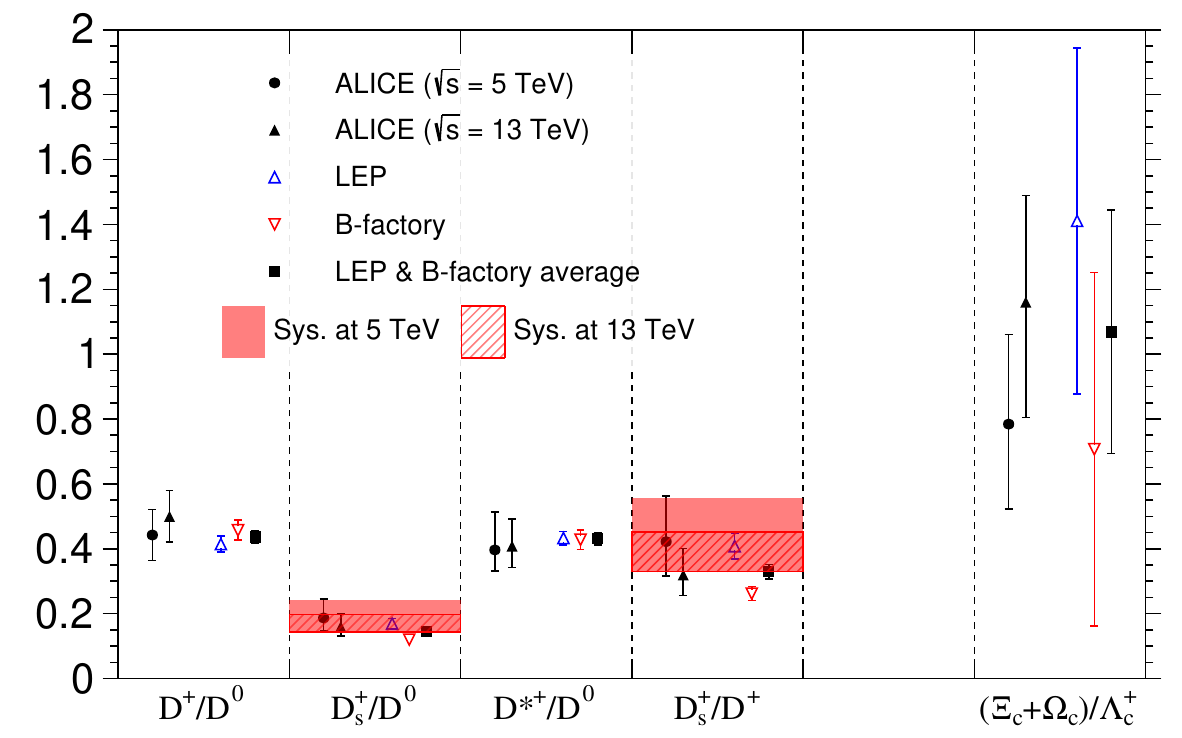}
  \caption{Meson-to-$D^0$ and baryon-to-$\Lambda_c^+$ ratios of the fragmentation fractions.} \label{subfig:sysUnc_fRat}
 \end{subfigure} \\[0.6cm]
 \begin{subfigure}[b]{1.\linewidth}
  \centering
  \includegraphics[width=0.75\textwidth]{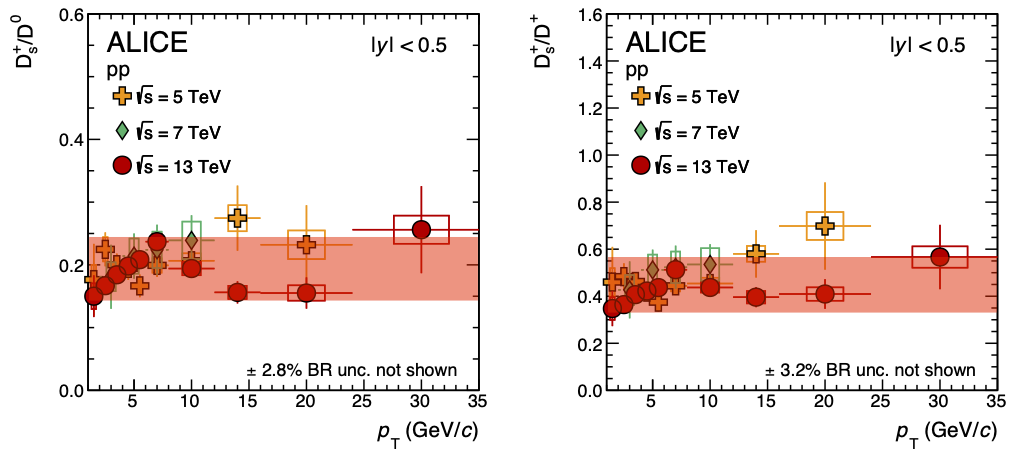}
  \caption{Ratios of $D_s^+$ to $D^0$ (left) and $D^+$ (right), with figures adapted from \cite{ALICE_cFragFrac_13TeV}. The red bands refer to the systematic uncertainties assigned at $\sqrt{s} =$ 5 TeV.} \label{subfig:sysUnc_DsToD}
 \end{subfigure}
 \caption{Additional uncertainty (red bands) assigned to account for a possible $p_T$ dependence of $D_s^+/D^0$ and $D_s^+/D^+$.} \label{fig:sysUnc_ftilde}
\end{figure*}

For the $\tilde{f}$ uncertainties, an additional systematic uncertainty was assigned to account for deviations from the assumption that meson-to-meson and baryon-to-baryon ratios are consistent between $e^+e^-$ and $pp$ collisions and independent of kinematics. Precise measurements in beauty production from LHC experiments \cite{CMS_Bmesons_13TeV} show that $B_s^0/B^+$ has a moderate but clear $p_T$ dependence at low $p_T$, and is asymptotically flat at high $p_T$. 
No precise measurement is available to show such a clear $p_T$ dependence yet for $D$ mesons (see \cite{ALICE_cFragFrac_13TeV}). Therefore, an additional uncertainty was assigned to account for a possible $p_T$ dependence of the $D_s^+/D^0$ and $D_s^+/D^+$ ratios, by covering the ALICE uncertainties of $0.14 < D_s^+/D^0 < 0.24$ and $0.33 < D_s^+/D^+ < 0.56$, as shown in Figure \ref{subfig:sysUnc_fRat}. This uncertainty covers well also the ratios measured as a function of $p_T$ (see Figure \ref{subfig:sysUnc_DsToD}).
Since no statistically significant deviations from the assumption of rapidity
independence were observed in the later fits to charm data in this work,
intermittently covering rapidities from 0 to 4.5, nor in the
more continuous rapidity coverage of the preliminary results in \cite{bph_c7TeV},
no further systematic uncertainties are assigned for this. 

In addition to $\tilde{f}$ functions for the ground states, $\tilde{f}$ for $D^{*+}$ was also derived in order to allow the extrapolation of CMS $D^{*+}$ measurements \cite{bph_c7TeV}. As will be explained in the next section, the same $F_{MS} (p_T)$ can be applied for $D^{*+}$, and $\tilde{f}_{D^{*+}}$ is defined as
\begin{equation} \label{eq:ftilde_dstar}
 \tilde{f}_{D^{*+}}(p_T) = f_{D^{*+}}^{uni} F_{MS}(p_T),
\end{equation}
which is shown by the green points in Figure \ref{fig:fTilde}.

\section{Data-driven FONLL} \label{sec:ddFONLL}

For the total charm-quark production cross section, charm-hadron measurements in a constrained kinematic range should be extrapolated (and interpolated, depending on the available measurements) to the full kinematic range, and converted from hadron level to quark level. For the approach advocated in this report, the FONLL perturbative theory is taken as the starting point for the parametrization
of the extrapolation function to be used for this purpose, in particular for
its functional shape in unmeasured regions.
This theory was chosen since it provides the highest
order double differential charm cross-section predictions for $pp$ as a function of transverse momentum ($p_T$) and rapidity ($y$) available to date, to order NLO+NLL. 

In the original FONLL approach \cite{fonll1, fonll2} 
%With the non-perturbative fragmentation function and fraction,
the double-differential (in $p_T$ and $y$) single-inclusive (the other charm hadron from a $c\bar c$ pair is integrated over) cross section for the production
of a particular charm hadron $H_c$ is parametrized as
\begin{equation} \label{eq:dhqhXsec}
 d\sigma_{H_c}^{FONLL} = f_{H_c}^{uni} \cdot (d\sigma_c \otimes D_{c\rightarrow H_c}^{\text{NP}}), \quad d\sigma_c = f_i f_j \otimes d\hat{\sigma}_{ij}, 
\end{equation}
where $f_{H_c}^{uni}$ is the `universal' fragmentation fraction previously extracted mainly from $e^+e^-$ data as explained in the previous section. $d\sigma_c$ is the differential cross section for charm-quark production $d\hat\sigma_{ij}$ folded with the parton density functions (PDFs) $f_i$ at perturbative order NLO+NLL according to the QCD factorization theorem. $D_{c\rightarrow H_c}^{\text{NP}}$ is the non-perturbative fragmentation function (distribution of the fraction of the charm-quark momentum transferred to the charm hadron) which is factorized out from $d\sigma_c$, again using the QCD factorization theorem. The calculation is done in the so-called general-mass heavy-flavour scheme, i.e. the charm-quark mass is appropriately accounted for in the NLO matrix elements.
The convolution symbol $\otimes$ indicates that distributions will be convoluted
during the cross-section integration, while the $\cdot$ symbol indicates scalar
multiplication. The cross section $\Delta \sigma$ integrated over a bin ($\Delta p_T,\Delta y$) in
hadron $p_T$ and $y$ is thus given by
\begin{equation} \label{eq:hqhXsec}
 \Delta\sigma_{H_c}^{FONLL}(\Delta p_T, \Delta y) = f_{H_c}^{uni} \cdot \int_{\Delta p_T,\Delta y} d\sigma_c \otimes D_{c\rightarrow H_c}^{\text{NP}} dp_T dy. 
\end{equation}
The QCD theory parameters entering this calculation are the strong coupling
constant $\alpha_s$, the charm-quark pole mass $m_c$, as well as the related QCD
renormalization and factorization scales $\mu_r$ and $\mu_f$. The central
reference scale is defined to be $\mu_0=\sqrt{m_c^2 + p_{Tc}^2}$. Further
parameters arise from the measurement-based parametrizations of the PDFs
and of the fragmentation function $D^{NP}$. The value of $\alpha_s(m_Z)$
is chosen and fixed to be consistent with the one of the PDF used (see below),
and evolved to the renormalization scale within the FONLL code. 

To phenomenologically account for the non-universal charm fragmentation, the only formal change to the theory parametrization, central to our new approach, is the replacement of the universal fragmentation fraction $f^{uni}$ by the binned $p_T$-dependent hadron production fraction $\tilde{f}$ derived in the previous section. I.e. Eq. \eqref{eq:hqhXsec} is modified to:
\begin{align} \label{eq:modFonll}
 \Delta\sigma_{H_c}^{\text{ddFONLL}}&(\Delta p_T, \Delta y) \\ \nonumber
 &\equiv \tilde{f}_{H_c}(p_T) \cdot \int_{\Delta p_T,\Delta y} d\sigma_{c} \otimes D_{c \rightarrow H_c}^{NP} dp_Tdy.
\end{align}
For reasons explained below, we call this the {\sl data driven FONLL} (ddFONLL) approach.
This relies on an empirical parametrization of experimentally measured cross sections without assuming any particular factorization or fragmentation breaking model in theory.

Note that this parametrization asymptotically converges to the original FONLL
one at high charm transverse momenta, through the definition of $\tilde f$
detailed in the previous section.
The theory treatment therefore remains consistent e.g. with the one
for high-$p_T$ charm jets, without our modification, i.e. many of the
previous high-$p_T$ charm-jet results obtained from LHC data (not treated here)
may remain unmodified based on the result of this work. Also, the theory
remains fully consistent with previous studies of $e^+e^-$ data, for which
$\tilde f_{e^+e^-} \equiv f^{uni}$ by definition.
However, all previous charm cross section extrapolations may be modified
through the procedure presented in this work. 

A priori there is no theoretical reason why the non-universal $p_T$ dependence
of $\tilde f$ in $pp$ collisions should factorize from a potential non-universal $p_T$ dependence
of $D^{NP}$ in Eq.~(\ref{eq:modFonll}). Nevertheless, this ansatz has been chosen here for three reasons.
One is practical on the experimental side: the
available data do not have enough precision to allow a potential unfolding
of the two effects. The second is practical from the theory side: The FONLL core
calculation does not foresee a nonperturbative variation of the fragmentation
function beyond the choice of the quark and hadron types for the respective
parametrization. The third and most
important one is phenomenological: We eventually find very good agreement
with all data in the $\sqrt{s}$ range 5 TeV - 13 TeV within uncertainties
(also see next section).
However, in order to allow more parametric freedom at this point, while we
choose the Kartvelishvili parametrization of the fragmentation function $D^{NP}$,
we do {\sl not} apply any external constraint on the corresponding
Kartvelishvili parameter $\alpha_K$, e.g. from $e^+e^-$ collisions, as it is
usually done.
Rather, in our approach, this parameter will only be constrained from the
$pp$ data themselves. We then only verify that the resulting parameter comes
out in a `reasonable' range when compared to $e^+e^-$ values.  

For the reasons outlined in the introduction, the a priori uncertainties
of the FONLL core theory for charm production are very large compared to
the measurement uncertainties of the available data. They are dominated by
QCD scale and low-$x$ PDF uncertainties.
Since our approach is phenomenological, for the purpose of the total charm
cross-section extrapolations, we will thus use data to
empirically constrain {\sl all} the parameters mentioned above, either from
external data constraints (PDFs, $\alpha_s$, $\tilde f$) or from a direct
fit to the data which we would like to extrapolate. The procedure also includes a full
data-driven treatment of all corresponding uncertainties.
This will greatly reduce the extrapolation uncertainties relative to an a-priori
theory-driven treatment, which is anyway no longer possible once the assumption
of charm-fragmentation universality is abandoned.  
From the description of this approach it is obvious that $ddFONLL$ is no
longer to be treated as
a theory prediction, but as a phenomenological parametrization relevant for
the purpose for which it was designed.

The approach is thus to get the best possible description of the data in the
regions in which they are measured, and to determine uncertainties such that
a reliable extrapolation into unmeasured regions becomes possible. 

In order to implement this approach, a $\chi^2$ scan was introduced for four of the QCD parameters; the two theory scales ($\mu_f$ and $\mu_r$), the charm mass ($m_c$) and the $\alpha_K$. In other words, the parameters describing data best are determined by a $\chi^2$ calculation defined by 
\begin{equation} \label{eq:chi2_calc_extrap}
 \sum_{\text{data bins}} \frac{(\text{ddFONLL}-\text{data})^2}{\text{statistical unc.}^2 + \text{systematic unc.}^2}.
\end{equation}
The four-dimensional minimum of this scan then defines the
ddFONLL central values, while appropriately chosen multi-dimensional
$\chi^2$ contours define their uncertainties.
Because this procedure is equivalent to a fit, we also sometimes refer to it as
the ddFONLL fit.
Other uncertainties are added externally as described below.

Since the FONLL calculation \cite{fonll1,fonll2} uses the FONLL general mass variable flavour number scheme (FONLL GM-VFNS), the ideal PDF for this work would be the VFNS version of the PROSA PDF \cite{prosa2019}. This set was co-fitted to ALICE and LHCb charm data and derived with a low-$x$ gluon parametrization of the rapidity dependence (only) of LHC charm data in different regions of $p_T$. Thus this PDF is {\sl not} affected by the non-universality of charm fragmentation. Unfortunately, however, only the central value for this PDF is available, while uncertainties are available only for the 3-flavour fixed-flavour (FFNS) version \cite{prosa2019}. Fortunately, it turns out that the older VFNS CTEQ6.6 PDF \cite{cteq} happens to be consistent with the PROSA PDF for both VFNS central value and FFNS uncertainty. We thus pragmatically use this PDF as a proxy for the PROSA\_VFNS PDF with uncertainties. In the application of PDF sets to this extrapolation via LHAPDF \cite{lhapdf}, the starting scale $Q_{min}$ is defined to be 1.3 GeV \cite{lhapdf_cteq66} for the CTEQ6.6 PDF set. Therefore, we excluded values of $\mu_f < 1.3$ GeV in the $\chi^2$ scan for the phase space down to $p_T = 0$ GeV with $m_c = 1.3$ GeV (the minimum $m_c$ considered in the extrapolation).
The associated value of the strong coupling constant is $\alpha_s(M_Z)=0.118$.

The {ddFONLL} fit was applied to $D^0$ measurements at $\sqrt{s} =$ 5 and 13 TeV in $pp$ collisions, in the same way as in an earlier preliminary evaluation \cite{eps2023_pos, moriond2024_Achim}. The $D^0$ measurements at $\sqrt{s} =$ 5 TeV were obtained from the ALICE~\cite{ALICE_Dmesons_5TeV}, CMS~\cite{CMS_D0_5TeV} and LHCb~\cite{LHCb_Dmesons_5TeV} experiments, of which the kinematic ranges covered are listed in Table \ref{tb:d05TeV_exps}.
\begin{table}
 \begin{center}
  \caption{The kinematic ranges covered by the ALICE~\cite{ALICE_Dmesons_5TeV}, CMS~\cite{CMS_D0_5TeV} and LHCb~\cite{LHCb_Dmesons_5TeV} experiments for $D^0$ measurements at $\sqrt{s} =$ 5 TeV.} \label{tb:d05TeV_exps}
  \renewcommand{\arraystretch}{1.5}
  \begin{tabular}{|c|c|c|}
   \hline
   ALICE & $|y| < 0.5$ & $0 < p_T < 36$ GeV \\
   \hline
    CMS  & $|y| < 1.0$ & $2 < p_T < 100$ GeV \\
   \hline 
         & $2.0 < y < 2.5$ & $0 < p_T < 10$ GeV \\
         & $2.5 < y < 3.0$ & $0 < p_T < 10$ GeV \\
    LHCb & $3.0 < y < 3.5$ & $0 < p_T < 10$ GeV \\
         & $3.5 < y < 4.0$ & $0 < p_T < 9$ GeV \\
         & $4.0 < y < 4.5$ & $0 < p_T < 6$ GeV \\
   \hline
  \end{tabular}
 \end{center}
\end{table}
However the ALICE and CMS measurements have overlapping cross-sections within $|y| < 1$. For the integrated cross section over $|y| < 1$, the CMS measurement covers $\sim 40\%$ while the ALICE measurement covers $\sim 50\%$ with much better precision. Furthermore, the contribution of the CMS measurement in the range $36 < p_T < 100$ GeV is negligible for the total charm cross section. In other words, the ALICE measurement alone already covers the maximum of the cross section in the range $|y| < 1$. Therefore, only the ALICE and LHCb measurements were considered in this fit.
The $D^0$ measurements at $\sqrt{s} =$ 13 TeV were taken from ALICE~\cite{ALICE_cFragFrac_13TeV} and LHCb~\cite{LHCb_Dmesons_13TeV}.
For the 13 TeV data, the kinematic ranges covered by each experiment are listed in Table \ref{tb:d013TeV_exps}. 
% 13 TeV
\begin{table}
\begin{center}
\caption{The kinematic ranges covered by ALICE~\cite{ALICE_cFragFrac_13TeV} and LHCb~\cite{LHCb_Dmesons_13TeV} for $D^0$ measurements at $\sqrt{s} =$ 13 TeV.} \label{tb:d013TeV_exps}
\renewcommand{\arraystretch}{1.5}
\begin{tabular}{|c|c|c|}
\hline
ALICE & $|y| < 0.5$ & $p_T > 0$ GeV \\
\hline
     & $2.0 < y < 2.5$ & $0 < p_T < 15$ GeV \\
     & $2.5 < y < 3.0$ & $0 < p_T < 15$ GeV \\
LHCb & $3.0 < y < 3.5$ & $0 < p_T < 15$ GeV \\
     & $3.5 < y < 4.0$ & $0 < p_T < 11$ GeV \\
     & $4.0 < y < 4.5$ & $0 < p_T < 7$ GeV \\
\hline
\end{tabular}
\end{center}
\end{table}
A detailed set of multi-dimensional $\chi^2$ tables can be found elsewhere \cite{Yewonthesis}.

\begin{figure*}
 \begin{center}
  \includegraphics[width=0.19\textwidth]{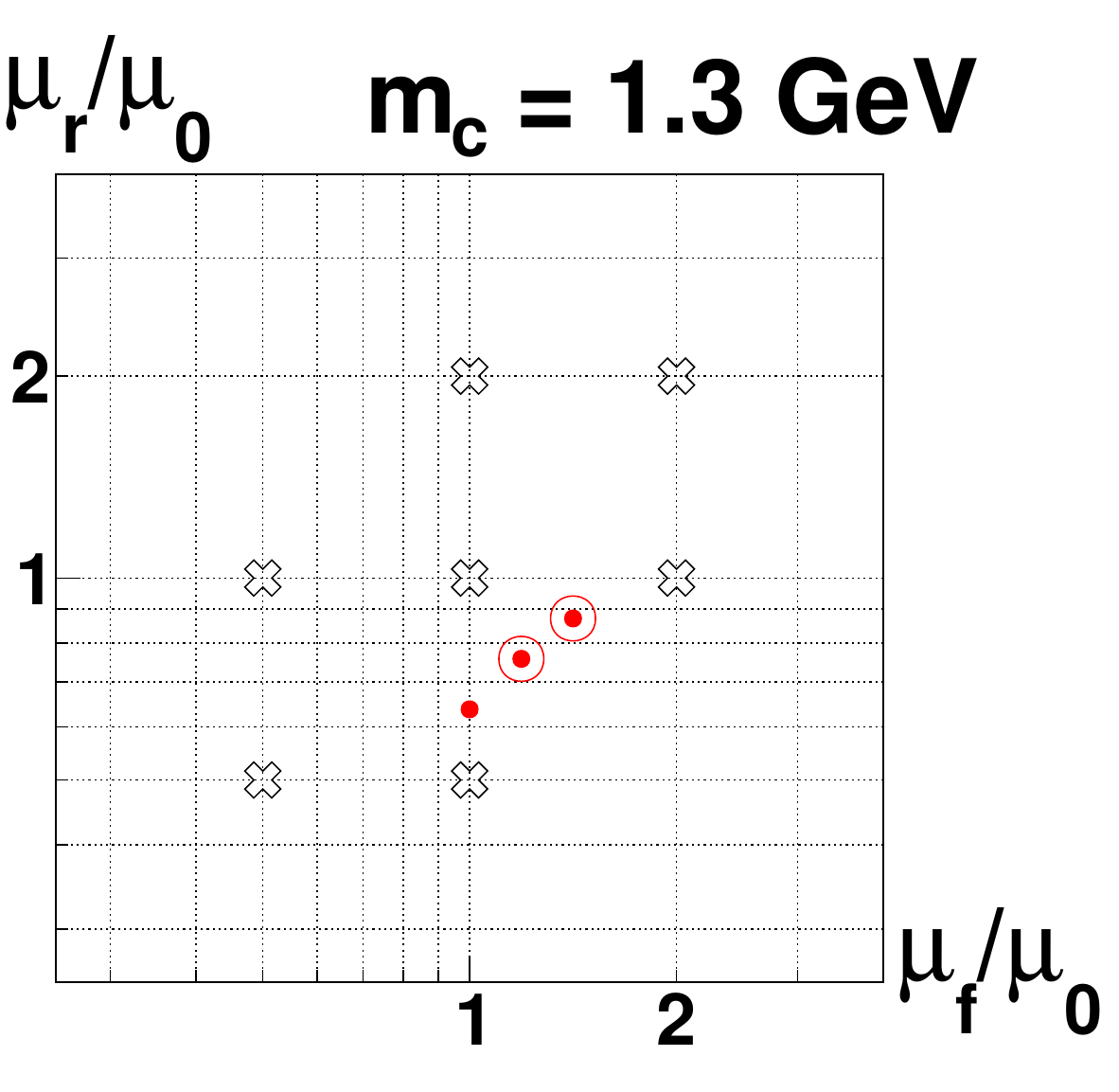} 
  \includegraphics[width=0.19\textwidth]{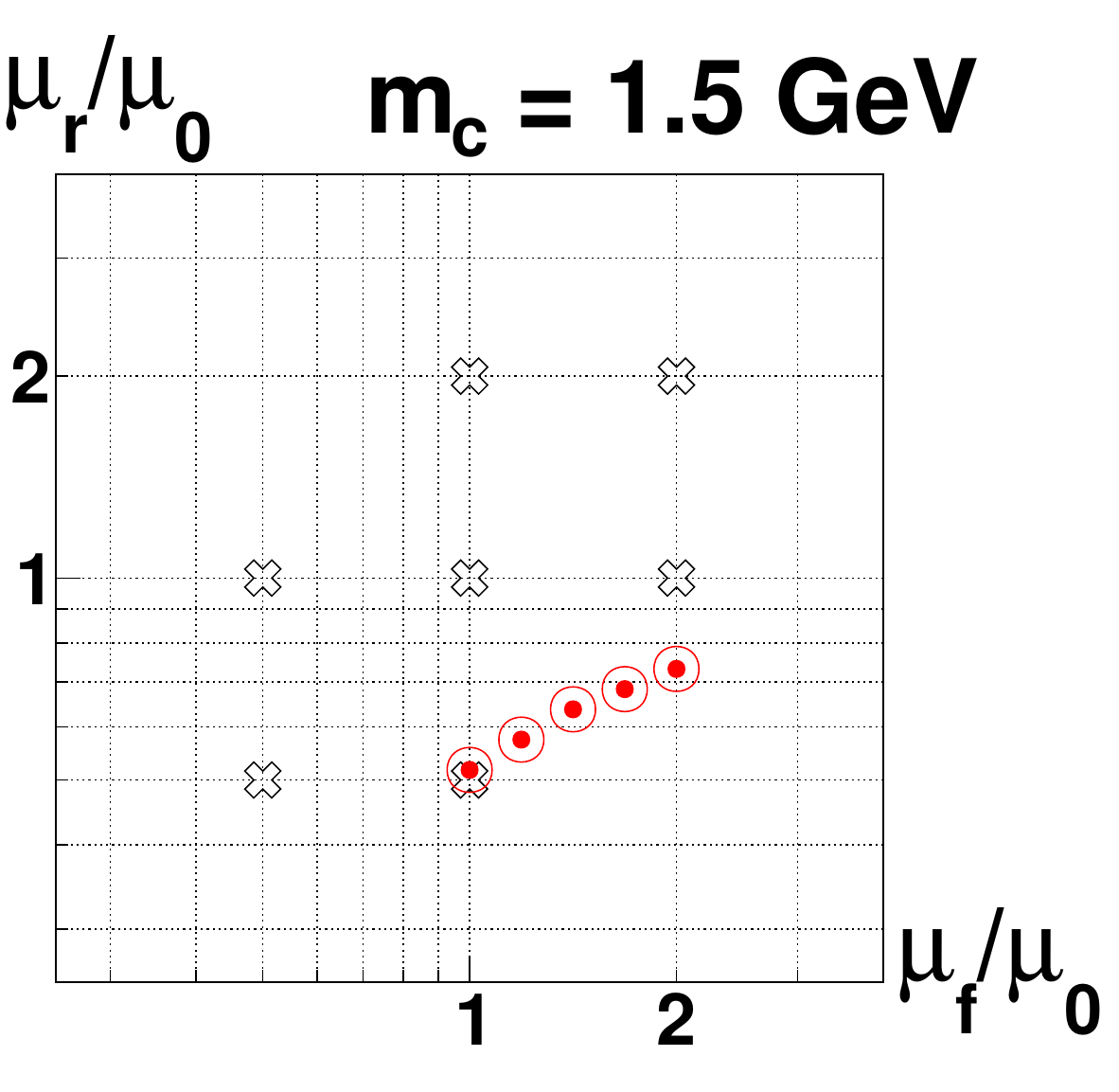} 
  \includegraphics[width=0.19\textwidth]{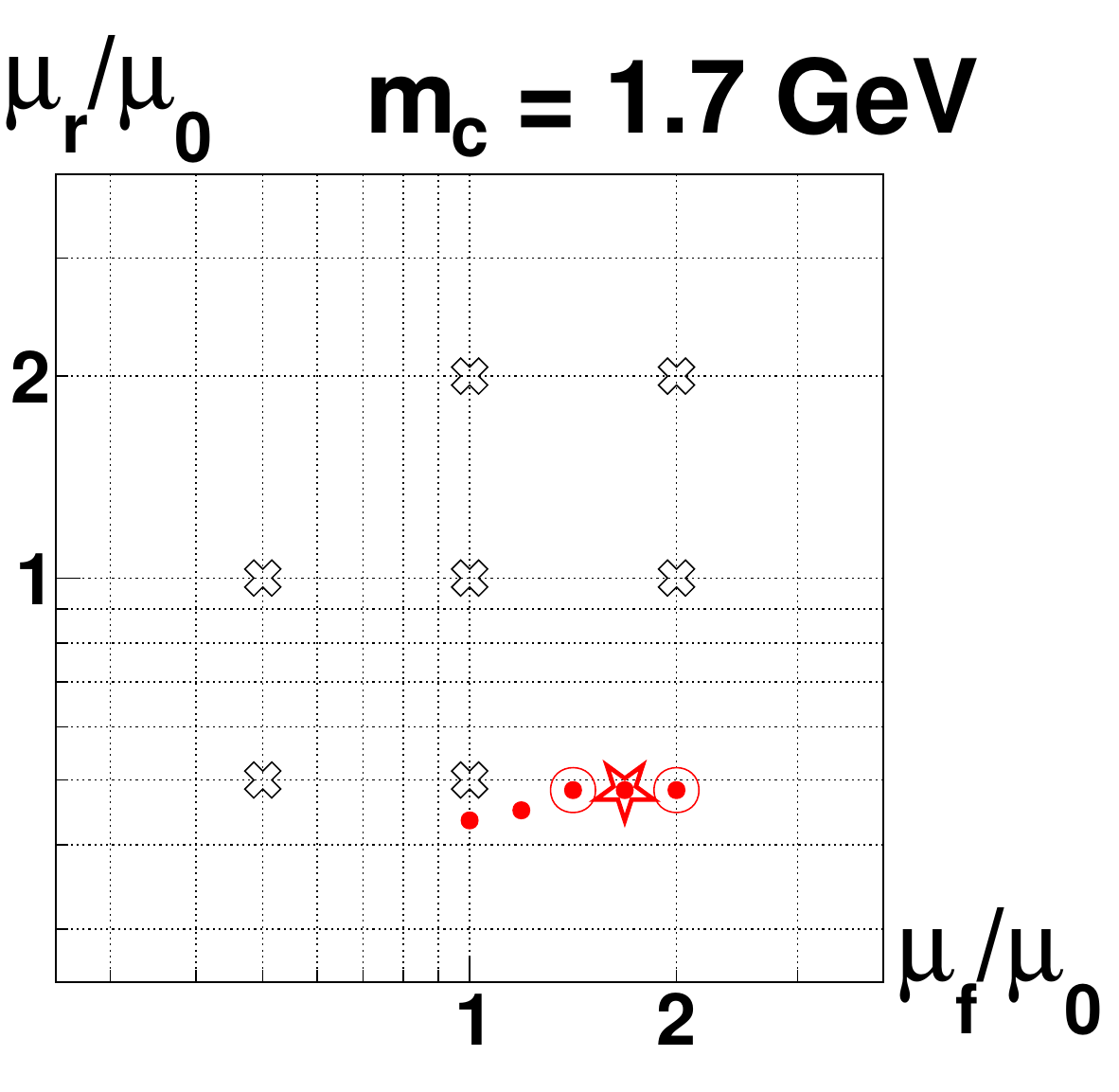} 
  \includegraphics[width=0.19\textwidth]{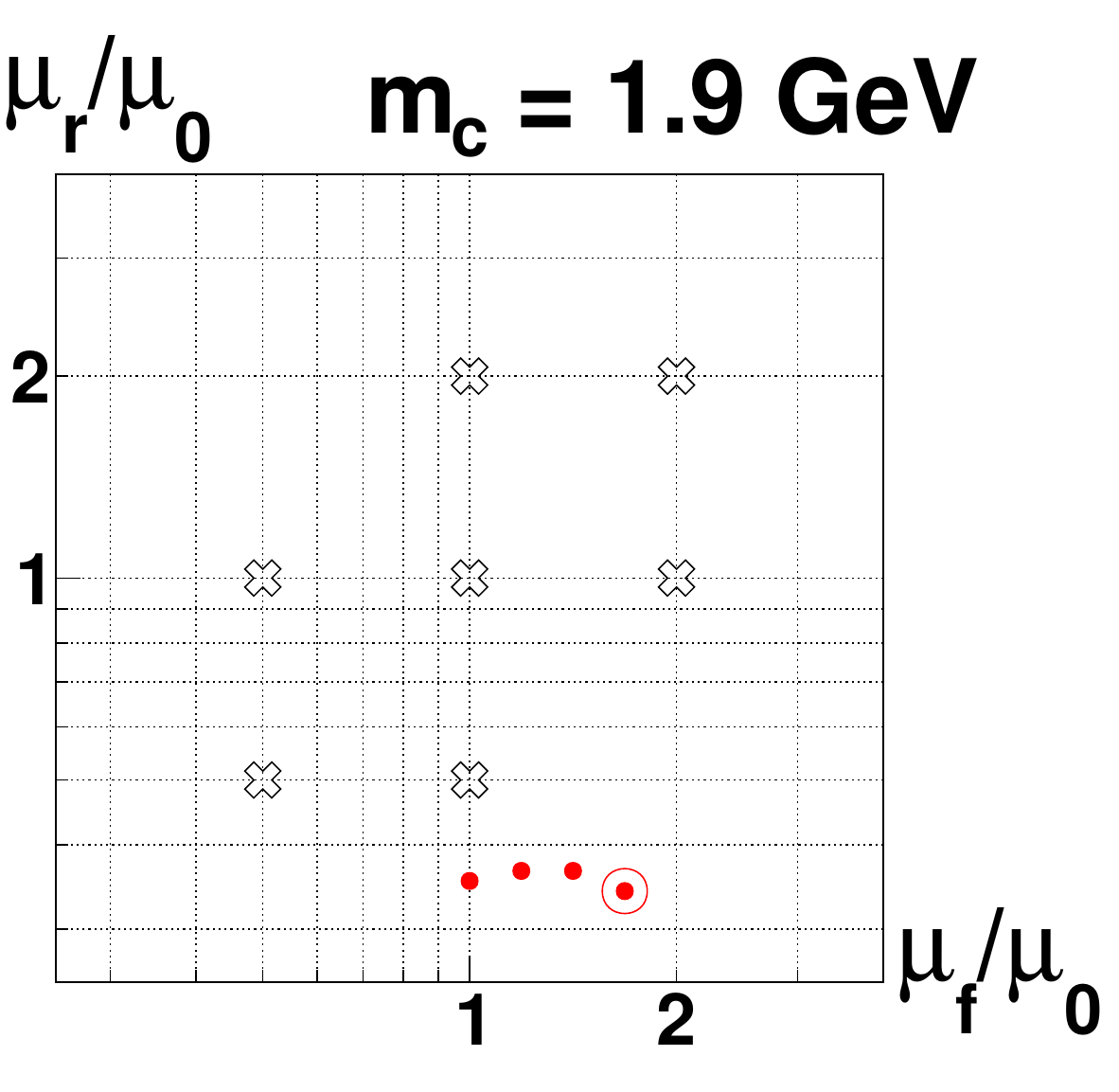} 
  \includegraphics[width=0.19\textwidth]{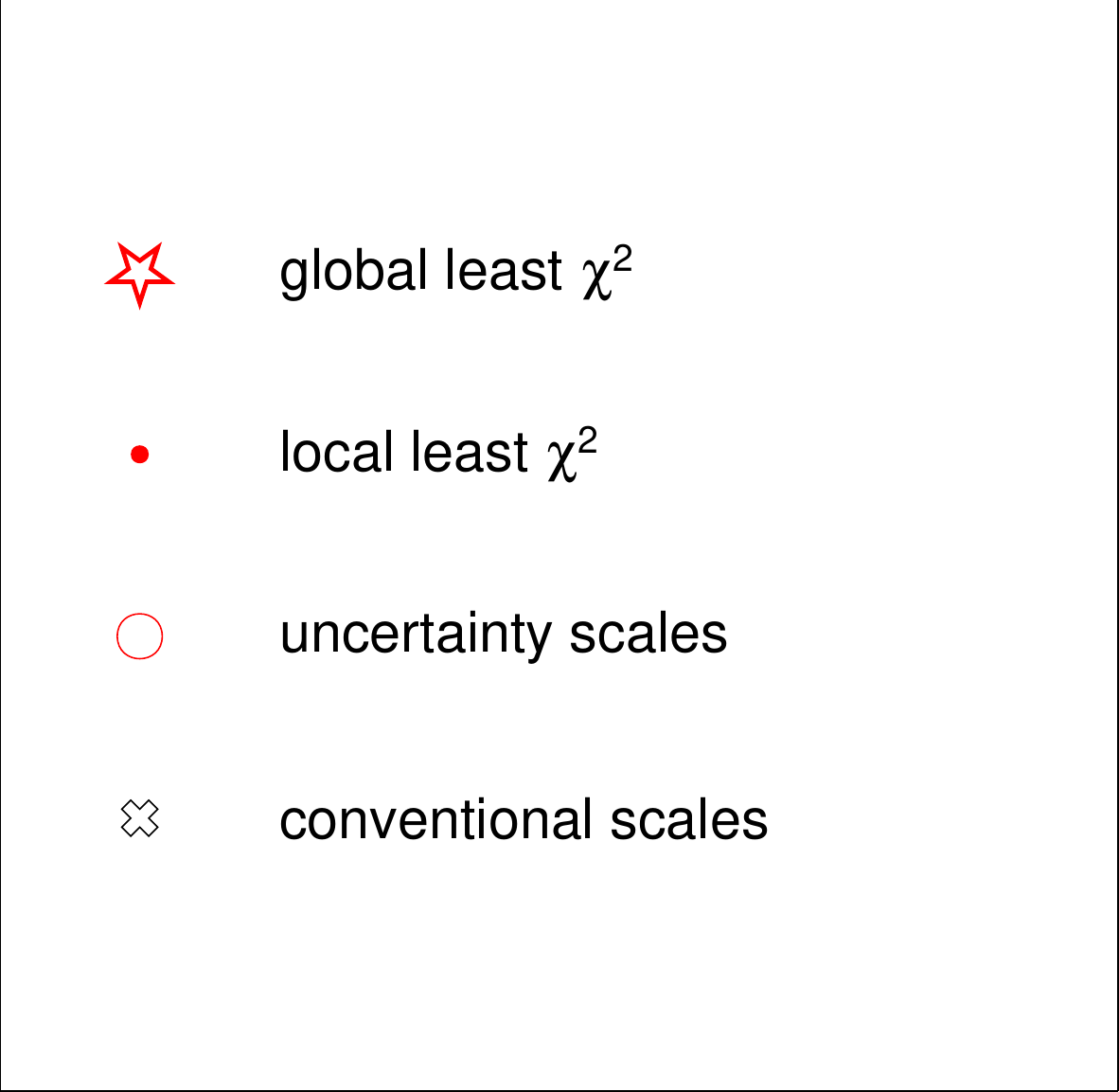} \\  
  \includegraphics[width=0.19\textwidth]{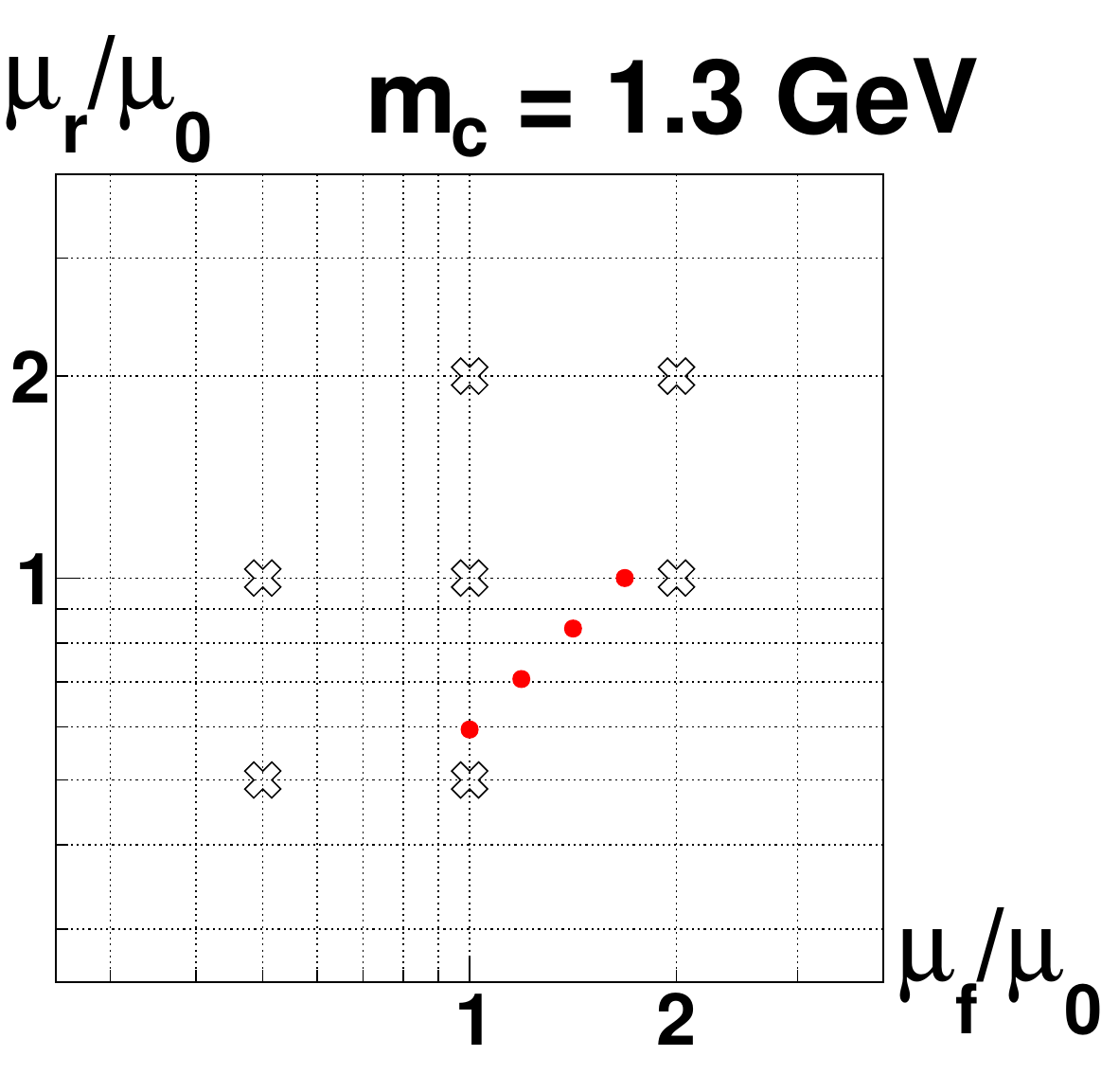} 
  \includegraphics[width=0.19\textwidth]{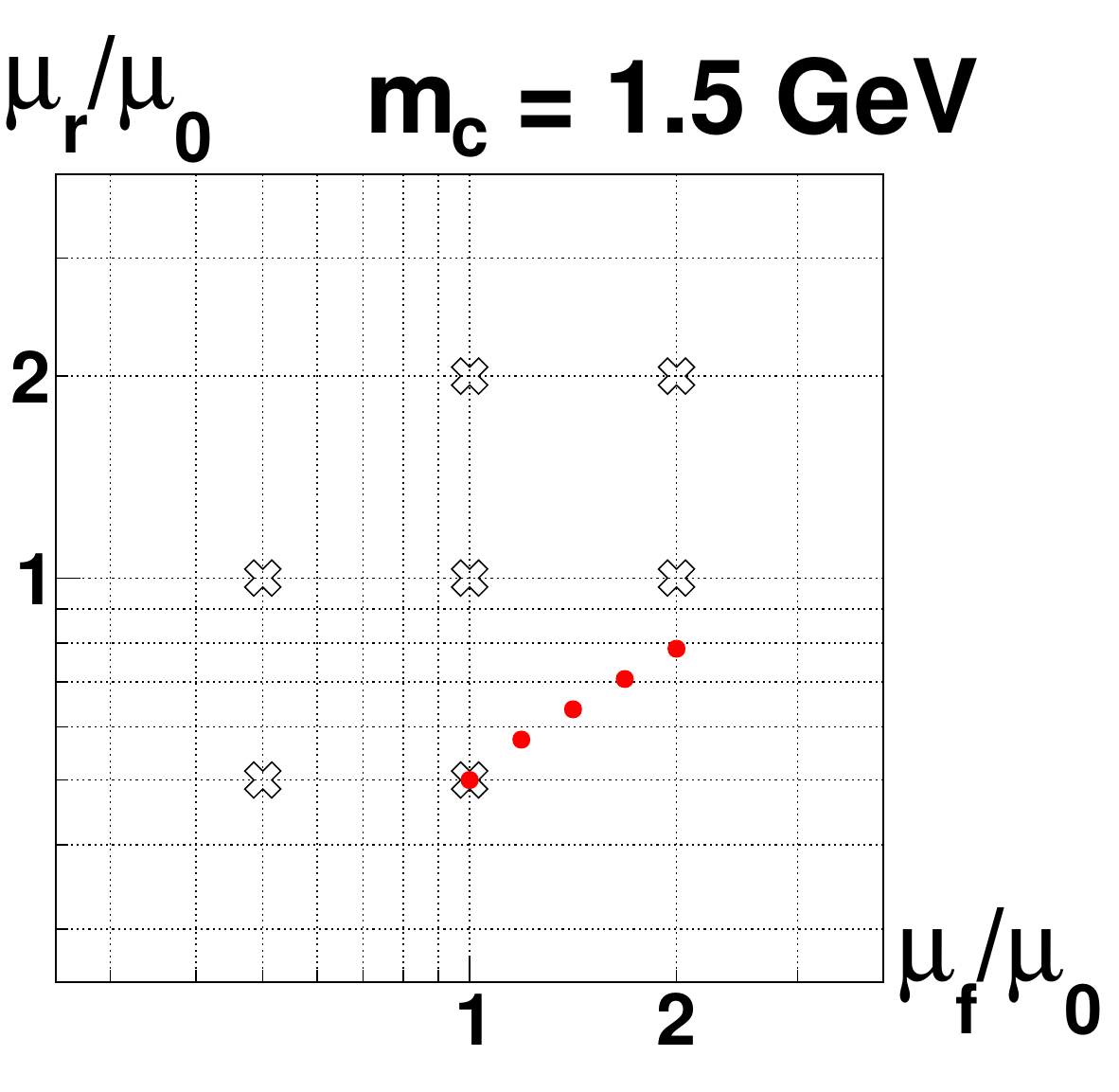}
  \includegraphics[width=0.19\textwidth]{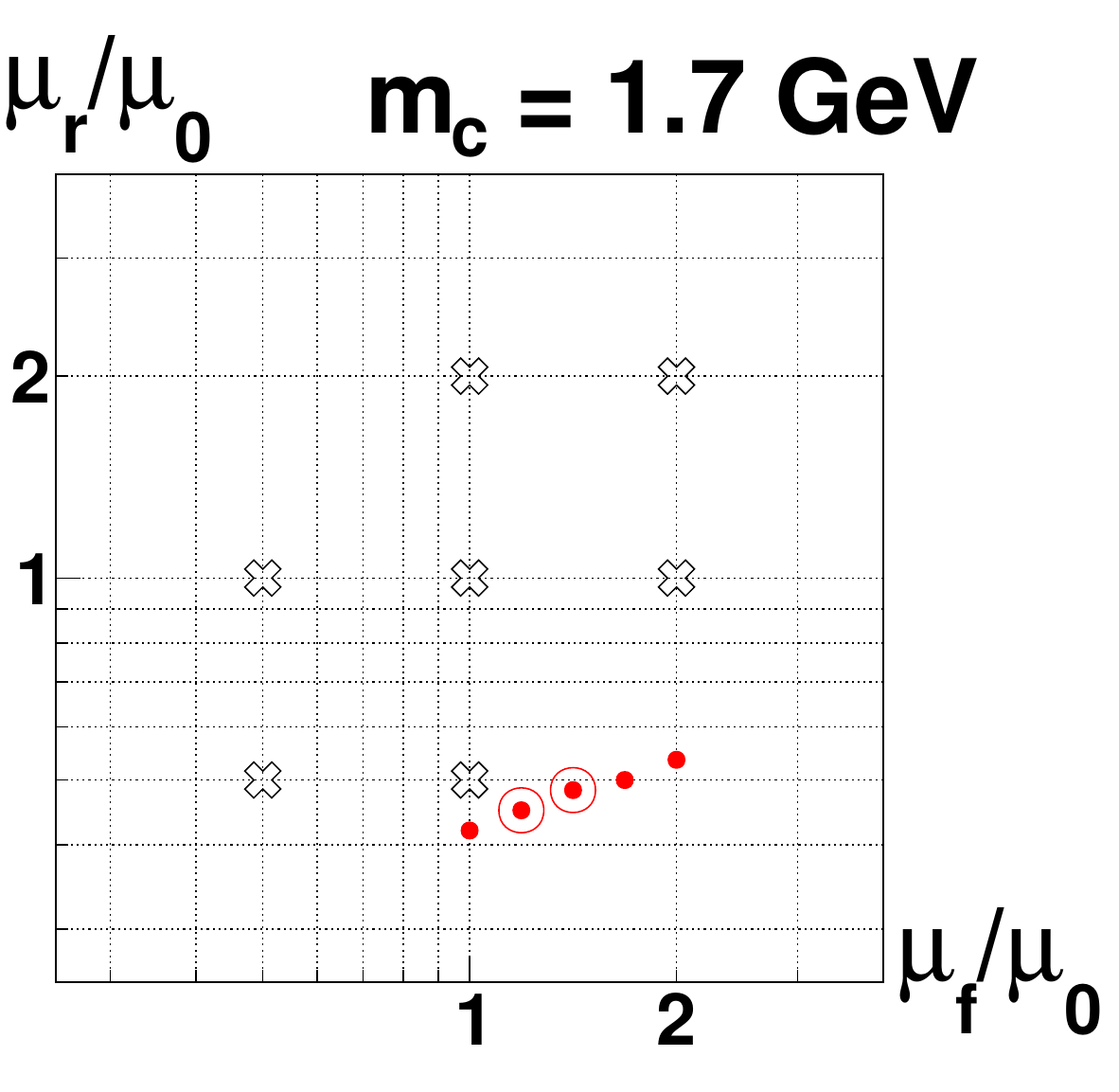}
  \includegraphics[width=0.19\textwidth]{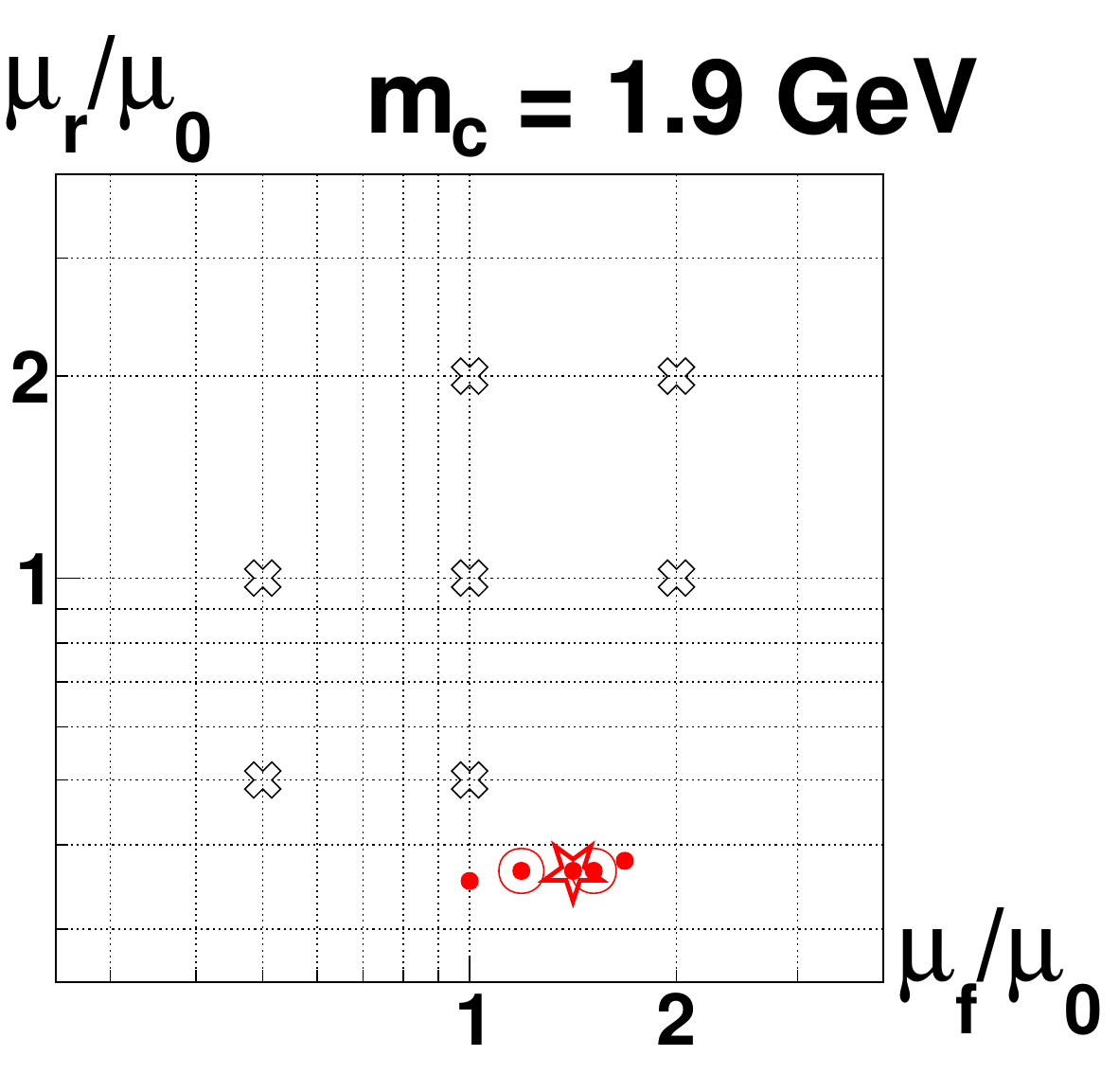}
  \includegraphics[width=0.19\textwidth]{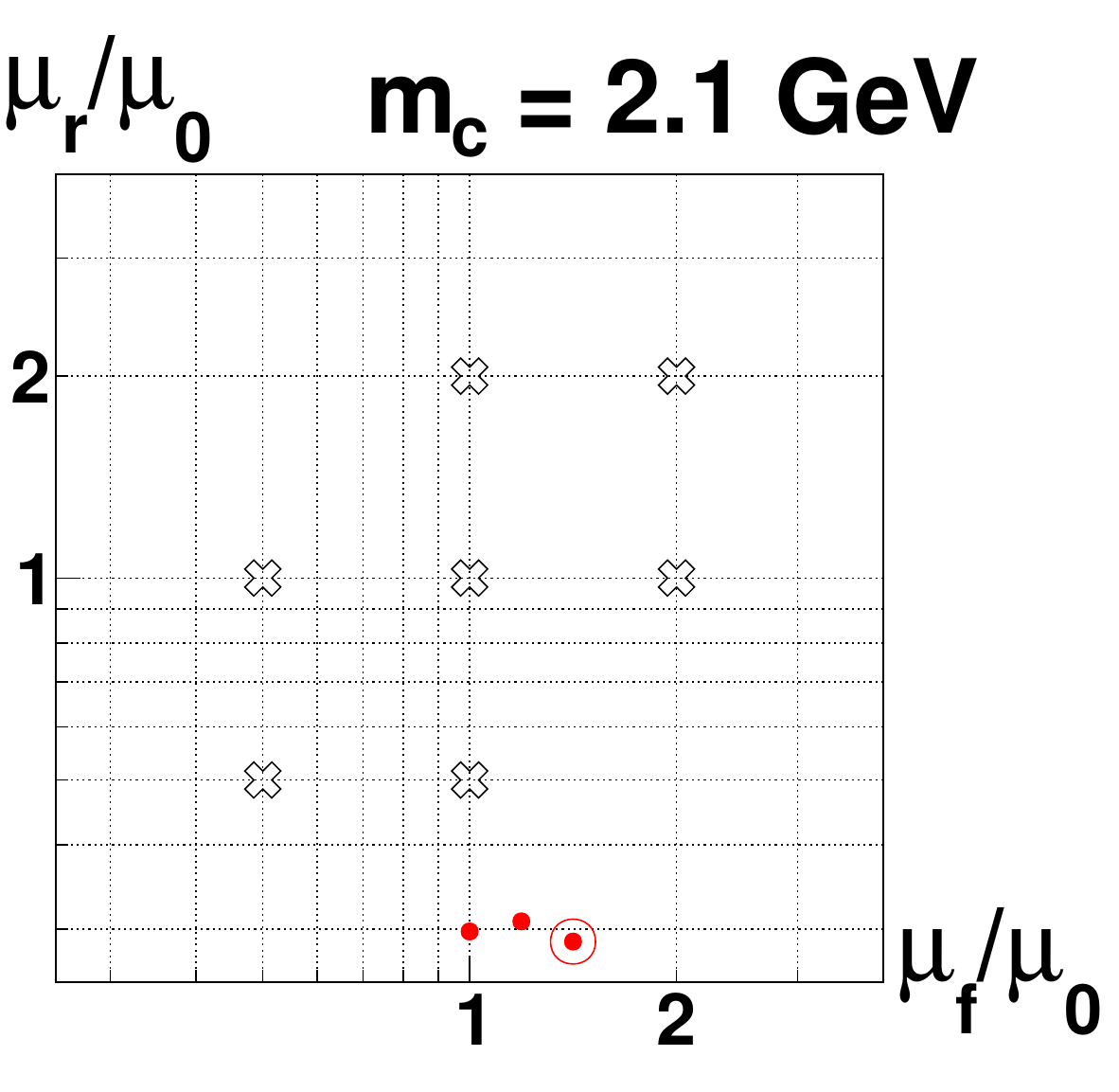}
 \caption{The scales giving the best description of the $D^0$ measurements at $\sqrt{s} =$ 5 TeV (the first row) and 13 TeV (the second row), respectively. The local least $\chi^2$ results are shown by the circle point symbols, while the global least $\chi^2$ for each $\sqrt{s}$ is marked by the additional star symbol. The scales entering the uncertainty evaluation are marked by additional outer circle symbols. The cross mark symbols indicate the conventional 7-point scale sets for theory.} \label{fig:scales_chosen}
 \end{center}
\end{figure*}

The fitted best parameters and uncertainty parameters obtained from these tables are shown in Figure \ref{fig:scales_chosen}, where the least $\chi^2$ results of 3-dimensional fits with $\mu_f$, $\mu_r$ and $\alpha_K$ (i.e., the local least $\chi^2$ results) were projected onto the 2-dimensional coordinates ($\mu_f$, $\mu_r$) with fixed $m_c$.
% figure of chosen scales
The best parameters, which were determined by the least $\chi^2$ of the 4-dimensional fits (i.e., the global least $\chi^2$), are marked by a star. The parameters within the four-dimensional uncertainty contour, which were determined by\footnote{In order to ensure a $\chi^2$/ndof of 1, an S-factor \cite{ParticleDataGroup:2024cfk} of 1.46 (1.16) was applied at $\sqrt{s} =$ 5 (13) TeV.} $\Delta \chi^2 < 4.7 \sim 1\sigma$, are marked by an additional outer circle. The cross marks indicate the conventional 7-point \cite{theorySc} scale variation sets which are generally used for calculation of QCD theory uncertainties 
and are shown in the figure as a reference for the fitted parameters.
The best parameters including their uncertainty ranges (edges of the 4D uncertainty ellipsoids) are summarized in Table \ref{tb:bestPars}.
\begin{table}
 \begin{center}
 \caption{The best parameters used for the ddFONLL parametrization. The parentheses indicate the ranges of the uncertainty parameters, which were used to calculate the so-called $\chi^2$ uncertainties of ddFONLL.} \label{tb:bestPars}
 \renewcommand{\arraystretch}{1.5}
\begin{tabular}{|c|c|c|c|}
  \hline
                   & $\sqrt{s} =$ 5 TeV & $\sqrt{s} =$ 13 TeV \\
  \hline
   $\mu_f$/$\mu_0$ & 1.68 (1.00 - 2.00) & 1.41 (1.19 - 1.52) \\
  \hline
   $\mu_r$/$\mu_0$ & 0.48 (0.34 - 0.93) & 0.37 (0.29 - 0.48) \\
  \hline
   $m_c$ [GeV]     & 1.7 (1.3 - 1.9)    & 1.9 (1.7 - 2.1) \\
  \hline
   $\alpha_K$      & 9 (6 - 28)         & 6 (5 - 9) \\
  \hline
 \end{tabular}
 \end{center}
\end{table}
The uncertainties turn out to be reasonably consistent with the conventional scales and $m_c$ values for perturbative QCD theory and the reference $\alpha_K$ values based on $e^+e^-$ data \cite{fragfunc_bcfy2}, although the renormalization scale comes out on the low side. Furthermore, it was observed from this study that there are significant correlations between the two optimized theory scales especially at lower $m_c$ (see Figure \ref{fig:scales_chosen}).

\begin{figure*}
 \begin{center} 
  \includegraphics[width=0.24\textwidth]{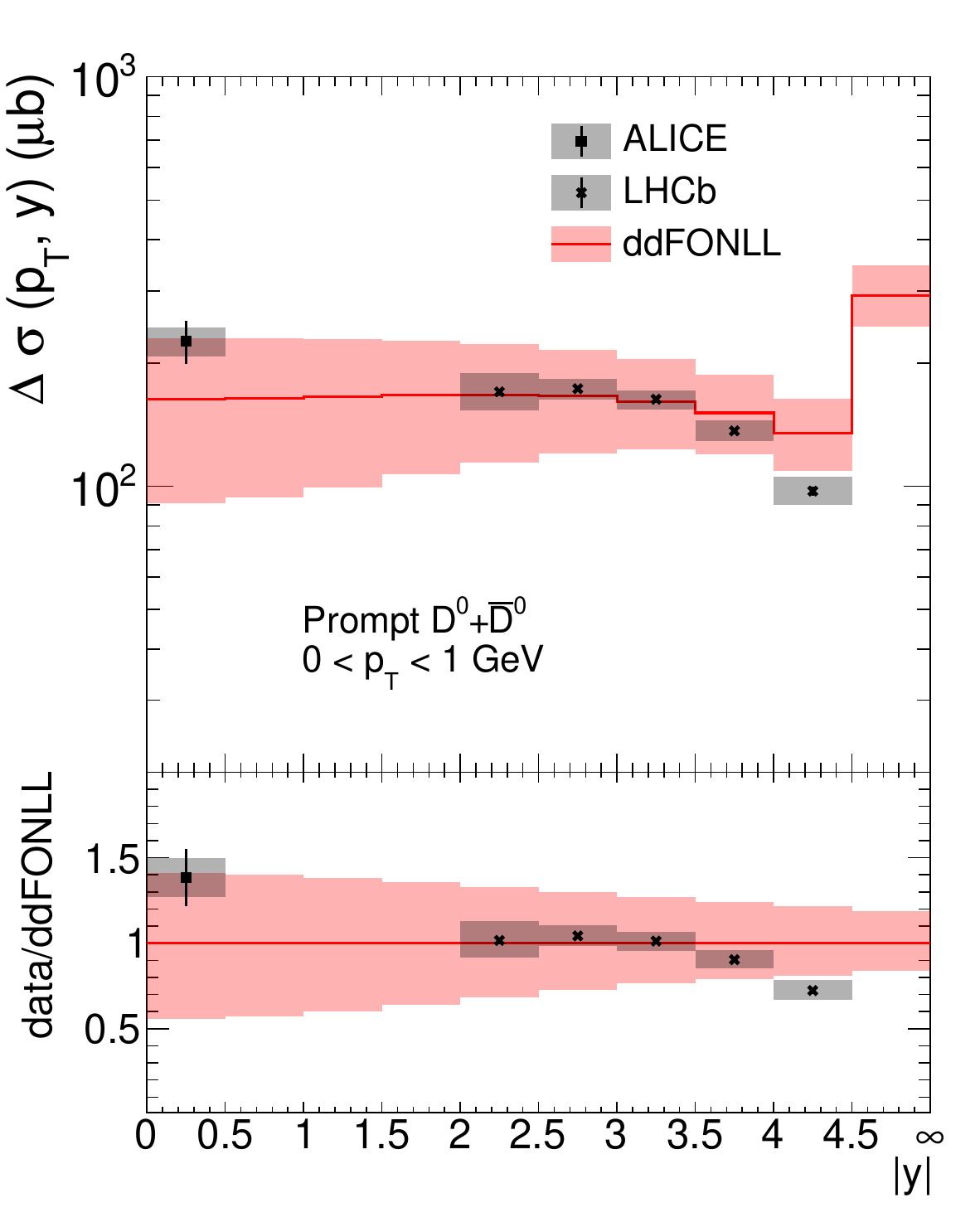}
  \includegraphics[width=0.24\textwidth]{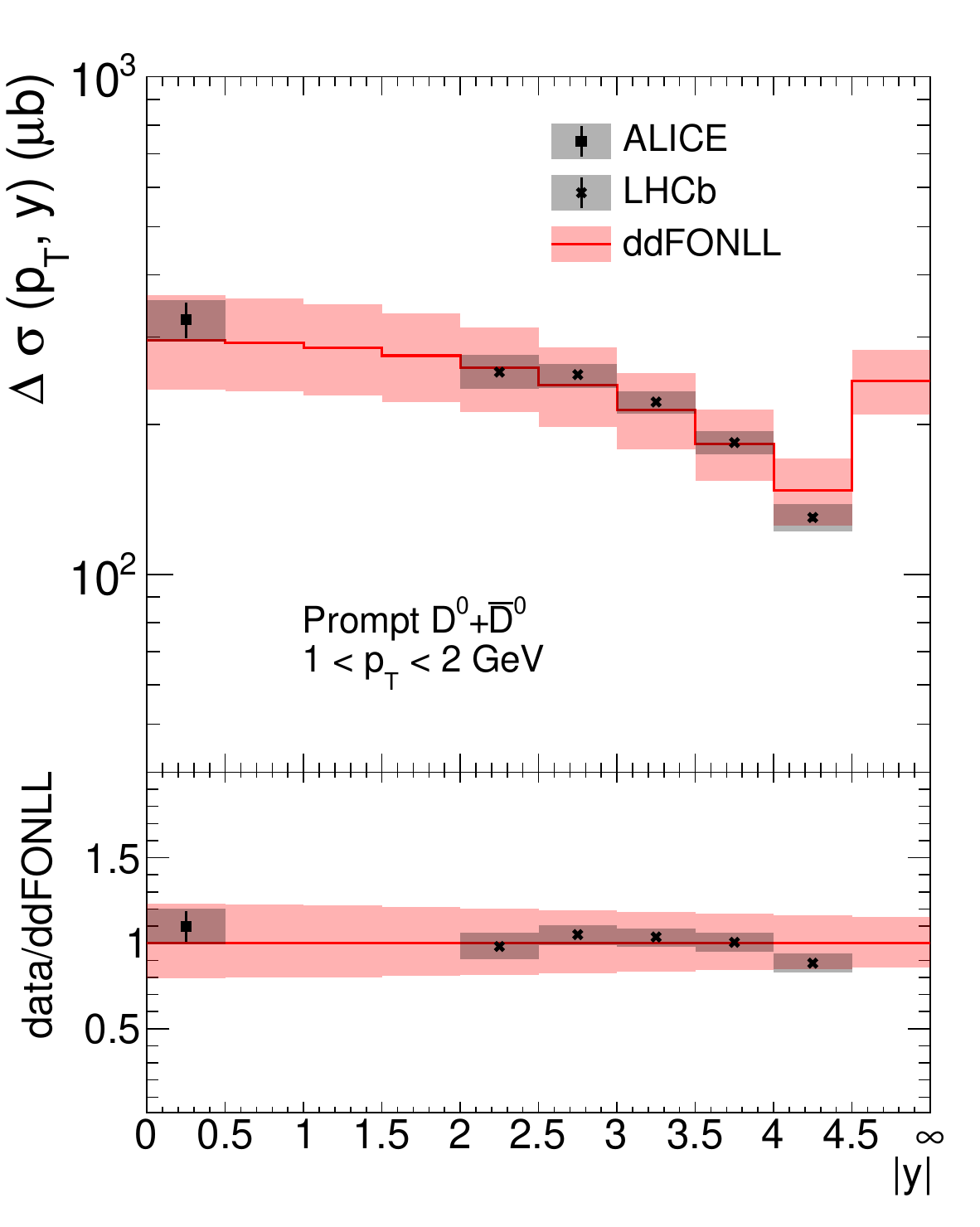}
  \includegraphics[width=0.24\textwidth]{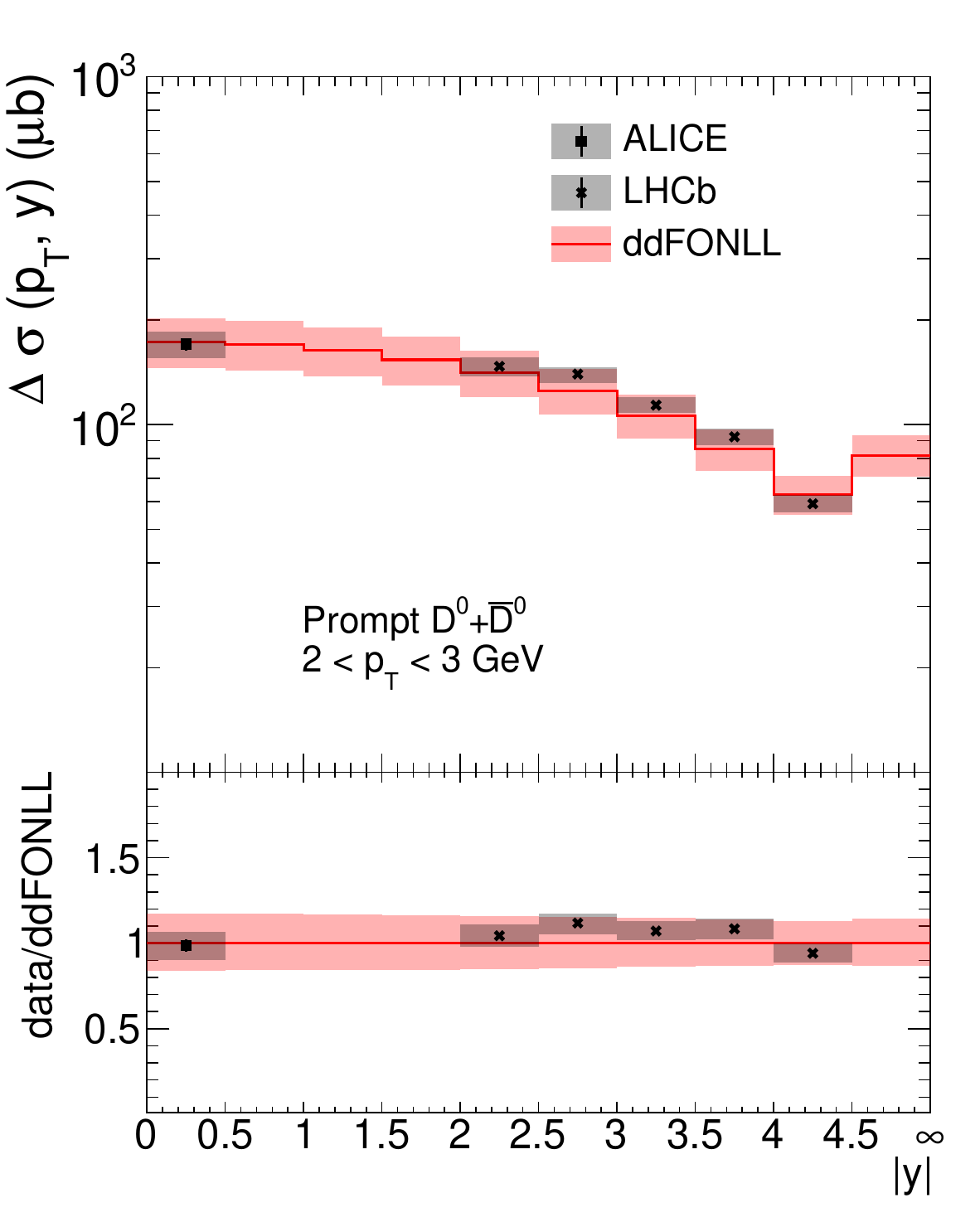}
  \includegraphics[width=0.24\textwidth]{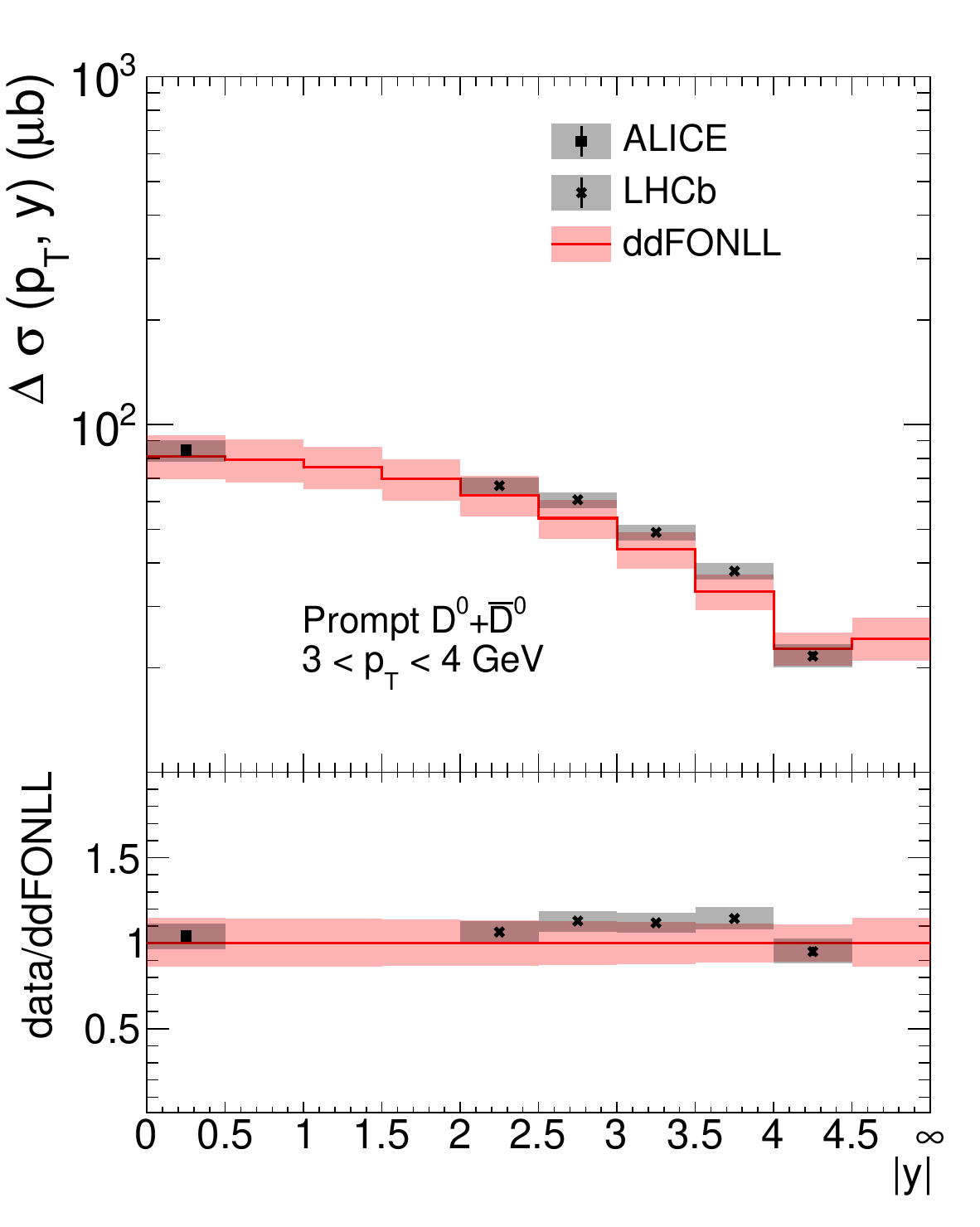}
  \includegraphics[width=0.24\textwidth]{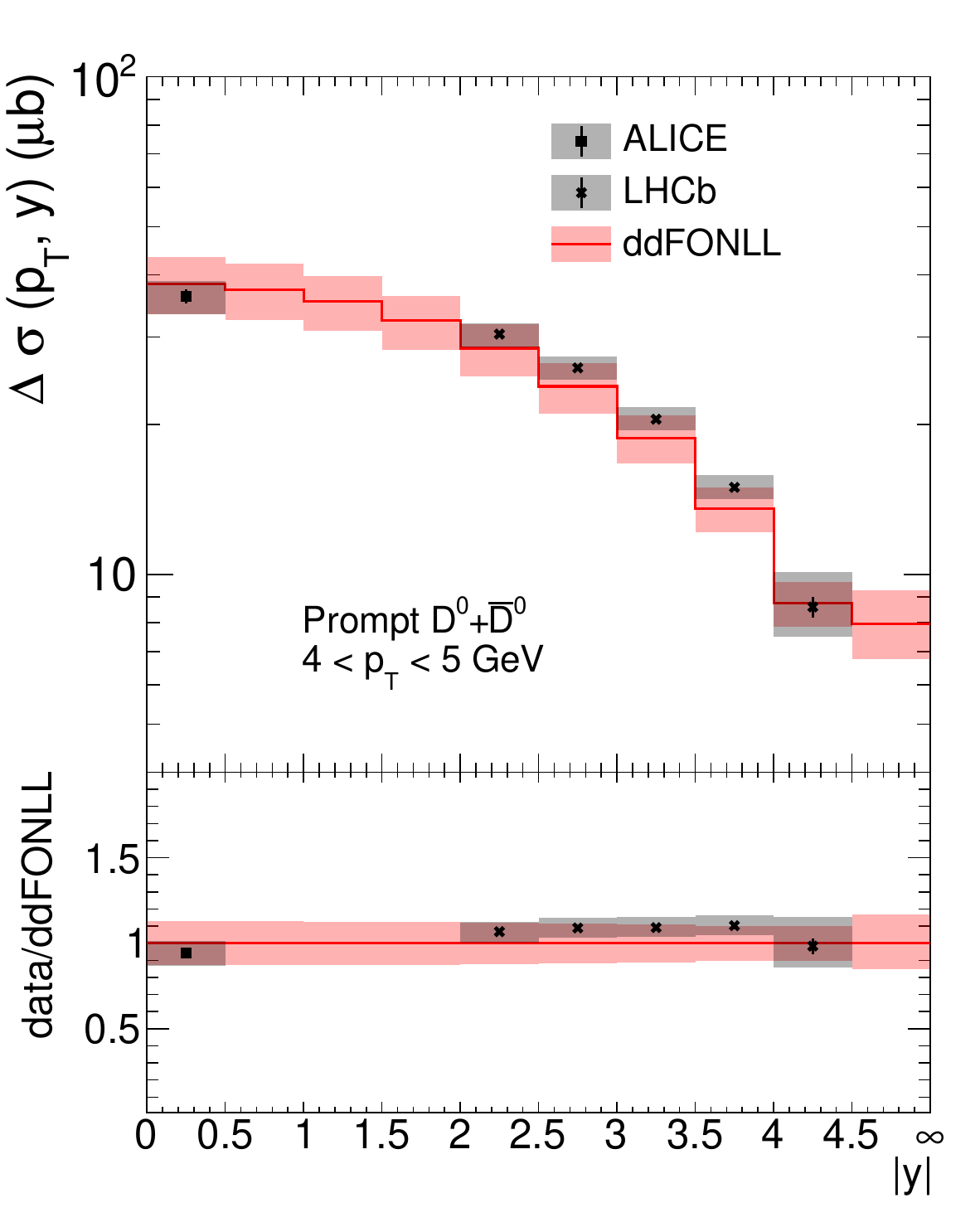}
  \includegraphics[width=0.24\textwidth]{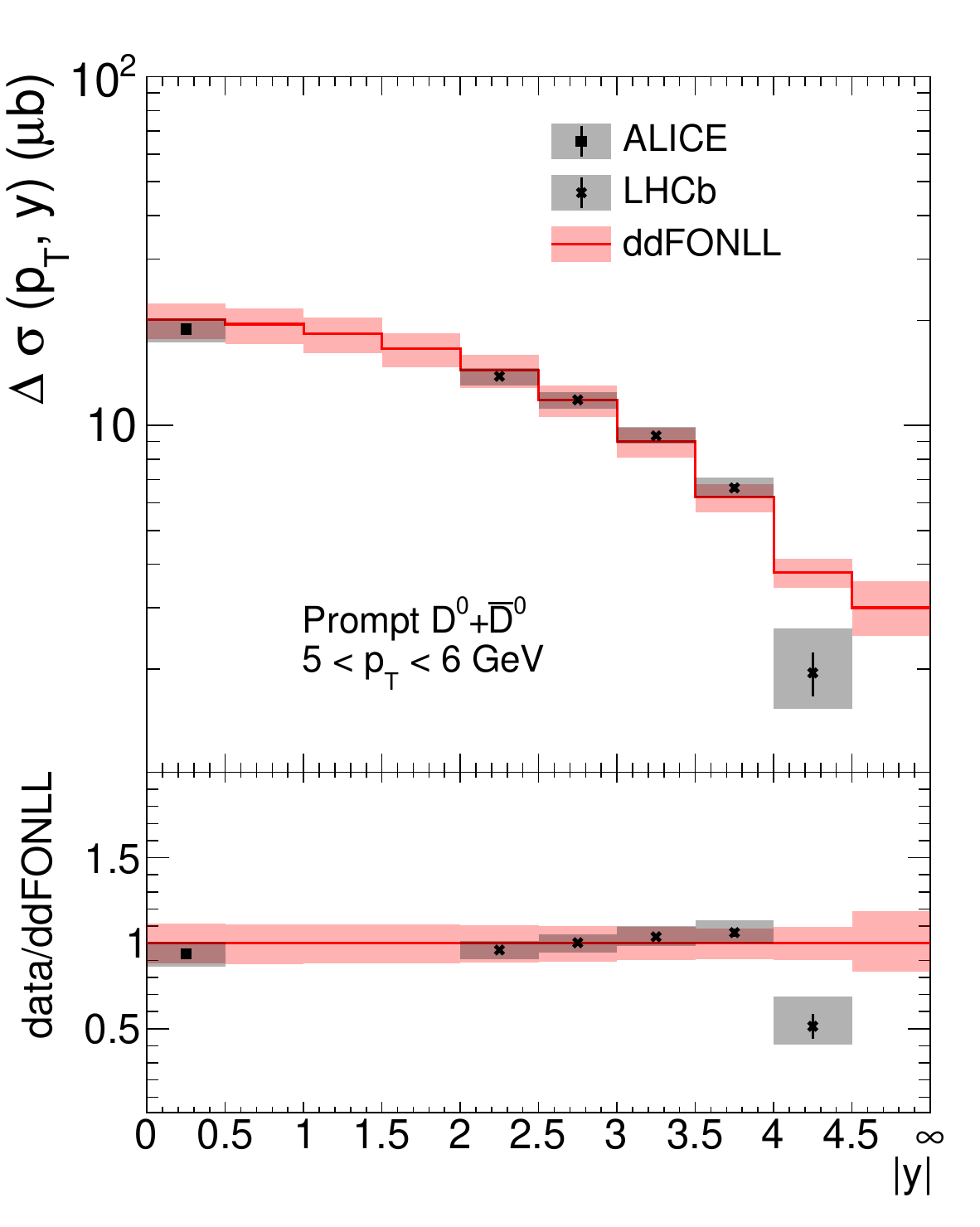}
  \includegraphics[width=0.24\textwidth]{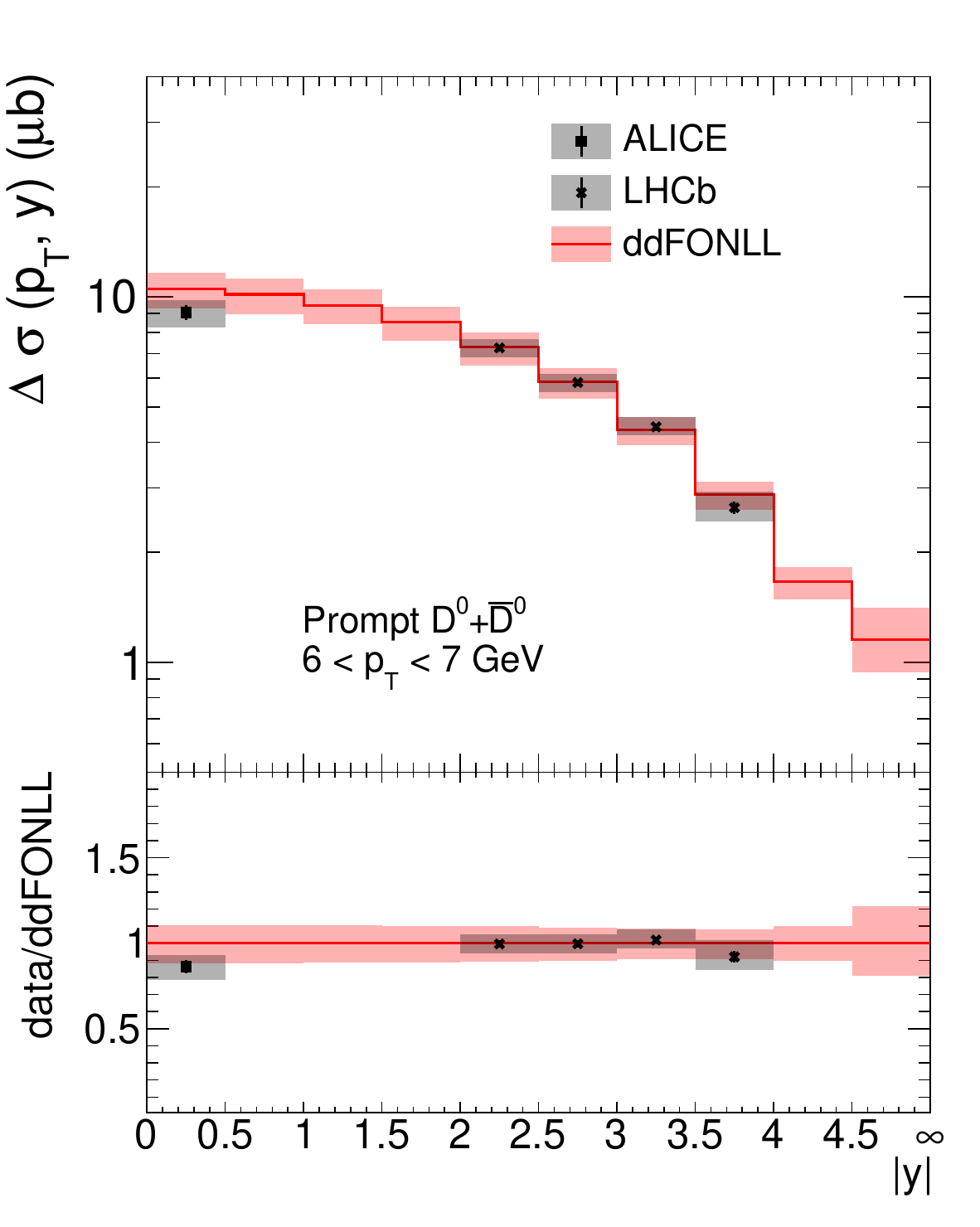}
  \includegraphics[width=0.24\textwidth]{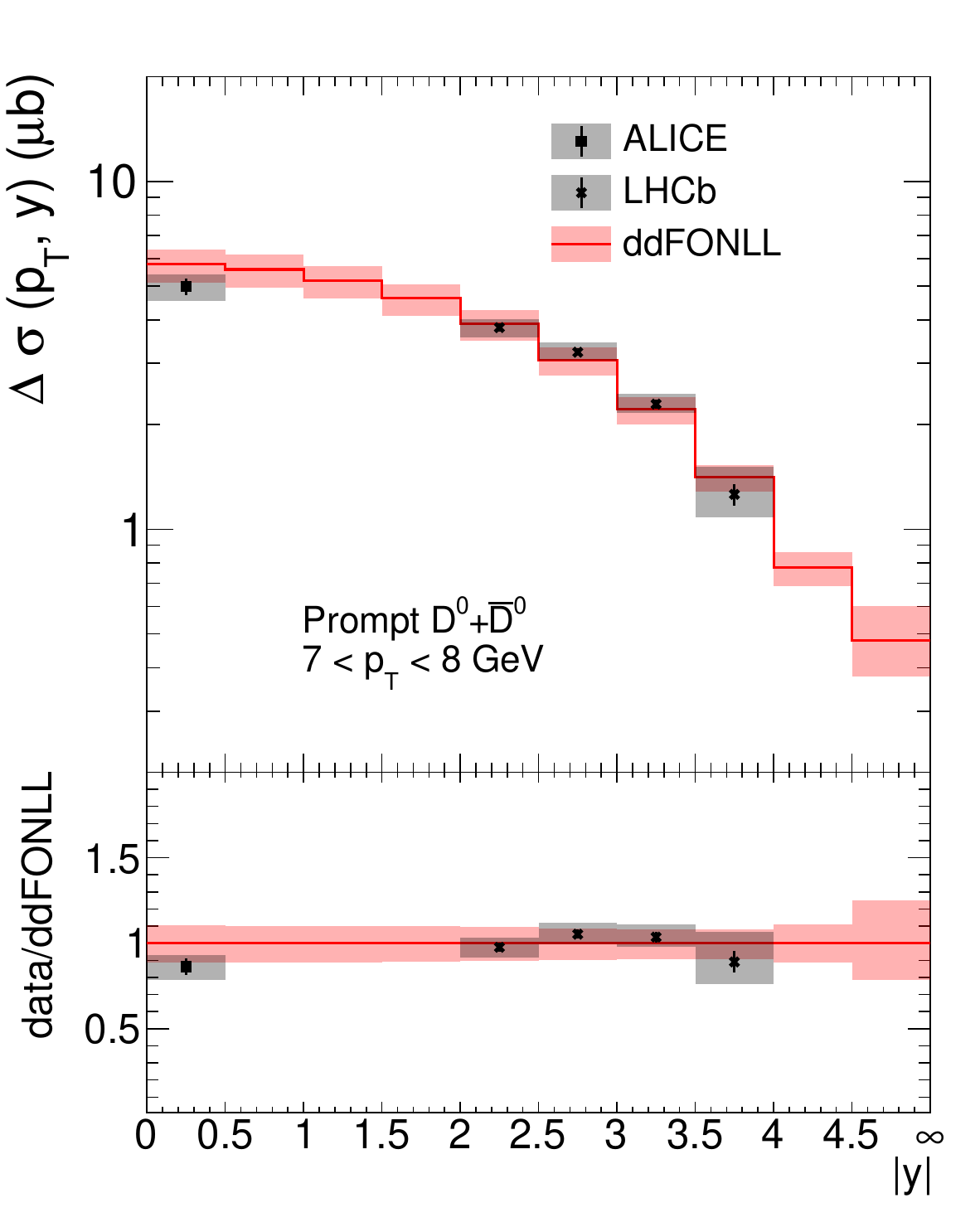}
  \includegraphics[width=0.24\textwidth]{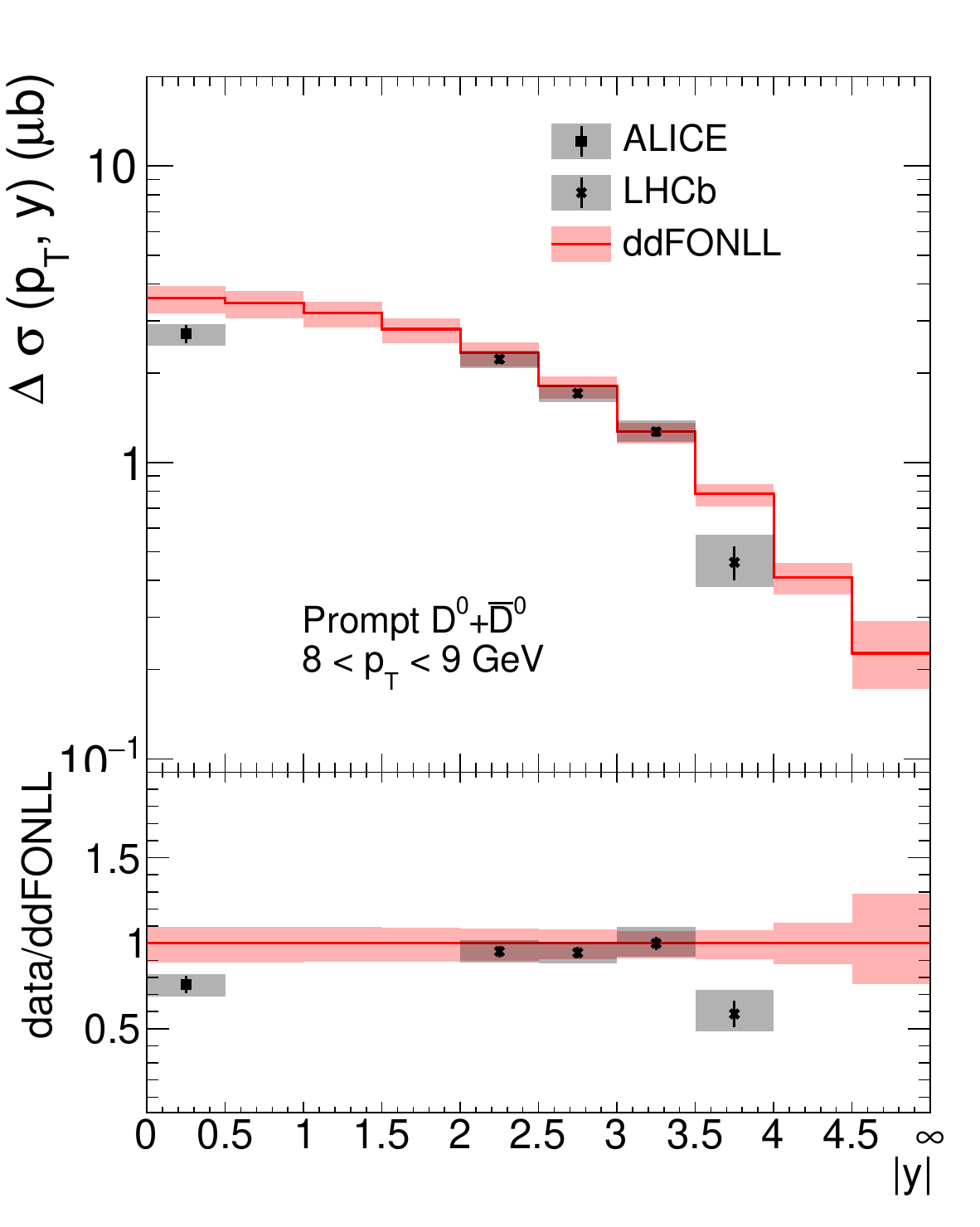}
  \includegraphics[width=0.24\textwidth]{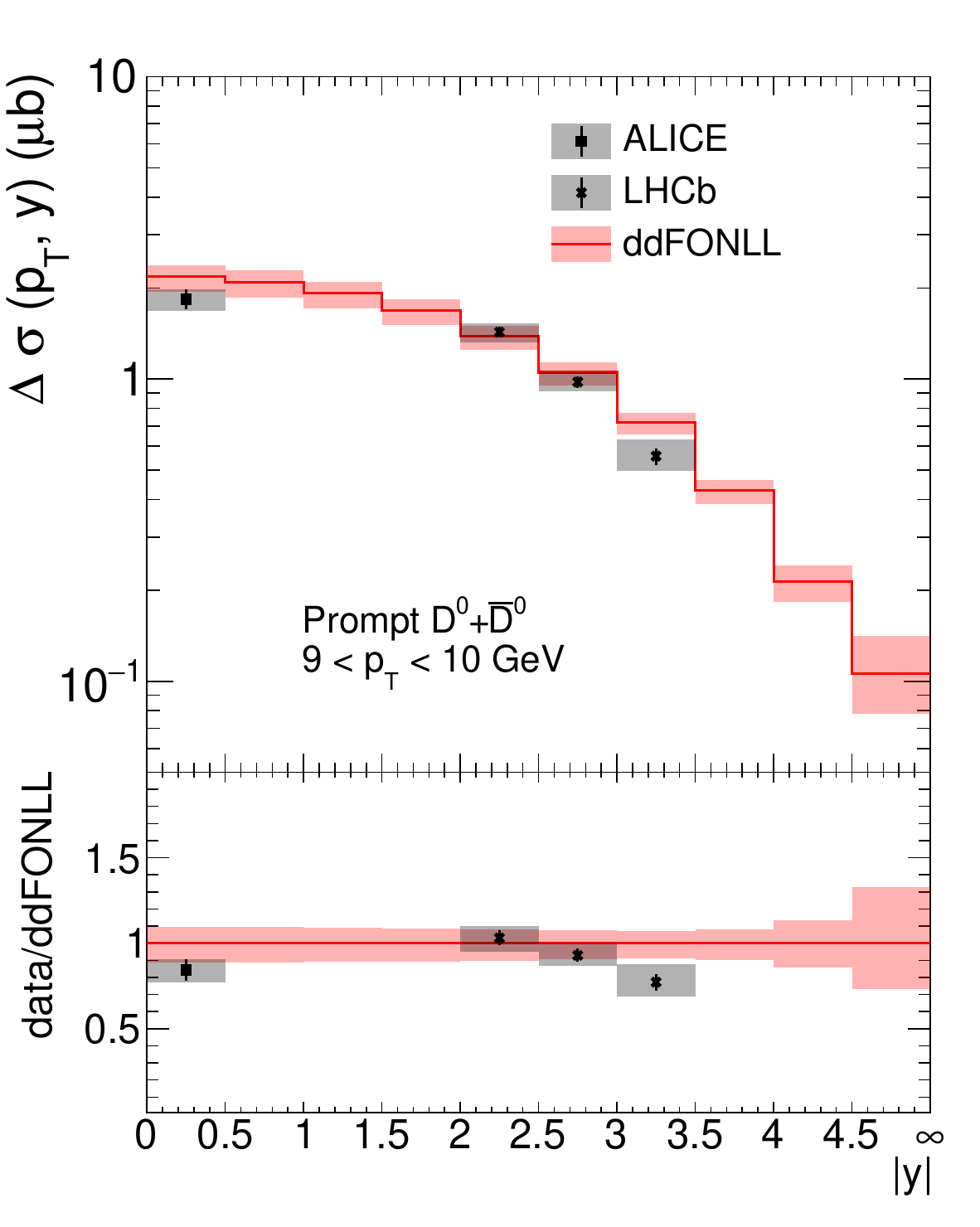}
  \includegraphics[width=0.24\textwidth]{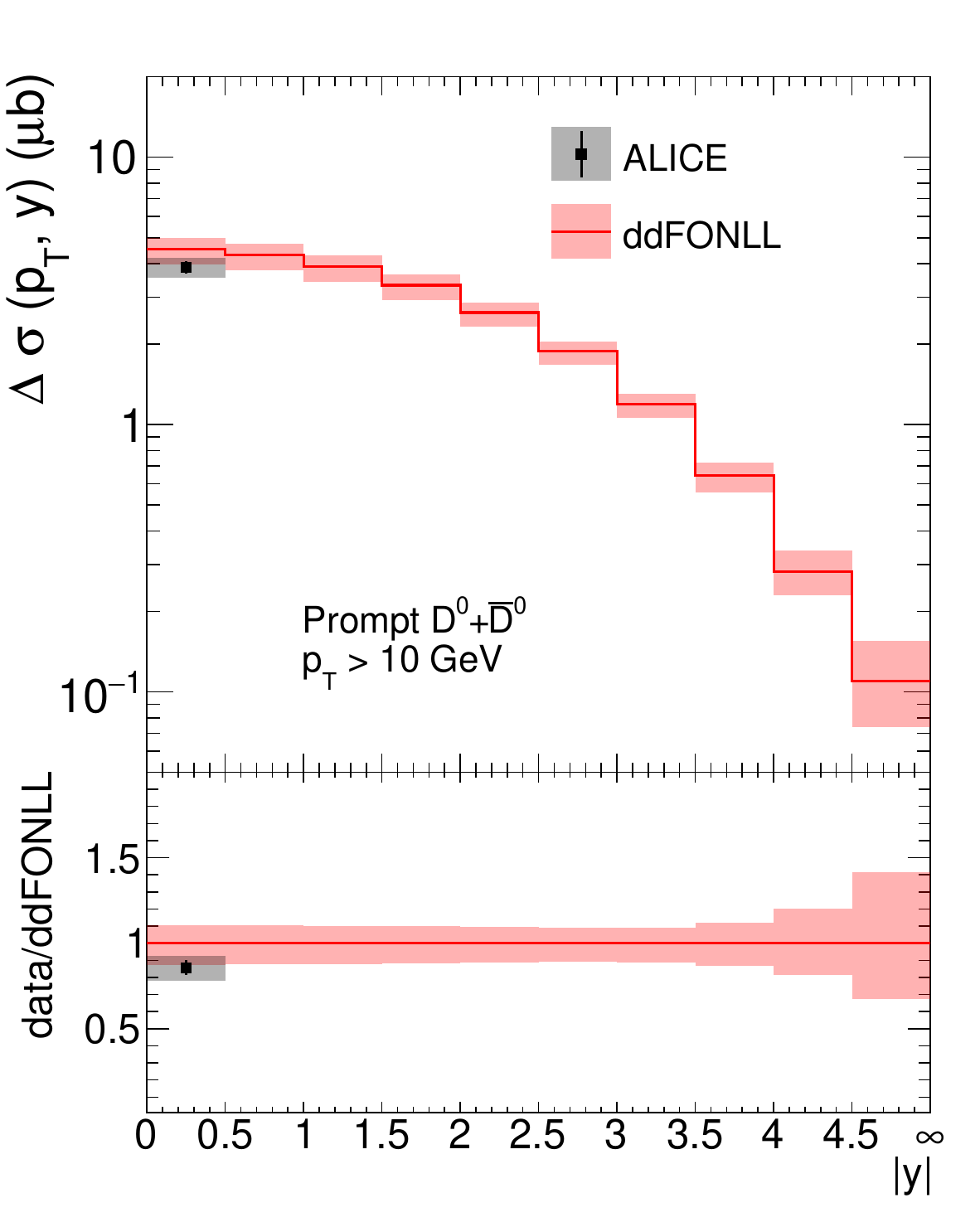}
 \end{center}
 \caption{$D^0+\overline{D}^0$ cross sections at $\sqrt{s} =$ 5 TeV as a function of $|y|$. The red bands of ddFONLL show the total uncertainty (CTEQ6.6 PDF $\oplus$ $\tilde{f}$ $\oplus$ $\chi^2$).} \label{fig:ddFONLL_d05TeV_yDiff_totUnc}
\end{figure*}

\begin{figure*}
 \begin{center} 
  \includegraphics[width=0.24\textwidth]{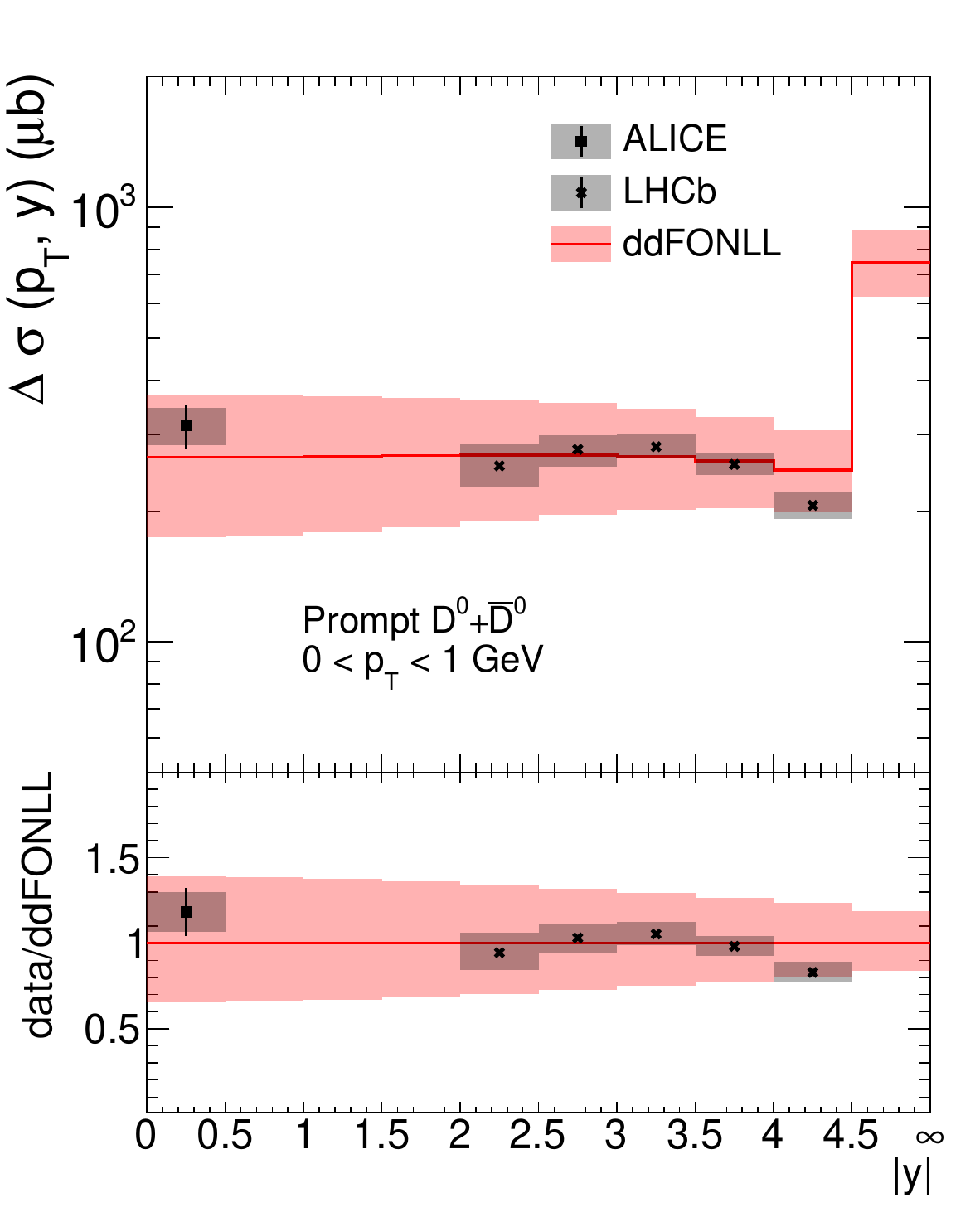}
  \includegraphics[width=0.24\textwidth]{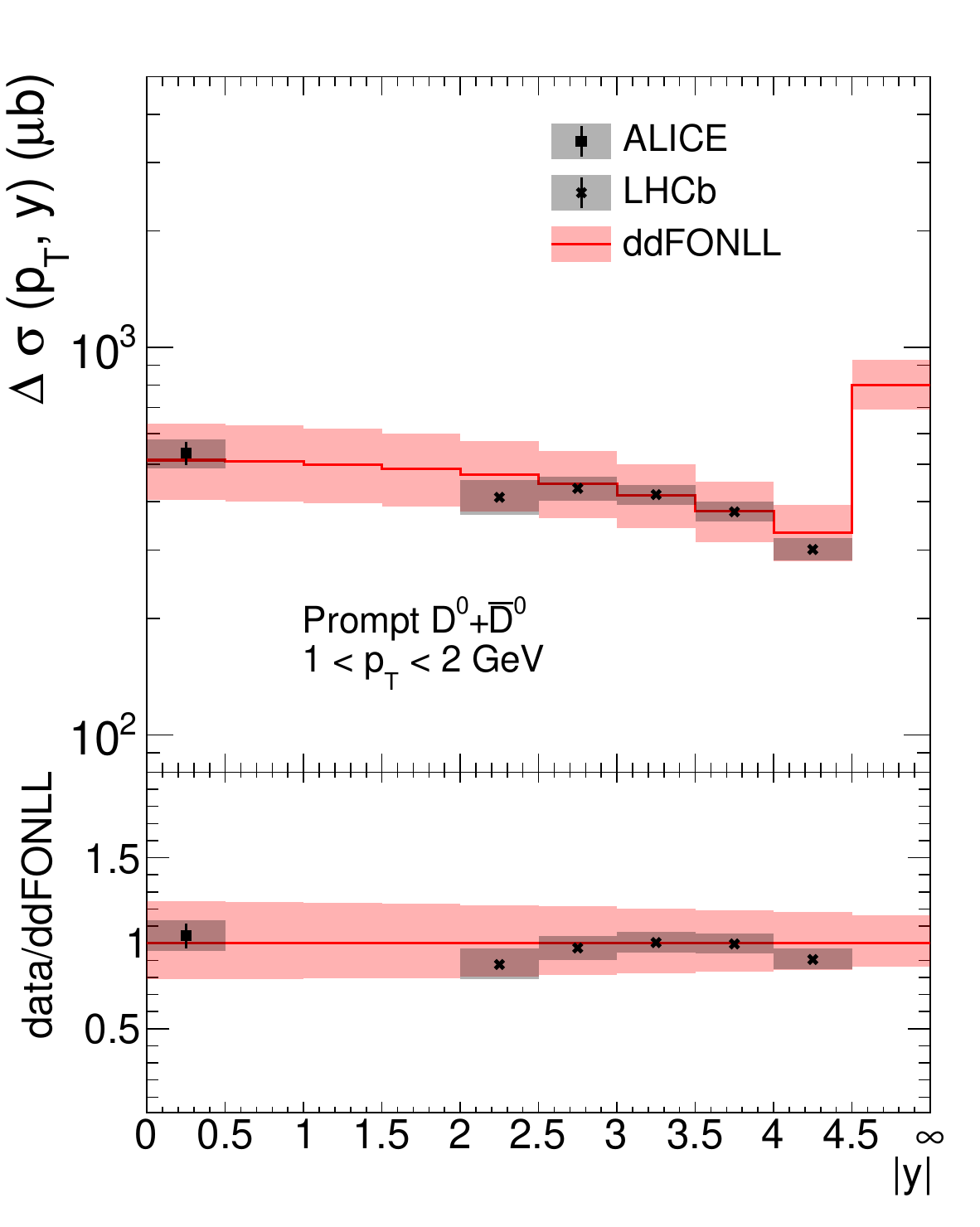}
  \includegraphics[width=0.24\textwidth]{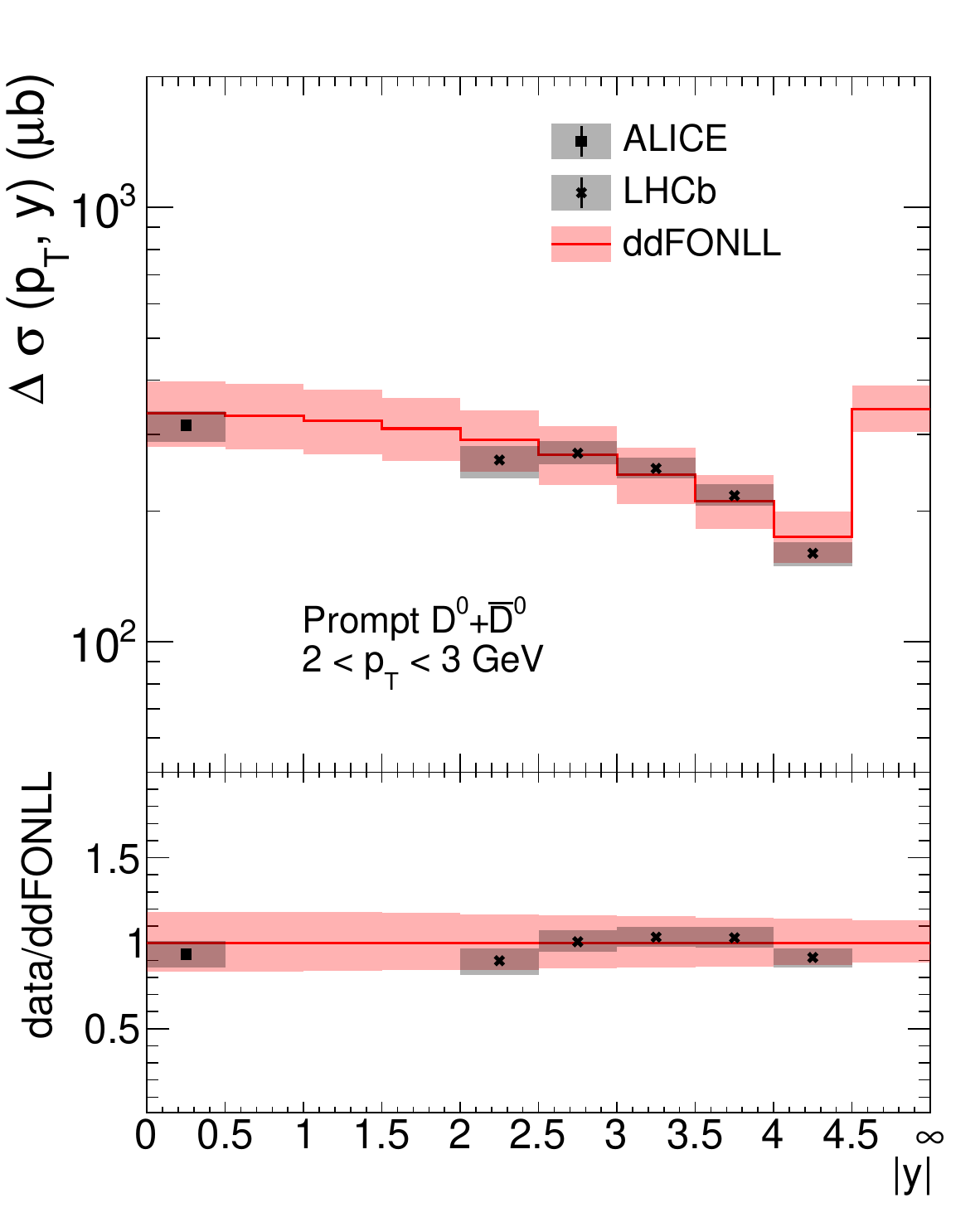}
  \includegraphics[width=0.24\textwidth]{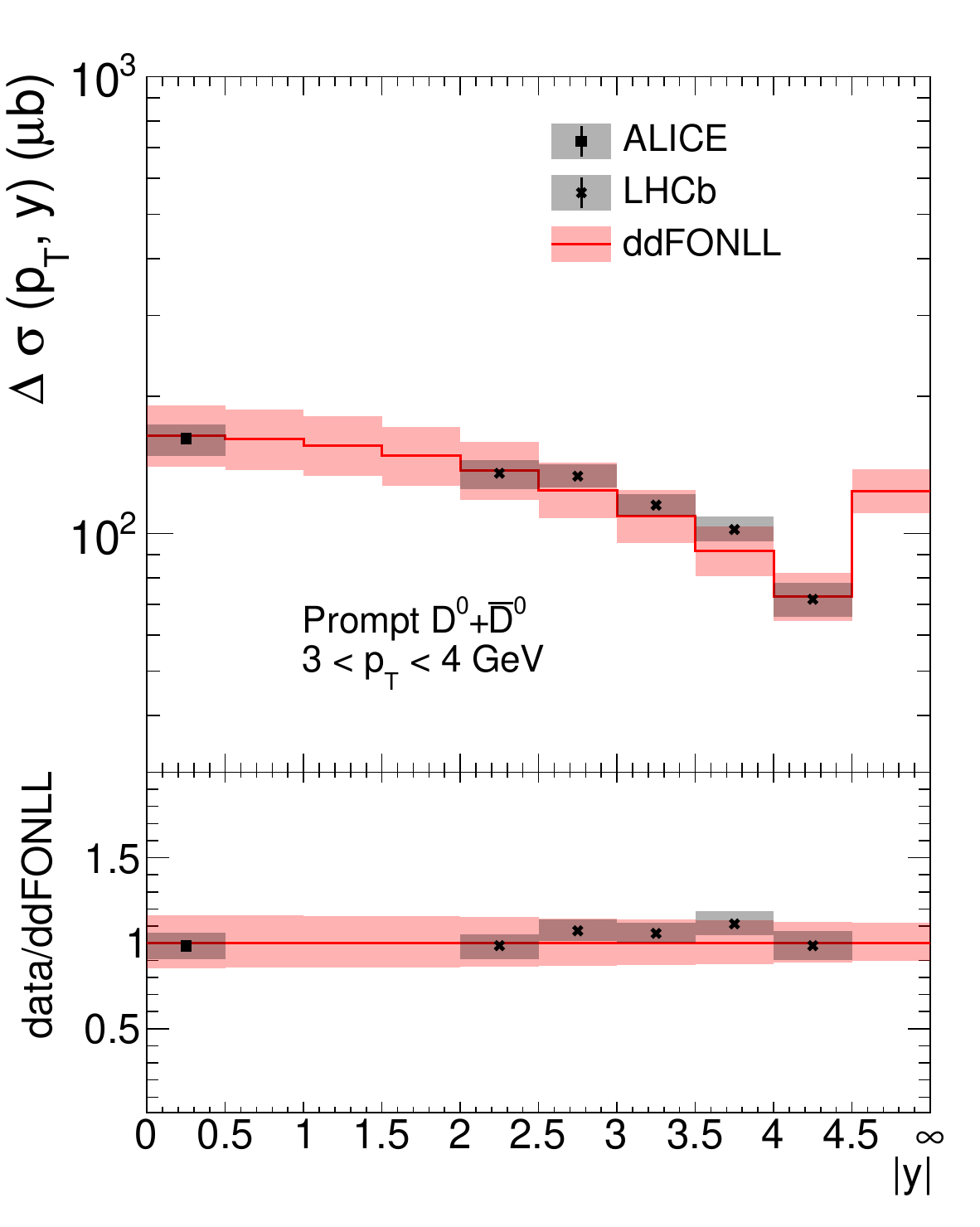}
  \includegraphics[width=0.24\textwidth]{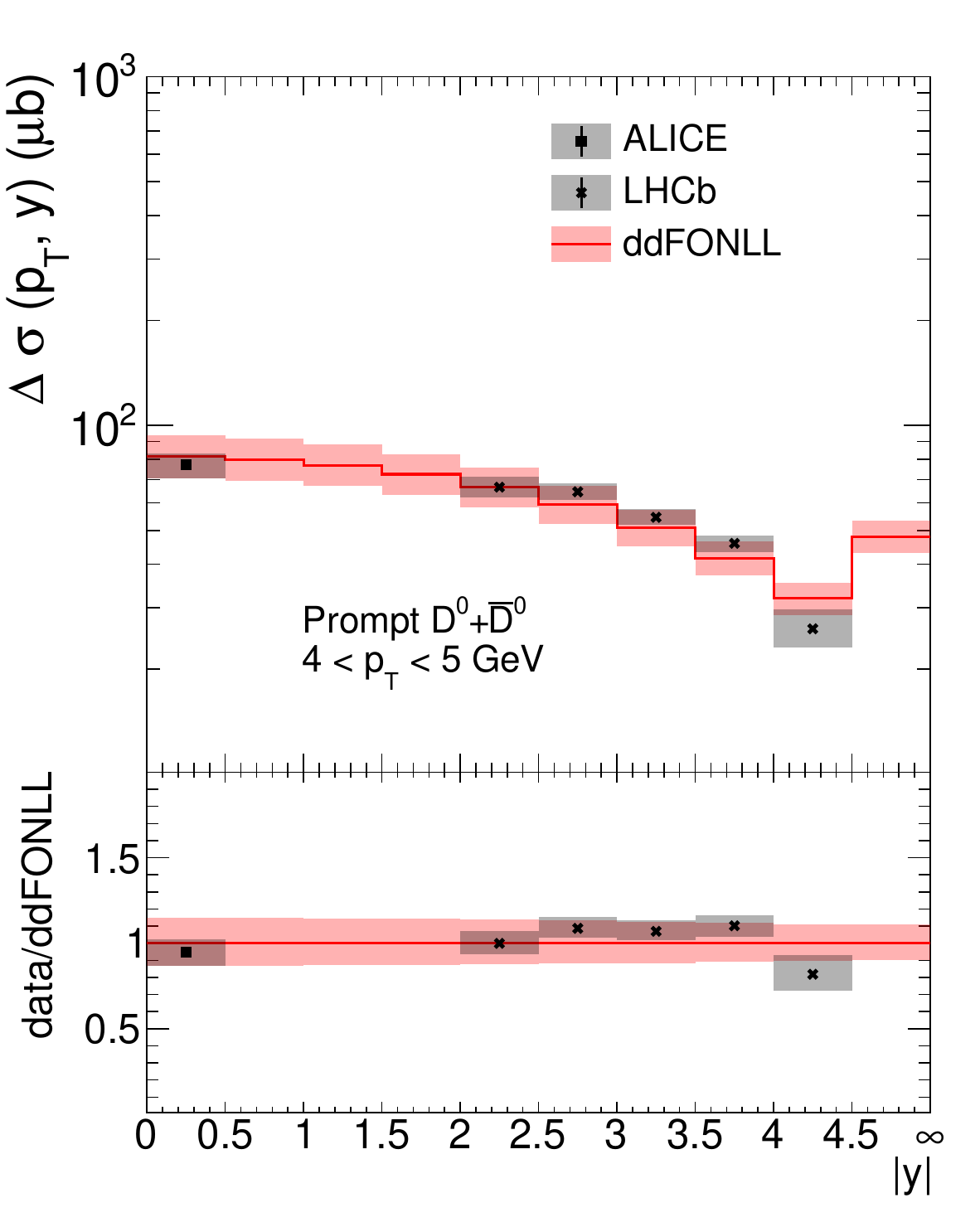}
  \includegraphics[width=0.24\textwidth]{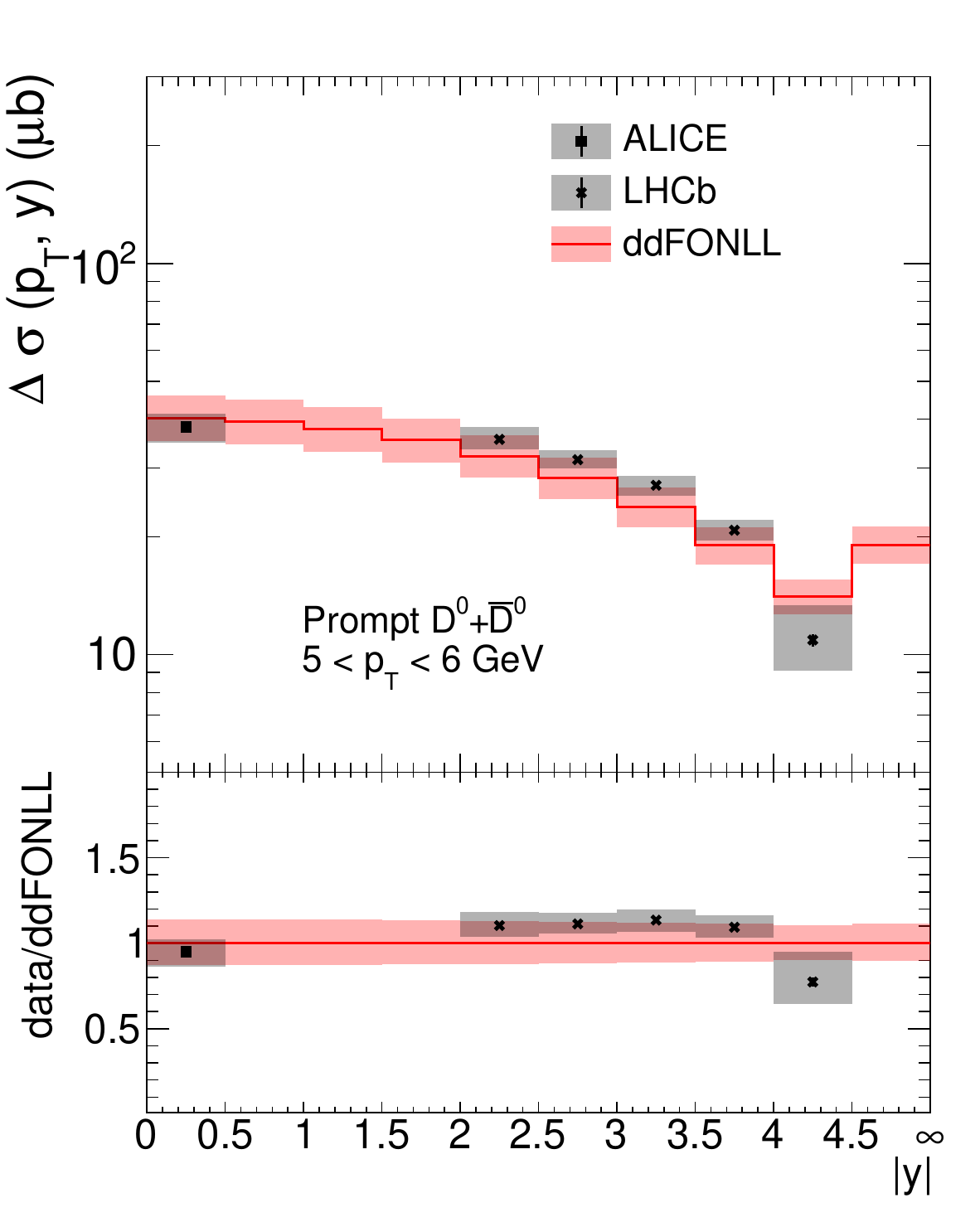}
  \includegraphics[width=0.24\textwidth]{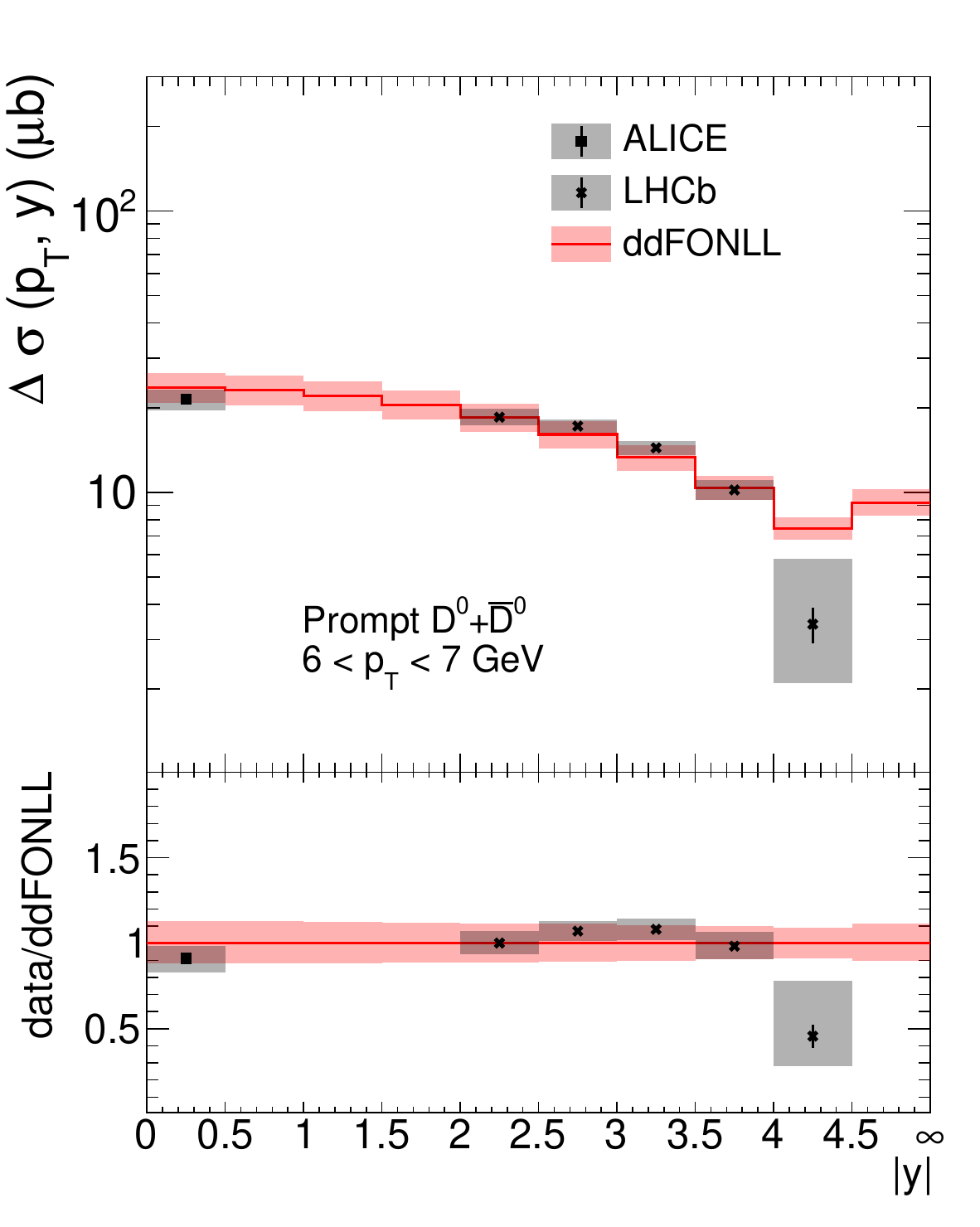}
  \includegraphics[width=0.24\textwidth]{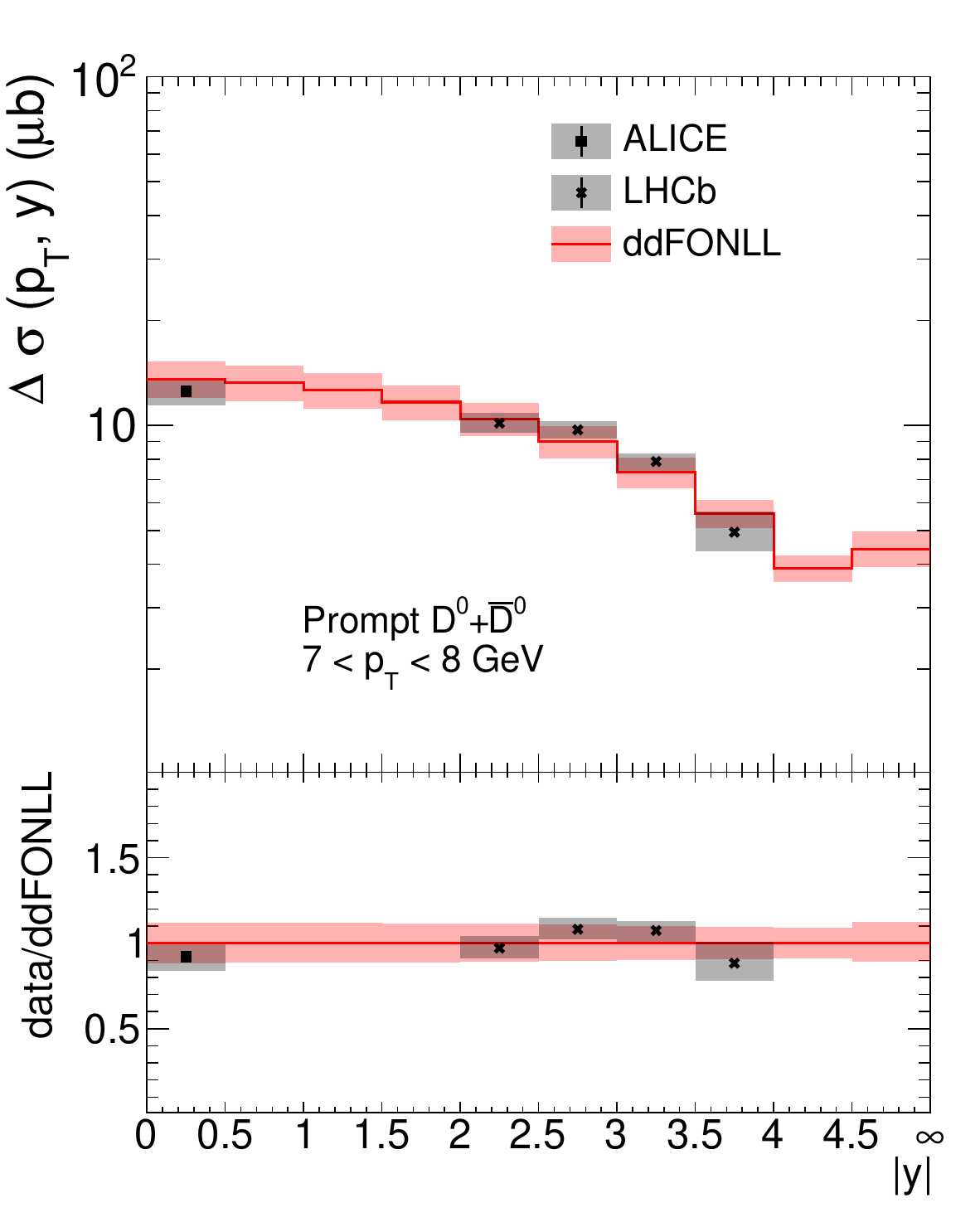}
  \includegraphics[width=0.24\textwidth]{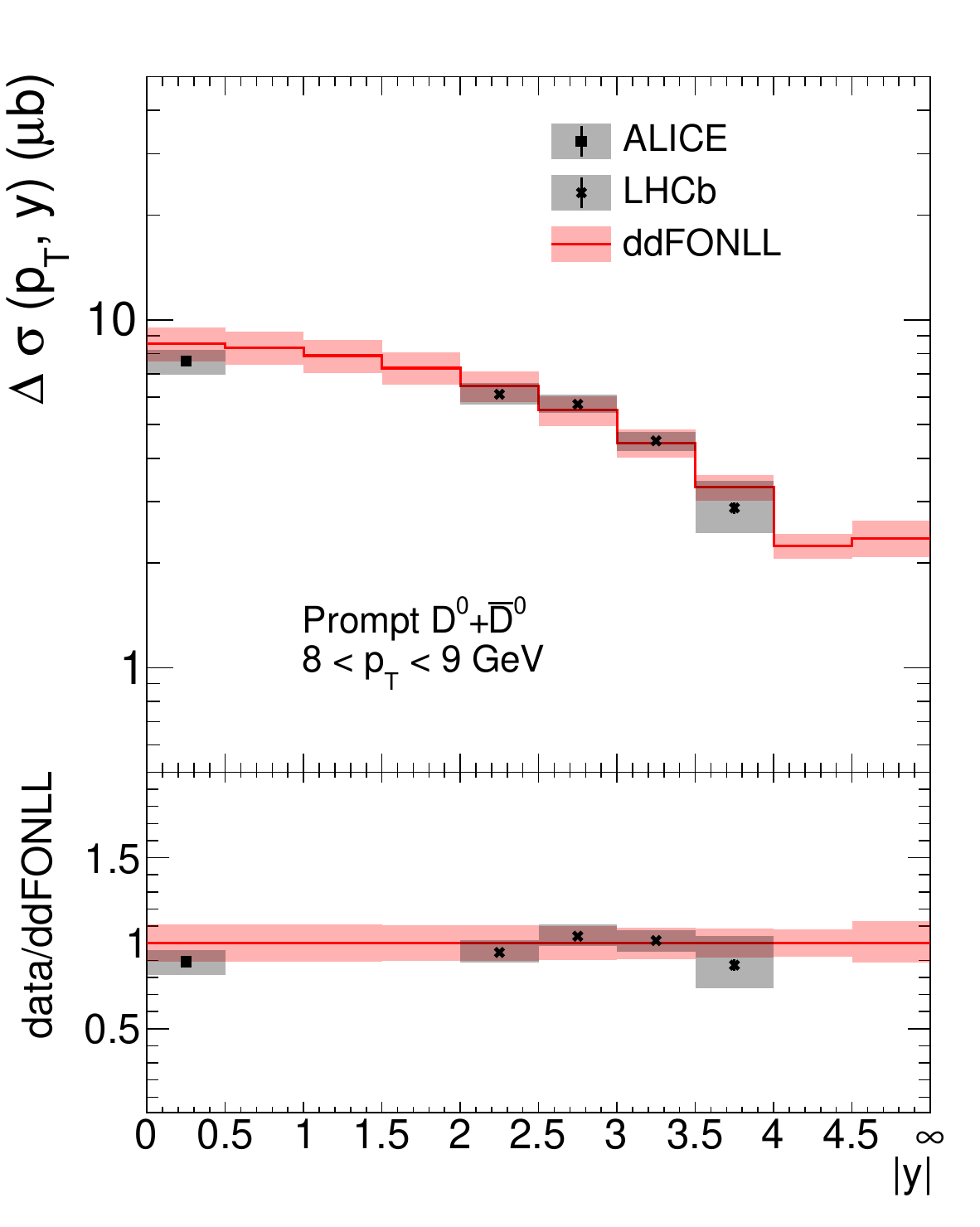}
  \includegraphics[width=0.24\textwidth]{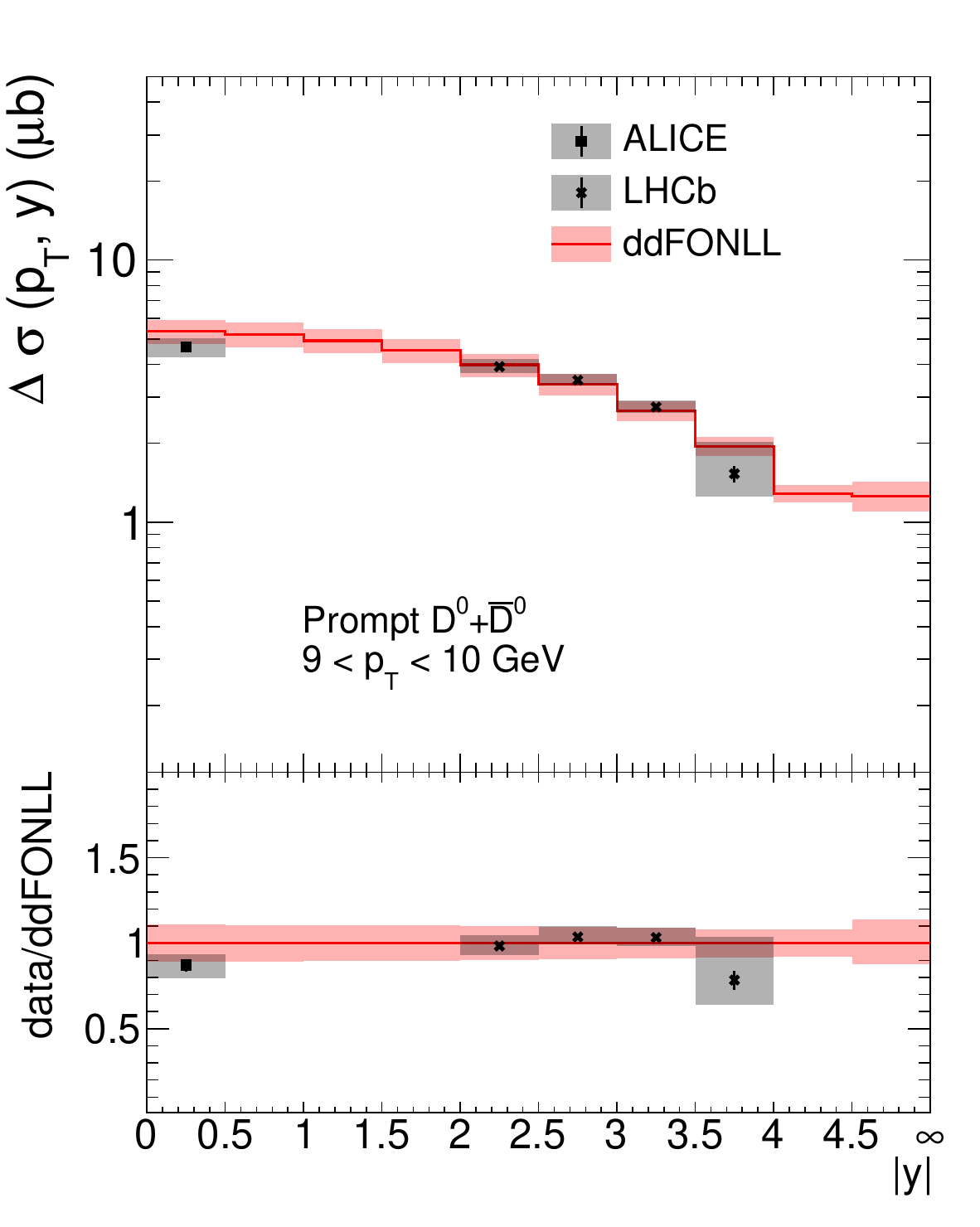}
  \includegraphics[width=0.24\textwidth]{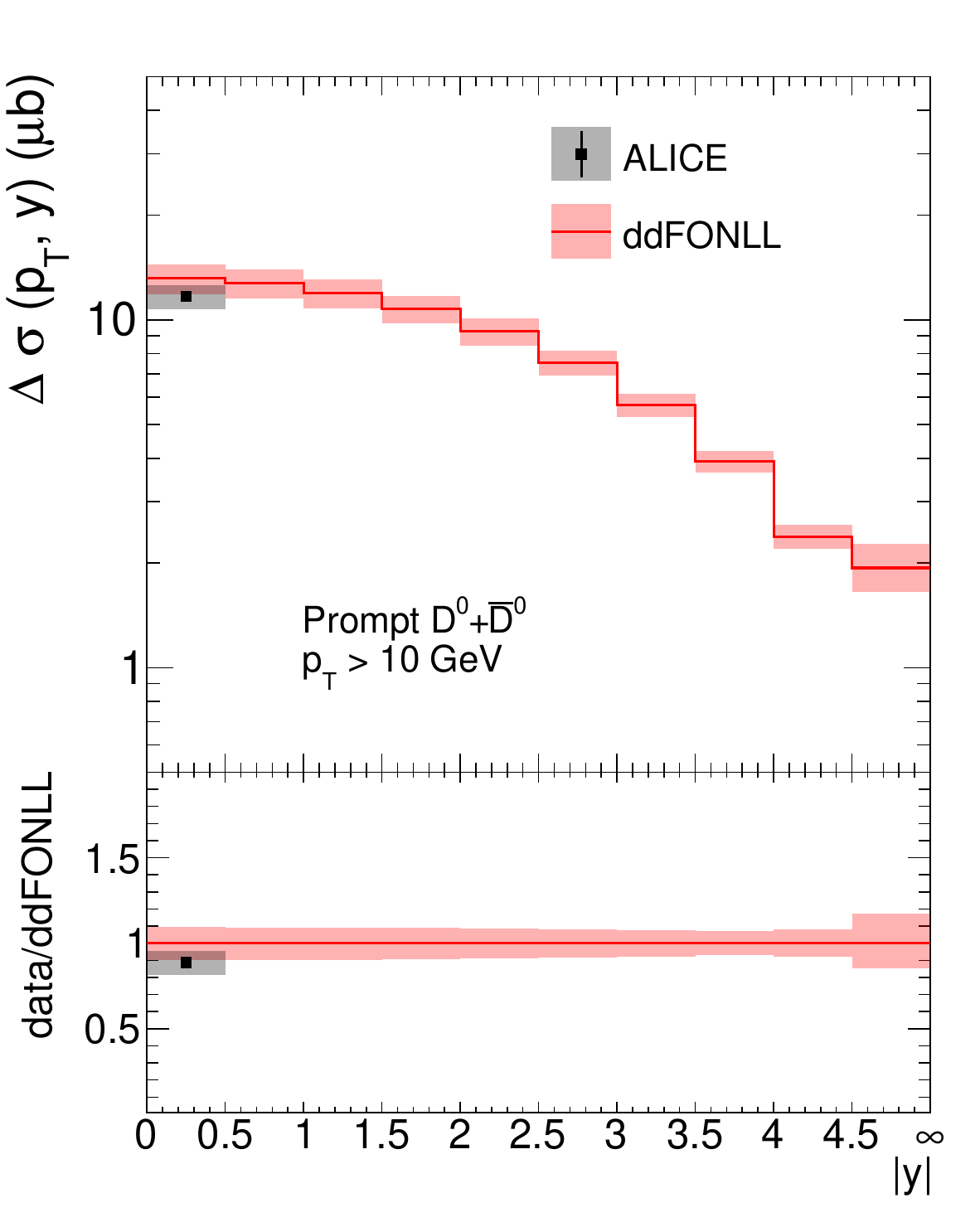}
 \end{center}
 \caption{$D^0+\overline{D}^0$ cross sections at $\sqrt{s} =$ 13 TeV as a function of $|y|$. The red bands of ddFONLL show the total uncertainty (CTEQ6.6 PDF $\oplus$ $\tilde{f}$ $\oplus$ $\chi^2$).} \label{fig:ddFONLL_d013TeV_yDiff_totUnc}
\end{figure*}

The ddFONLL parametrization is then defined by $d\sigma_{H_c}^{\text{ddFONLL}}(\mu_f^b, \mu_r^b, m_c^b, \alpha_K^b)$, where $\mu_f^b$, $\mu_r^b$, $m_c^b$ and $\alpha_K^b$ are the best parameters and the scan uncertainty band includes all the parameter sets within the $\Delta \chi^2$ ellipsoid.
The ddFONLL parametrization and the measurements used as input for them are shown as a function of $p_T$ in Figure \ref{fig:ddFONLL_d05TeV_yDiff_totUnc}, %\ref{fig:ddFONLL_dstar7TeV_yDiff_totUnc}
and \ref{fig:ddFONLL_d013TeV_yDiff_totUnc} for 5 %, \textcolor{red}{7 \cite{bph_c7TeV}}
 and 13 TeV, respectively. 
The ddFONLL uncertainties include the uncertainties of $\tilde{f}$, the $\chi^2$ scan (i.e., $\mu_f, \mu_r, m_c$, and $\alpha_K$ uncertainties), and the PDFs, added in quadrature. 
Note that the ddFONLL parametrization describes the data well in the
full phase space, which is consistent with the assumption of rapidity independence of the fragmentation (Assumption~\ref{ast:nonUni2}).
% D0 5 TeV
% D0 13 TeV

\section{Validation and application to different final states}
\label{sect:validation}

\begin{figure*}
 \begin{center} 
  \includegraphics[height=0.35\textheight]{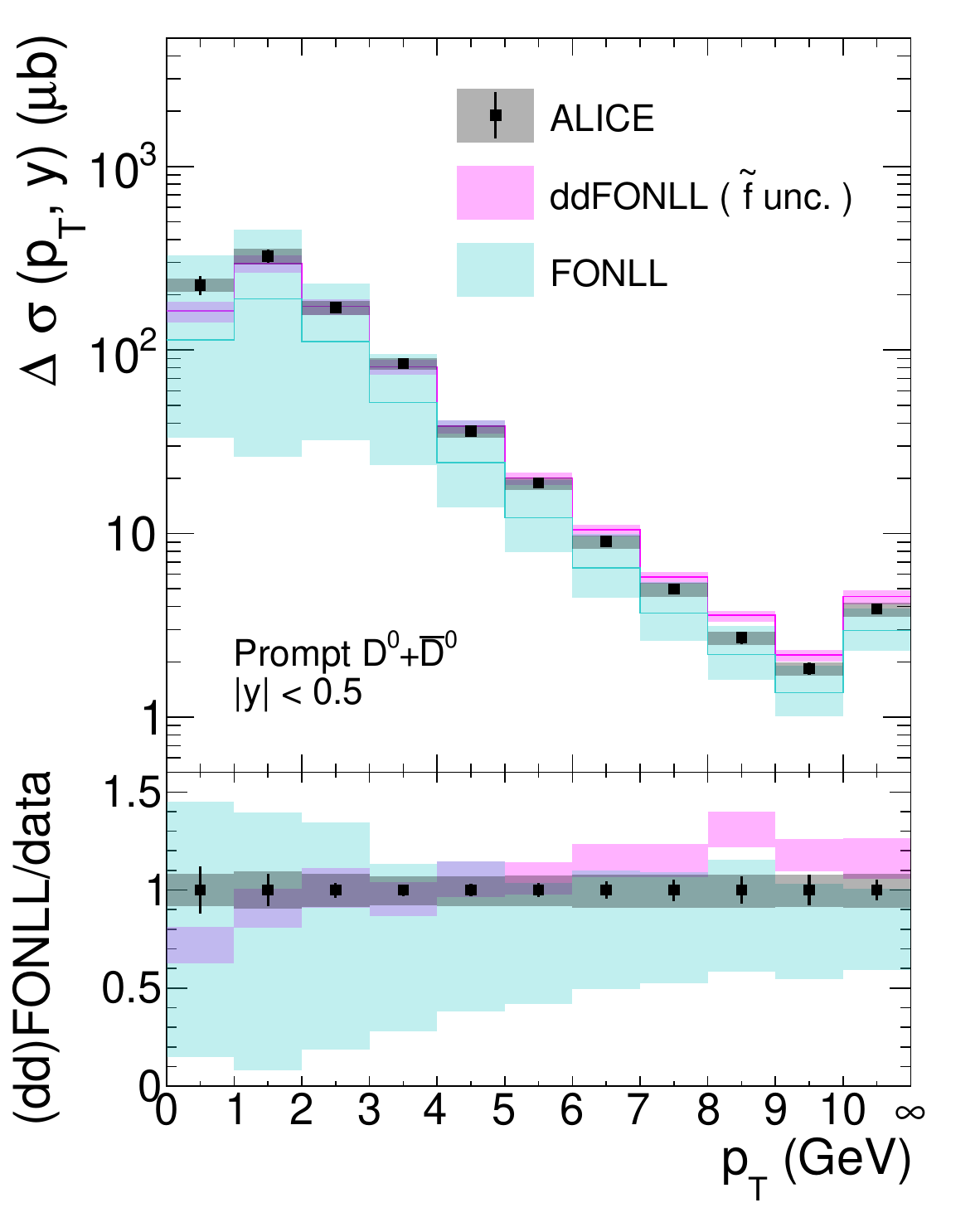}
  \includegraphics[height=0.35\textheight]{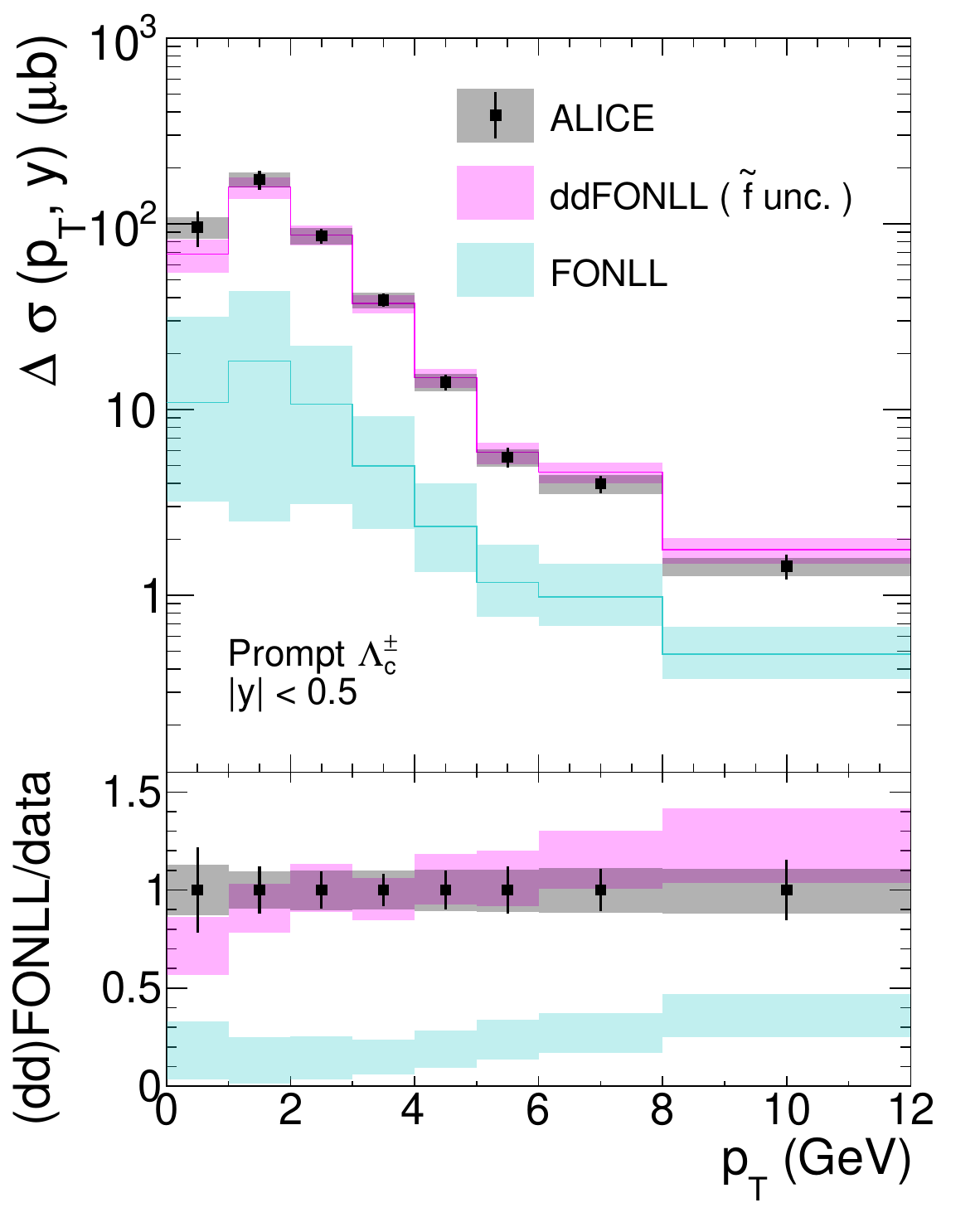}
 \end{center}
 \caption{$D^0+\overline{D}^0$ (left) and $\Lambda_c^{\pm}$ (right) cross sections at $\sqrt{s} =$ 5 TeV as a function of $p_T$ at central rapidity. The ddFONLL parametrization with $\tilde{f}$ uncertainty but still w/o PDF uncertainties (the pink band) describes both the $D^0$ and $\Lambda_c^+$ data well.} \label{fig:ddFONLL_lambdac5TeV}
\end{figure*}

\begin{figure*}
 \begin{center} 
  \includegraphics[height=0.35\textheight]{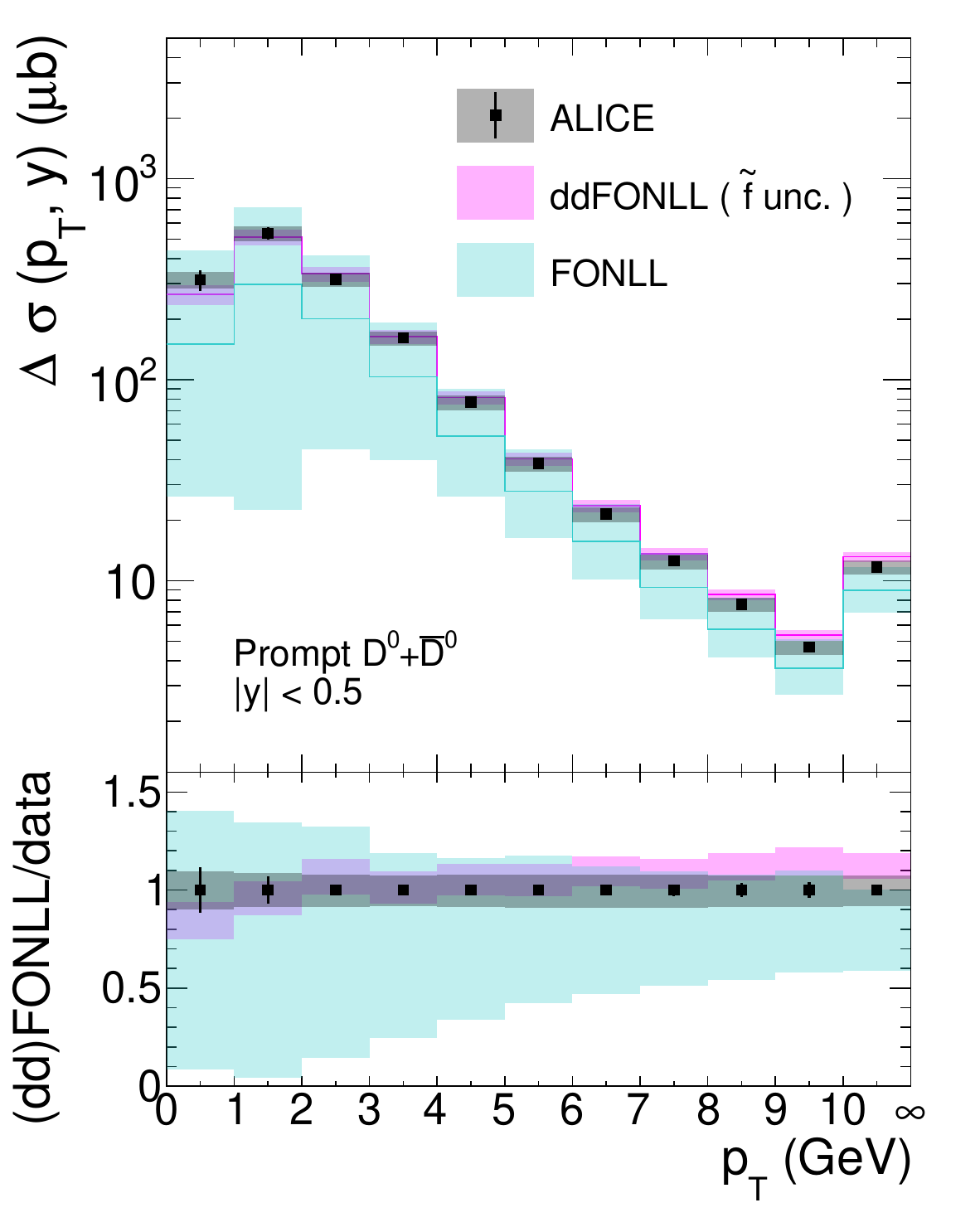}
  \includegraphics[height=0.35\textheight]{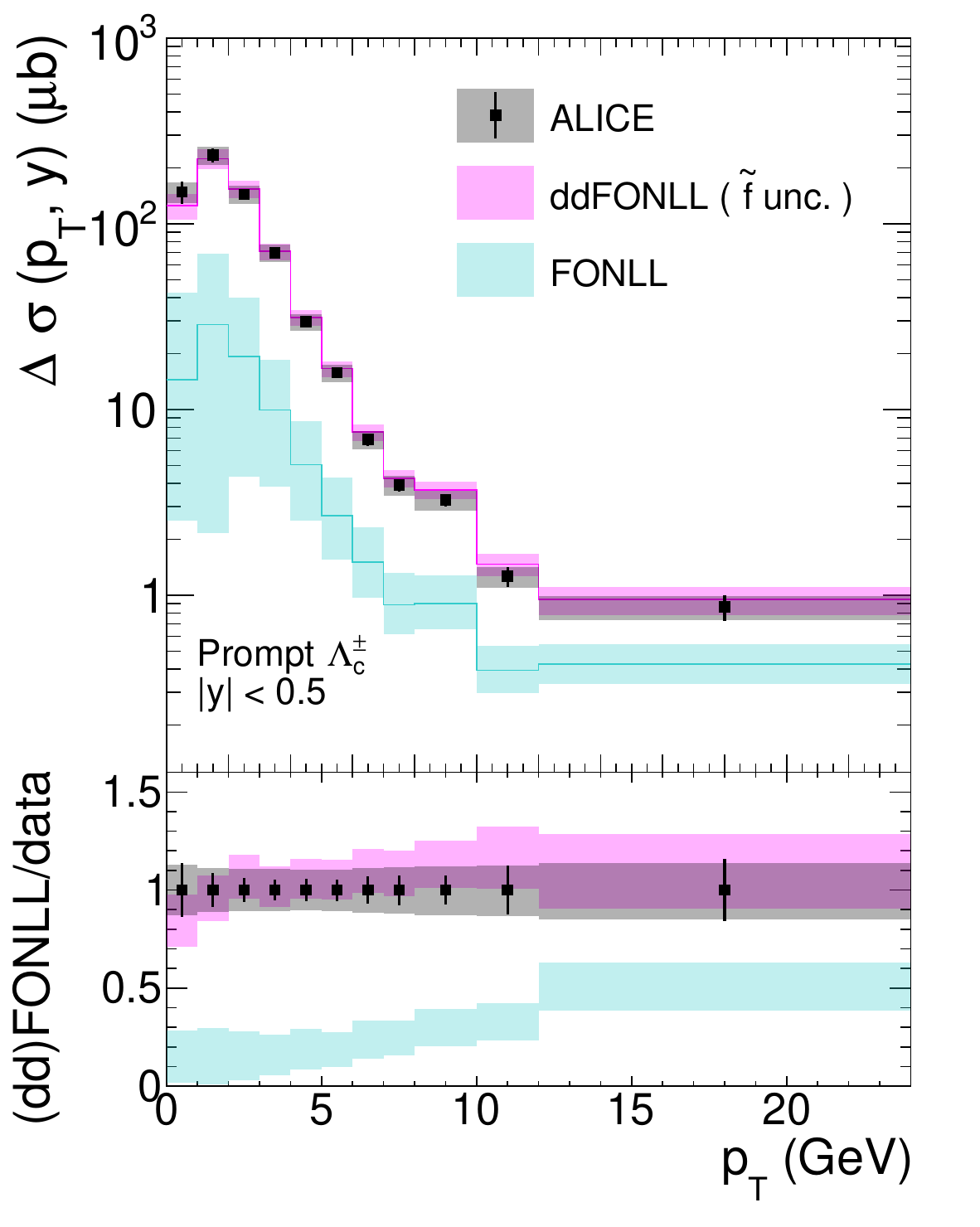}
 \end{center}
 \caption{$D^0+\overline{D}^0$ (left) and $\Lambda_c^{\pm}$ (right) cross sections at $\sqrt{s} =$ 13 TeV as a function of $p_T$ at central rapidity. The ddFONLL parametrization with $\tilde{f}$ uncertainty but still w/o PDF uncertainties (the pink band) describes both the $D^0$ and $\Lambda_c^+$ data well.} \label{fig:ddFONLL_lambdac13TeV}
\end{figure*}

As a cross check, the ddFONLL parametrizations for the $\Lambda_c^+$ spectrum were compared with the ALICE measurements at $\sqrt{s} =$ 5 and 13 TeV. The $\Lambda_c^+$ ddFONLL parametrization was derived with the best parameters of the fit of the $D^0$ measurements at $\sqrt{s} =$ 5 or 13 TeV, but applying $\tilde{f}_{\Lambda_c^+}$ (the blue points in Figure \ref{fig:fTilde}) instead of $\tilde{f}_{D^0}$, which is shown by the pink band in Figure \ref{fig:ddFONLL_lambdac5TeV} and \ref{fig:ddFONLL_lambdac13TeV}.
% 5 TeV
% 13 TeV
The original FONLL theory (the blue band), which is based on the universality assumption, totally disagrees with the measurements in the $\Lambda_c^+$ comparison, while the ddFONLL parametrization describes both the $D^0$ and $\Lambda_c^+$ measurements well, as it should do by construction.
This means that if the $\Lambda_c$ measurements would be used in the extrapolation to the total cross section in Section \ref{sec:totXsec} instead of the $D^0$ data, the result would remain consistent, although with larger uncertainties. 

\begin{figure*}
 \begin{center} 
  \includegraphics[height=0.35\textheight]{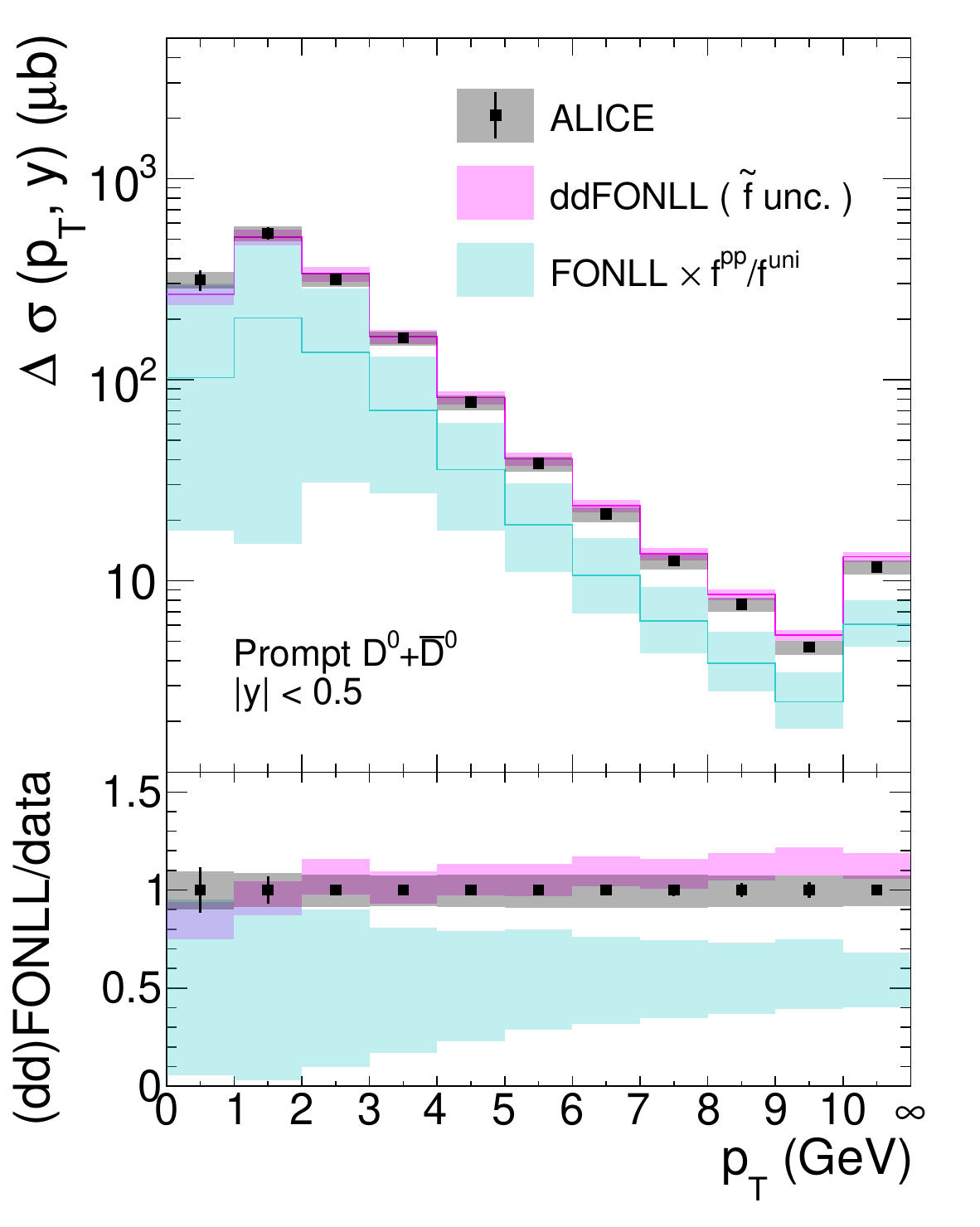}
  \includegraphics[height=0.35\textheight]{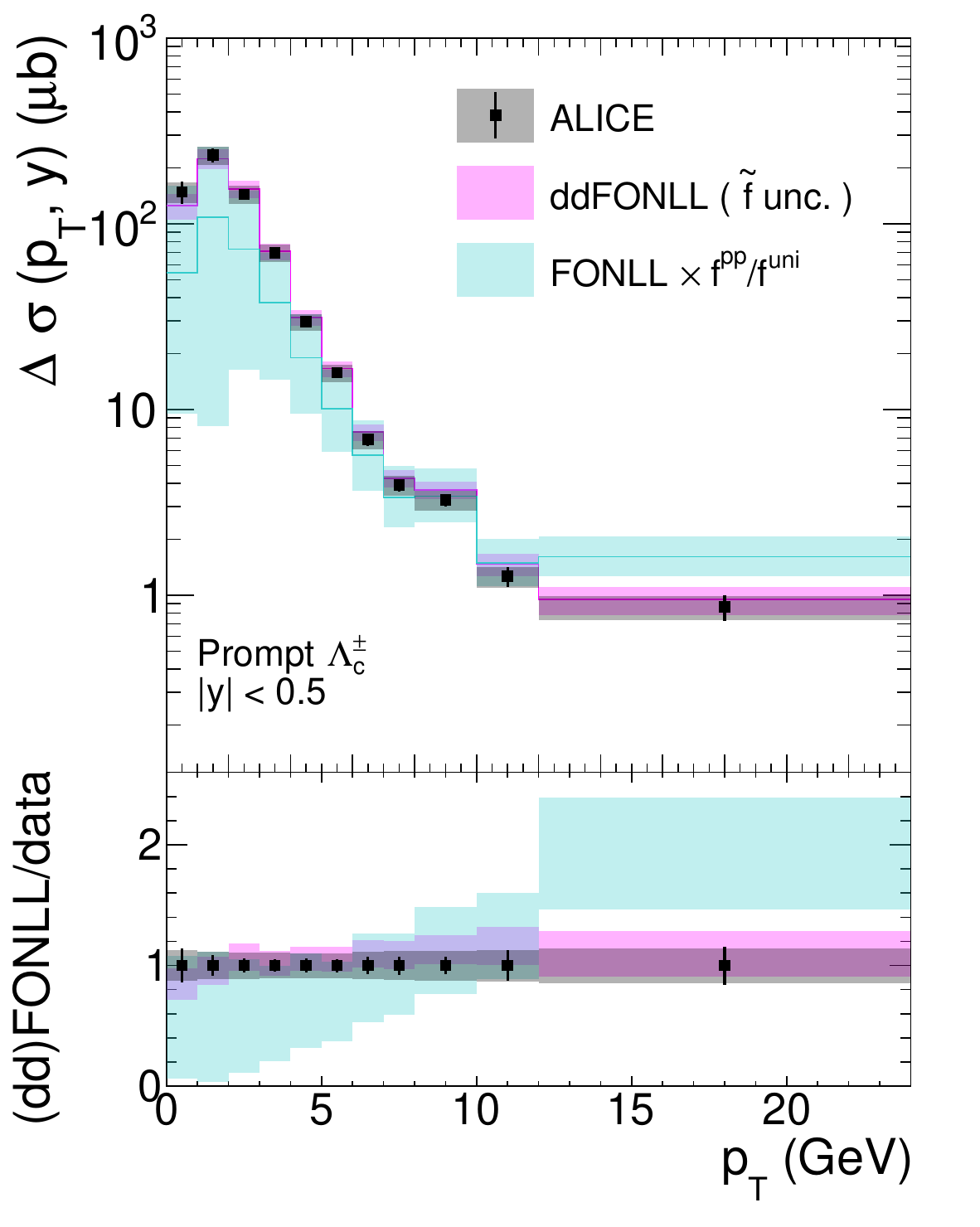}
 \end{center}
 \caption{$D^0+\overline{D}^0$ (left) and $\Lambda_c^{\pm}$ (right) cross sections at $\sqrt{s} =$ 13 TeV as a function of $p_T$ at central rapidity, compared to FONLL with a simple $p_T$-independent rescaling according to the average fragmentation fractions $f^{pp}$ as measured by ALICE. Despite much larger uncertainties, the FONLL prediction with average
fragmentation fraction rescaling describes both the $D^0$ and $\Lambda_c^+$ data less well than the $p_T$-dependent ddFONLL parametrization (same pink band as in previous figure).} \label{fig:ddFONLL_lambdac13TeValt}
\end{figure*}

As an alternative to the $p_T$ dependent $\tilde f$ correction in Fig.~\ref{fig:fTilde},
it has been suggested to try a simple rescaling of the FONLL prediction by
the average fragmentation fractions as measured by ALICE
(Fig.~\ref{fig:fragfrac_ee_ep}).  As can be seen from Fig. \ref{fig:ddFONLL_lambdac13TeValt}, this would improve the average normalization
of the predicted $\Lambda_c$ cross section compared to data but would still
somewhat disagree in shape, i.e. undershoot at low $p_T$ and overshoot at
high $p_T$ (see Fig. \ref{fig:ddFONLL_lambdac13TeValt}).
The prediction for $D^0$ would still agree at low $p_T$ within uncertainties,
but no longer at high $p_T$
(see Fig. \ref{fig:ddFONLL_lambdac13TeValt}).
In addition to disagreeing with the high $p_T$ charm data, this would also make
the predicted high $p_T$ charm behaviour inconsistent with LEP and with the
standard heavy flavour treatment in high $p_T$ charm jet tagging at LHC.
This simpler option is therefore not a fully viable alternative. 
% 13 TeV alternative

A further cross check for Assumptions~\ref{ast:nonUni1} and \ref{ast:nonUni2} can be obtained by comparing the ddFONLL parametrization also with $\Xi_c^0$ measurements, e.g. at $\sqrt{s} =$ 5 TeV from ALICE~\cite{ALICE_Xc_5TeV}.
This successful cross check is documented in Ref.~\cite{Geiser:2024qsq}. 

To extend the usage of the ddFONLL approach to $D^*$ final states in measurements such as those in \cite{ALICE_Dmeson_2p76TeV,ALICE_DmesonRatios_5TeV,ALICE_Dmeson_7TeV,ALICE_Dmeson_7TeV_2,ALICE_cTotXsec_7TeV,LHCb_Dmesons_5TeV,LHCb_Dmesons_7TeV,LHCb_Dmesons_13TeV,ATLAS_Dmeson_7TeV,CMS_Dmesons_13TeV,bph_c7TeV},
a set of $\tilde f$ functions is needed also for $D^*$ final states. 

\begin{figure}
 \begin{center}
  \includegraphics[width=0.45\textwidth]{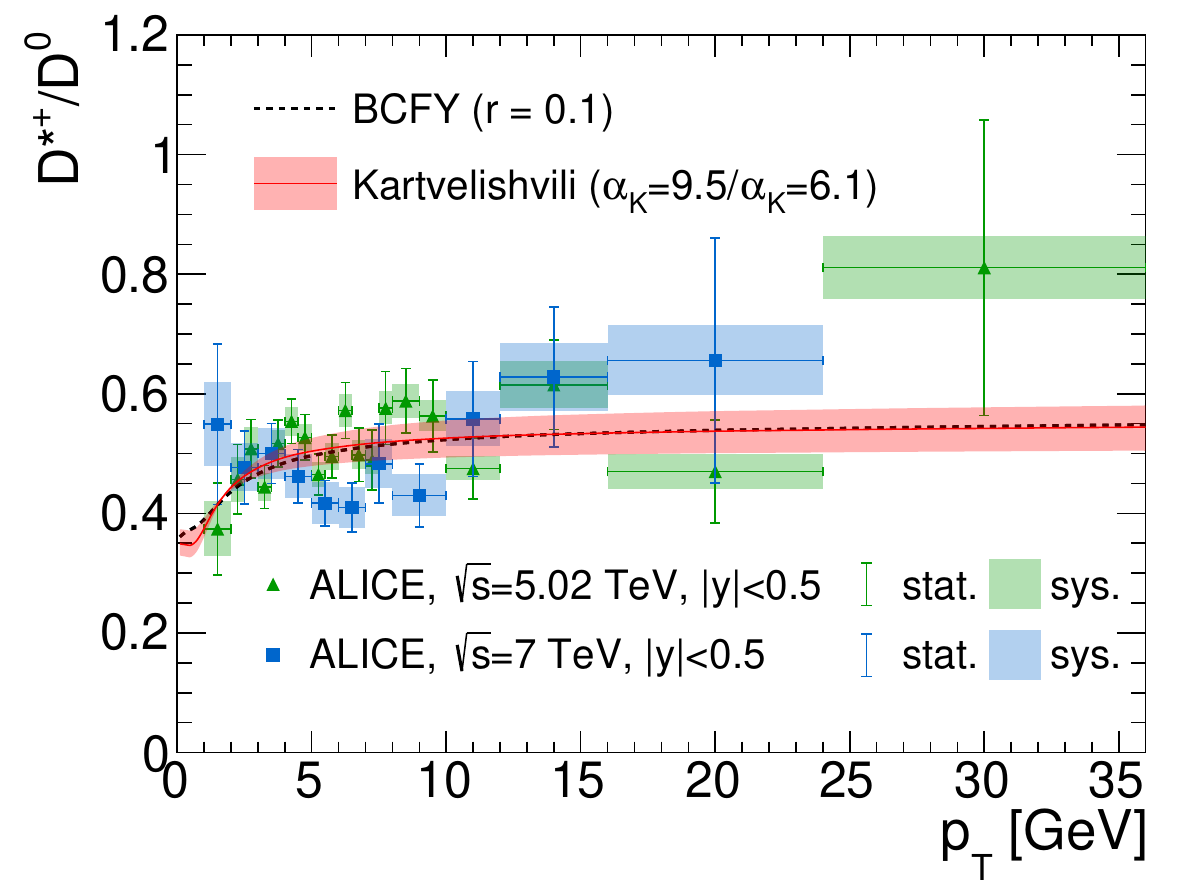}
 \end{center}
 \caption{$D^*/D^0$ comparison between ALICE measurements \cite{ALICE_DmesonRatios_5TeV} at $\sqrt{s} =$ 5~TeV (the green triangle points) and 7~TeV (the blue square points) and FONLL predictions (the black dashed and red solid line).} \label{fig:dstarToD0}
\end{figure}

To obtain these, $D^{*+}$/$D^0$ ratio measurements from ALICE at $\sqrt{s} =$ 5 and 7~TeV~\cite{ALICE_DmesonRatios_5TeV} were compared with a prediction extracted based on $e^+e^-$ data, as shown in Figure \ref{fig:dstarToD0}.
The dashed curve is the prediction which was obtained by FONLL using the BCFY functions for $D^{*+}$ and $D^0$ \cite{fragfunc_bcfy2, fonll2}.
The comparison between the ALICE measurements and the prediction shows consistency within the measurement uncertainties. This means that the same $F_{MS} (p_T)$ as for the ground states in (\ref{eq:fTilde_D0})ff can be applied for $D^{*+}$,
\begin{equation} \label{eq:fTilde_Dstar}
 \tilde{f}_{D^{*+}}(p_T) = f_{D^{*+}}^{uni}F_{MS}(p_T).
\end{equation}

In Figure~\ref{fig:dstarToD0}, the FONLL prediction derived with the Kartvelishvili function \cite{fragfunc_kart}, which is defined as
\begin{equation} \label{eq:kartFunc}
 D^{\text{NP}}(x) = (\alpha_K + 1)(\alpha_K + 2)x^{\alpha_K}(1-x),
\end{equation}
is also shown by the red curve.
The $\alpha_K$ values were determined for $D^{*+}$ and $D^0$, respectively, by comparing the FONLL predictions derived using the Kartvelishvili function to the ones derived with the BCFY functions, tuned to LEP data. As a result, $\alpha_K =$ 9.5 ($D^{*+}$) and 6.1 ($D^0$) were used in Figure \ref{fig:dstarToD0}, which again gives consistent results compared to the ALICE measurements.
These are then the values for $\alpha_K$ valid for LEP. The values from the
ddFONLL fits in Table \ref{tb:bestPars} turn out to be consistent with these. 

The extrapolation of $D^*$ data can thus also use the Kartvelishvili function instead of the BCFY one. Introducing $\alpha_K$ as a free parameter in the $\chi^2$ scan can deal with a possible $p_T$ dependent ratio of $D^{*+}$ to $D^0$, which is expected and allowed to vary even in the universality case, as shown in Figure \ref{fig:dstarToD0}.

\section{Total charm cross sections} \label{sec:totXsec}

The total $H_c$ cross section $\sigma^{\rm tot}_{H_c}$ was determined by taking all measurements where available, and ddFONLL was taken for the non-measured kinematic ranges:
\begin{align} \label{eq:sigma_Hc}
 \sigma^{\rm tot}_{H_c} = \ &\Delta \sigma_{H_c}^{\text{data}}(\text{measured phase space}) \\ \nonumber
 + &\Delta \sigma_{H_c}^{\text{ddFONLL}} (\text{unmeasured phase space}).
\end{align}
The total charm cross section is then obtained by dividing the total $H_c$ cross section by the fragmentation fraction of $H_c$ as measured from $pp$ collisions (denoted by $f_{H_c}^{pp}$): 
\begin{equation} \label{eq:sigma_ccbar}
\sigma^{\rm tot}_{c\bar{c}} = \frac{\sigma^{\rm tot}_{H_c}}{f_{H_c}^{pp}}.
\end{equation}
At the moment of writing this report, $f_{H_c}^{pp}$ was measured only at $\sqrt{s} =$ 5 and 13 TeV from ALICE, of which the latest numbers can all be found in \cite{ALICE_cFragFrac_13TeV}.

Eventually, the values for $\sigma^{\rm tot}_{c\bar{c}}$ were determined at $\sqrt{s} =$ 5 and 13 TeV with Eqs.~\eqref{eq:sigma_Hc} and \eqref{eq:sigma_ccbar}, of which the results can be found in Tables \ref{tb:dsigma_data} and \ref{tb:dsigma_data_d013TeV}.
% dsigma table for 5 TeV
\begin{table*}
 \begin{center}
  \caption{The integrated $D^0+\overline{D}^0$ cross section ($\Delta\sigma_{D^0+\overline{D^0}}$) and the total charm cross section ($\sigma_{c\bar{c}}$) at $\sqrt{s} =$ 5 TeV with $f_{D^0}^{pp} = 0.391^{+0.030}_{-0.041}$.} \label{tb:dsigma_data}
  \renewcommand{\arraystretch}{1.5}
  \begin{tabular}{|l|r|r|l|}
   \hline
              &                   & [GeV]          & $\Delta\sigma_{D^0+\overline{D^0}}$ [mb] \\ 
   \hline
   ALICE      & $0.0 < |y| < 0.5$ & $0 < p_T < 36$ & $0.88^{+0.08}_{-0.08}$ \\                  
   \hline
   LHCb       & $2.0 < |y| < 2.5$ & $0 < p_T < 10$ & $0.70^{+0.06}_{-0.05}$ \\                
   \cline{2-4}
              & $2.5 < |y| < 3.0$ & $0 < p_T < 10$ & $0.68^{+0.04}_{-0.04}$ \\                
   \cline{2-4}
              & $3.0 < |y| < 3.5$ & $0 < p_T < 10$ & $0.59^{+0.03}_{-0.03}$ \\              
   \cline{2-4}
              & $3.5 < |y| < 4.0$ & $0 < p_T < 9$  & $0.48^{+0.03}_{-0.03}$ \\               
   \cline{2-4}
              & $4.0 < |y| < 4.5$ & $0 < p_T < 6$  & $0.32^{+0.02}_{-0.02}$ \\               
   \cline{2-4}
              & $\Sigma$          & $\Sigma$       & $2.75^{+0.17}_{-0.17}$ \\
   \hline
   \multicolumn{3}{|l|}{ALICE+LHCb}                & $3.64^{+0.19}_{-0.19}$ \\                
   \hline
   %\multicolumn{3}{|l|}{ddFONLL complement}        & $2.95^{+0.31}_{-0.33}(\tilde{f})^{+0.52}_{-0.44}(\text{PDF})^{+0.10}_{-0.09}(\chi^2)$ \\
   ddFONLL complement
              & $0.5 < |y| < 2.0$ & $p_T > 0$      & $2.29^{+0.24}_{-0.25}(\tilde{f})^{+0.45}_{-0.38}(\text{PDF})$ \\
   \cline{2-4}
              & $|y| > 4.5$       & $p_T > 0$      & $0.66^{+0.07}_{-0.08}(\tilde{f})^{+0.07}_{-0.06}(\text{PDF})$ \\ 
   \cline{2-4}
              & $2.0 < |y| < 3.5$ & $p_T > 10$     & \\ 
   \cline{2-3}
              & $3.5 < |y| < 4.0$ & $p_T > 9$      & $[8.45^{+0.66}_{-0.77}(\tilde{f})^{+0.47}_{-0.47}(\text{PDF})]\times 10^{-3}$ \\
   \cline{2-3}
              & $4.0 < |y| < 4.5$ & $p_T > 6$      & \\ 
   \cline{2-4} 
              & $\Sigma$          & $\Sigma$       & $2.95^{+0.31}_{-0.33}(\tilde{f})^{+0.52}_{-0.44}(\text{PDF})^{+0.10}_{-0.09}(\chi^2)$ \\
   \hline
   \hline
   $\sigma^{\rm tot}_{c\bar{c}}$ [mb] & \multicolumn{3}{l|}{$8.43 ^{+0.25}_{-0.25}\text{(data)} ^{+0.40}_{-0.42}(\tilde{f}) ^{+0.67}_{-0.56}\text{(PDF)} ^{+0.13}_{-0.12}(\chi^2) ^{+0.88}_{-0.65}(f^{pp})$} \\
                            & \multicolumn{3}{l|}{$8.43 ^{+1.21}_{-0.99}\text{(total)}$} \\
   \hline
  \end{tabular}
 \end{center}
\end{table*}
% dsigma table for the 13 TeV
\begin{table*}
 \begin{center}
  \caption{The integrated $D^0+\overline{D}^0$ cross section ($\Delta\sigma_{D^0+\overline{D^0}}$) and the total charm cross section ($\sigma_{c\bar{c}}$) at $\sqrt{s} =$ 13 TeV with $f_{D^0}^{pp} = 0.382^{+0.026}_{-0.045}$.} \label{tb:dsigma_data_d013TeV}
  \renewcommand{\arraystretch}{1.5}
  \begin{tabular}{|l|r|r|l|}
   \hline
              &                   & [GeV]          & $\Delta\sigma_{D^0+\overline{D^0}}$ [mb] \\
   \hline
   ALICE      & $0.0 < |y| < 0.5$ & $0 < p_T < 50$ & $1.50^{+0.14}_{-0.14}$ \\
   \hline
   LHCb       & $2.0 < |y| < 2.5$ & $0 < p_T < 15$ & $1.20^{+0.12}_{-0.11}$ \\
   \cline{2-4}
              & $2.5 < |y| < 3.0$ & $0 < p_T < 15$ & $1.25^{+0.09}_{-0.09}$ \\
   \cline{2-4}
              & $3.0 < |y| < 3.5$ & $0 < p_T < 15$ & $1.18^{+0.07}_{-0.07}$ \\
   \cline{2-4}
              & $3.5 < |y| < 4.0$ & $0 < p_T < 11$ & $1.04^{+0.07}_{-0.06}$ \\
   \cline{2-4}
              & $4.0 < |y| < 4.5$ & $0 < p_T < 7\ $& $0.78^{+0.06}_{-0.06}$ \\
   \cline{2-4}
              & $\Sigma$          & $\Sigma$       & $5.44^{+0.40}_{-0.38}$ \\
   \hline
   \multicolumn{3}{|l|}{ALICE+LHCb}                & $6.94^{+0.43}_{-0.41}$ \\ 
   \hline
   ddFONLL complement
              & $0.5 < |y| < 2.0$ & $p_T > 0$      & $4.24^{+0.38}_{-0.39}(\tilde{f})^{+0.87}_{-0.73}(\text{PDF})$ \\
   \cline{2-4}
              & $|y| > 4.5$       & $p_T > 0$      & $2.10^{+0.20}_{-0.21}(\tilde{f})^{+0.24}_{-0.20}(\text{PDF})$ \\ 
   \cline{2-4}
              & $2.0 < |y| < 3.5$ & $p_T > 15$     & \\ 
   \cline{2-3}
              & $3.5 < |y| < 4.0$ & $p_T > 11$     & $[3.63^{+0.20}_{-0.22}(\tilde{f})^{+0.19}_{-0.20}(\text{PDF})]\times 10^{-2}$ \\
   \cline{2-3}
              & $4.0 < |y| < 4.5$ & $p_T > 7$      & \\ 
   \cline{2-4} 
              & $\Sigma$          & $\Sigma$       & $6.38^{+0.52}_{-0.60}(\tilde{f})^{+1.12}_{-0.93}(\text{PDF})^{+0.18}_{-0.14}(\chi^2)$ \\  
   \hline
   \hline
   $\sigma^{\rm tot}_{c\bar{c}}$ [mb] & \multicolumn{3}{l|}{$17.43 ^{+0.56}_{-0.53}\text{(data)} ^{+0.76}_{-0.78}(\tilde{f}) ^{+1.47}_{-1.22}\text{(PDF)} ^{+0.24}_{-0.18}(\chi^2) ^{+2.05}_{-1.19}(f^{pp})$} \\
                            & \multicolumn{3}{l|}{$17.43 ^{+2.70}_{-1.96}\text{(total)}$} \\
   \hline
  \end{tabular}
 \end{center}
\end{table*}
In each table, the integrated fiducial cross-sections of $H_c+\overline{H}_c$ for data and ddFONLL also are shown, of which the sum gives $\sigma^{\rm tot}_{H_c}$ after dividing by 2 to average particle and anti-particle state, accounting for the fact that an `open' $c\bar c$ pair always produces one of each.

For the 5 TeV results, and for `closeby' 7 TeV results treated elsewhere \cite{bph_c7TeV}, the fragmentation fractions are proposed to be taken from the ALICE measurements at $\sqrt{s} =$ 5 TeV, which are $0.391^{+0.030}_{-0.041}$ and $0.155^{+0.043}_{-0.022}$ for $D^0$ and $D^{*+}$, respectively \cite{ALICE_cFragFrac_5TeV}\footnote{Here, for historical reasons, the fragmentation fractions at $\sqrt{s} =$ 5 TeV are still based on the earlier measurements \cite{ALICE_cFragFrac_5TeV} rather than the latest ones in \cite{ALICE_cFragFrac_13TeV}. The difference is very small.}. However, while the $D^0$ fragmentation fraction is suitable for direct use, the fragmentation fraction for $D^{*+}$ has very large uncertainties especially for the upper value ($\sim 28\%$). Thus, since it is shown in Figure \ref{fig:dstarToD0} that $D^{*+}/D^0$ is consistent between $pp$ and $e^+e^-$ collisions, $f_{D^{*+}}^{uni}/f_{D^0}^{uni}$ was taken to translate $f_{D^0}^{pp} = 0.391^{+0.030}_{-0.041}$ into $f_{D^{*+}}^{pp}$ instead of taking the direct measurement. Assuming the uncertainties to be fully uncorrelated, the result turns out to be $f_{D^{*+}}^{pp} = 0.168^{+0.015}_{-0.019}$, which is then proposed to also be used as the $D^*$ fragmentation fraction for the total charm cross section extraction from $D^*$ at $\sqrt{s} =$ 7 TeV. The fragmentation fraction of $D^0$ at $\sqrt{s} =$ 13 TeV was taken from the ALICE measurements at $\sqrt{s} =$ 13 TeV: $f_{D^0}^{pp} = 0.382^{+0.026}_{-0.045}$ \cite{ALICE_cFragFrac_13TeV}.

In total 5 different uncertainties were determined for the total charm cross section: data uncertainty (for the parts covered by fiducial cross section measurements),
and $\tilde{f}$, PDFs, $\chi^2$ and $f_{H_c}^{pp}$ uncertainty (for the parts covered by the ddFONLL extrapolation). 
The data uncertainties were calculated by treating statistical uncertainties as fully uncorrelated and systematic uncertainties conservatively as fully correlated for each experiment, while both were treated as fully uncorrelated to the other experiment. Eventually the uncertainties are quoted as the sum of statistical and systematic uncertainties in quadrature.
The $\tilde{f}$, PDF and $\chi^2$ uncertainties were propagated from the ddFONLL fits.
Then the total uncertainty was calculated by treating all the individual uncertainties as fully uncorrelated.
The result is
\begin{equation}
  \sigma^{\rm tot}_{c\bar{c}} (5\ {\rm TeV}) = 8.43 ^{+1.21}_{-0.99}\text{(total)\ {\rm mb}},
\end{equation}
$$ \sigma^{\rm tot}_{c\bar{c}} (13\ {\rm TeV}) = 17.43 ^{+2.70}_{-1.96}\text{(total)\ {\rm mb}}. $$
The total charm cross section is thus rising substantially with $pp$ center of mass energy, and constitutes a sizeable part of the
total inelastic $pp$ cross section. 
The breakdown of the uncertainties can be found in Table \ref{tb:dsigma_data} and \ref{tb:dsigma_data_d013TeV}.
The extrapolation factors, i.e. the ratios of $\sigma_{cc}^{\rm tot}$ to the measured part of the cross sections for ALICE+LHCb in Tables \ref{tb:dsigma_data} and \ref{tb:dsigma_data_d013TeV}, turn out to be 1.8 and 1.9 at $\sqrt{s} =$ 5 and 13 TeV, respectively.
This overall factor is composed of a factor 1.6 (1.6) from interpolation in $0.5<|y|<2$, a factor 1.2 (1.3) from extrapolation in $|y|>4.5$, and a minor contribution from the high $p_T$ region in $2<|y|<4.5$ at $\sqrt{s} =$ 5 (13) TeV. 
The preliminary CMS+LHCb result at 7 TeV, with larger rapidity coverage, has an extrapolation factor of only 1.4 \cite{bph_c7TeV}.  

\begin{figure}
 \centering
 \includegraphics[width=0.45\textwidth]{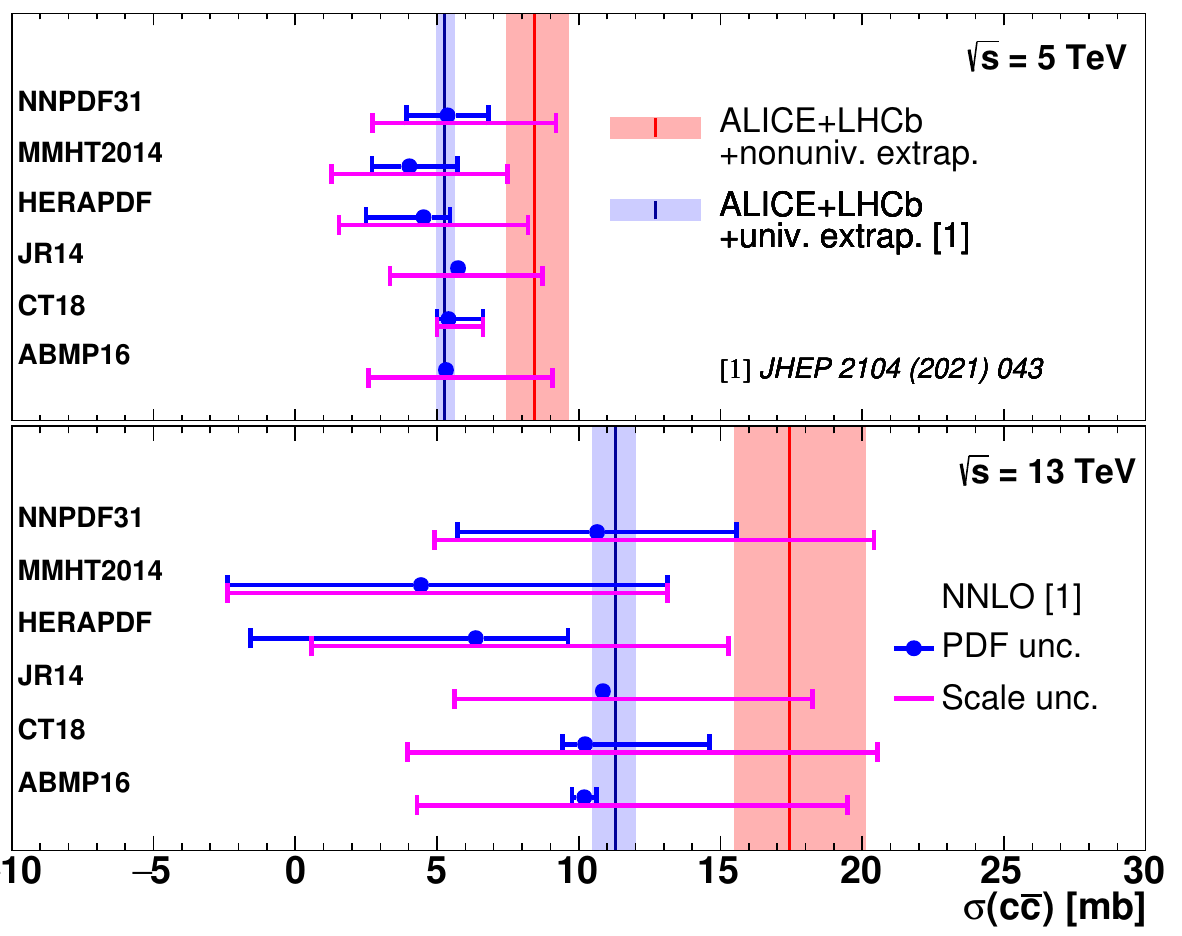}
 \caption{The total charm cross sections at $\sqrt{s} =$ 5 (top) %, 7 (middle)
   and 13 TeV (bottom), with figures adapted from \cite{nnloCharm1}. The vertical red bands are the total charm cross sections provided in this work.} \label{fig:totXsec_5_7_13TeV}
\end{figure}

These total charm-pair cross sections, obtained by extrapolating the $D^0$ cross sections at $\sqrt{s} =$ 5 and 13 TeV,
%and the $D^{*+}$ cross sections at $\sqrt{s} =$ 7 TeV with the non-universal charm fragmentation
are compared to NNLO QCD predictions with various PDF sets in Figure \ref{fig:totXsec_5_7_13TeV}.
The predictions in this figure, provided in \cite{nnloCharm1}, are based directly on NNLO theory at parton level.
The extrapolated measurements show good agreement with the upper bands of the theoretical uncertainties. 
Compared to the earlier measurement extrapolations, also provided in \cite{nnloCharm1} still with the assumption of fragmentation universality, which are shown by the blue bands in Figure \ref{fig:totXsec_5_7_13TeV}, the total charm cross sections at both center-of-mass-energies are increased significantly after the treatment of non-universal fragmentation. The increased uncertainties reflect the
current uncertainties of the LHC measurements entering the non-universality
treatment, and can be reduced again by future more precise measurements.
The results in this work thus supersede these previous extractions.

In the examples explicitly treated here the fiducial measurements (Tables \ref{tb:dsigma_data} and \ref{tb:dsigma_data_d013TeV}) extend all the way down to $p_T=0$ GeV, and thus the bulk of the $p_T$ dependence of the cross section is covered by measurement. Furthermore the extrapolation into nonmeasured rapidity regions can alternatively be treated in a $p_T$-averaged way.
Thus the differences between the central values of the ``universal'' and ``non-universal'' extrapolations in Figure \ref{fig:totXsec_5_7_13TeV} are effectively mainly driven by the difference between the $e^+e^-$ $f^{uni}_{D^0}$ value of $0.6141\pm0.0073$ used in \cite{nnloCharm1} and the measured average $f^{pp}_{D^0}$ value of $0.391^{+0.030}_{-0.041}$ ($0.382^{+0.026}_{-0.045}$) used here, giving an enhancement by a factor 1.57 (1.61) at $\sqrt{s}=$ 5 (13) TeV.
As can be seen from Tables \ref{tb:dsigma_data} and \ref{tb:dsigma_data_d013TeV}, the final effect of the ``high $p_T$'' extrapolation ($p_T > 6$ GeV or higher), strongly influenced by the treatment of the $p_T$ dependence, is numerically small since its relative contribution is small.
Overall, the central value for the total cross section increases by a factor 1.60 (1.54) at $\sqrt{s} =$ 5 (13) TeV.
In addition, in particular for the data driven uncertainty evaluation, needed for a reliable uncertainty estimate, there are effects from all the other treated ingredients.

This situation changes significantly when meson data are included that fill the gaps in the rapidity dependence, but partially lack measurements in the low $p_T$ region e.g. below 1 or even 2 GeV, as it is e.g. the case for the application of the method in \cite{bph_c7TeV}. Then the corrections for the nonmeasured low-$p_T$ regions, which dominate the extrapolation, change from the average $f^{pp}$ ratios to the somewhat higher low-$p_T$ only $\tilde{f}$ ratios illustrated in Figs. \ref{fig:Fms_Fby} and \ref{fig:fTilde}, which creates a noticeable effect also for the central value.
As illustrated in Figs. \ref{fig:ddFONLL_lambdac5TeV} and \ref{fig:ddFONLL_lambdac13TeV}, the change would be more drastic if the extrapolation would be applied to baryon measurements instead of meson measurements.
The advantage of the ddFONLL method is that (within uncertainties) the central result of the extrapolation does neither depend on the details of the choice or availability of the measured fiducial range, nor on the choice of the charm hadron final state being extrapolated. Of course, when fewer or less precise data are used, the absolute uncertainty of the extrapolation increases.

For the purpose of QCD sensitivity studies, we also compute a pseudodata point
for the total $pp$ charm cross section at $\sqrt{s} = 0.9$~TeV by simply
applying all the central ddFONLL parameters obtained at 5 TeV (Table \ref{tb:bestPars}) to FONLL calculations at 0.9 TeV and integrating the resulting cross
sections, assuming a similar fraction of the cross section to be measured.
Again for simplicity, the total relative uncertainty is assumed to be the same
as the one for the 5 TeV result, i.e. includes also a pseudouncertainty from
potential future measurements. 
The result is
\begin{equation}
  \sigma^{\rm tot}_{c\bar{c}, {\rm pseudo}} (0.9\ {\rm TeV}) = 1.67 \pm 0.24\ {\rm mb}.
\end{equation}
Given some indications of potential $\sqrt{s}$ dependence of heavy quark
production fractions \cite{LHCb:2021qbv}, which might still have to be
corrected for differences in the $p_T$ spectra,  it is of course not
clear so far whether such a simple extrapolation of ddFONLL parameters
should work over a wide $\sqrt{s}$ range. The hope is that this pseudopoint
might at some point be replaced
by an actual measurement at $\sqrt{s}=0.9$ TeV \cite{DPNOTE}.
For the moment this pseudopoint only serves for the purpose of QCD sensitivity
studies.

\section{Sensitivity to QCD parameters at NNLO} \label{sec:QCDsens}

The total charm cross section measurements obtained in this work at $\sqrt{s} =$ 5 and 13 TeV can now be used to constrain input parameters for genuine NNLO QCD calculations, specifically, the $\overline{\text{MS}}$ charm mass $m_c(m_c)$, as well as the proton PDFs. For sensitivity studies, the preliminary data point at 7 TeV \cite{bph_c7TeV}, as well as the pseudodata point at 0.9 TeV explained in the
previous section, can also be included.
The $\overline{\text{MS}}$ scheme is chosen because it was shown that using this mass definition improves convergence of the perturbative expansion of heavy-quark production total cross sections~\cite{Garzelli:2015psa}.

\begin{figure*}
	\centering
	\includegraphics[width=0.495\textwidth]{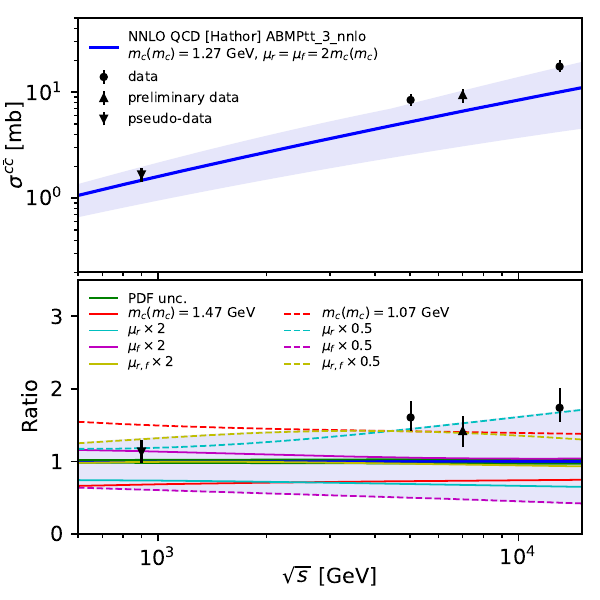}
	\includegraphics[width=0.495\textwidth]{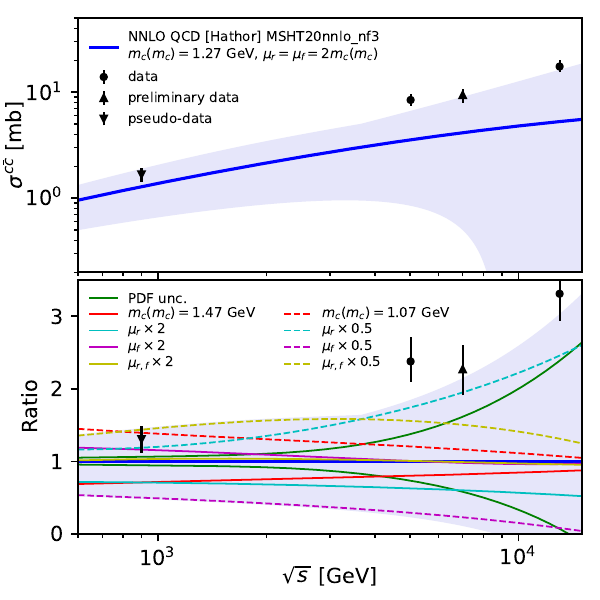}
	\caption{The total charm cross sections at NNLO QCD computed using ABMPtt\_3\_nnlo (left) and MSHT20nnlo\_nf3 (right) as a function $\sqrt{s}$. The lower panel displays the theoretical predictions and the data or pseudodata normalized to the central theoretical prediction.}
	\label{fig:thpred}
\end{figure*}

In Fig.~\ref{fig:thpred} the data and the NNLO QCD predictions for the total charm production cross section as a function of $\sqrt{s}$ are shown. The predictions are computed using the Hathor program~\cite{hathor} interfaced in xFitter~\cite{alekhin:2014irh}. For total top quark production calculations, it is known since a long time \cite{Langenfeld:2009wd,Dowling:2013baa} that a central scale choice of $m_t(m_t)$ gives a good description of the data with the $\overline{\text{MS}}$ running mass scheme. Also, in general, theoretical arguments exist \cite{hq_HERA,Geiser:2007tw} why choosing half the `natural' renormalization scale may often give appropriate estimates of central values and uncalculated higher order corrections, in particular for heavy flavour production. Here, partially motivated by technical (PDF) and convergence (closeness to $\Lambda_{QCD}$)
issues, the factorization and renormalization scales are set to $\mu_R=\mu_F=2m_c(m_c)=\mu_{0c}$ for the central calculation, i.e. on the high side of the relevant range. This may lead to an underestimate of the corresponding central values (also see e.g. the comparisons in \cite{CMS_Dmesons_13TeV} for the NLO single differential case), but should still cover the eventual true values within uncertainties.

To estimate the corresponding uncertainties, the scales are varied by a factor of two up or down according to the 7-point scale variation procedure. The proton PDFs are described by the ABMPtt\_3\_nnlo~\cite{Alekhin:2024bhs} or MSHT20nnlo\_nf3~\cite{msht} sets%
\footnote{The two other modern PDF sets by the CT~\cite{hou:2019efy} and NNPDF~\cite{NNPDF:2021njg} groups cannot be used to compute the predictions, because the former does not have eigenvectors for its variant with 3 light flavours and does not allow computing its PDF uncertainties, and the latter is available starting from $Q_{\textrm{min}}=1.65$~GeV only, which does not allow computing lower scale variation uncertainties.},
together with the corresponding uncertainties presented as eigenvectors. The number of light flavours is set to 3 consistently in the PDFs and in the matrix elements.
For each PDF set, the associated $\alpha_s(m_Z)$ value and $\alpha_s$ evolution is taken from LHAPDF \cite{lhapdf}.
The total theoretical uncertainty band is calculated by summing in quadrature the scale variation and PDF uncertainties. The $\overline{\text{MS}}$ charm mass is set to $m_c(m_c) = 1.27$~GeV \cite{ParticleDataGroup:2024cfk}. To illustrate the sensitivity to the charm quark mass, the latter is varied by $\pm 0.2$~GeV and the corresponding predictions are shown; however, this variation is not included into the total theoretical uncertainty band. In general, the theoretical uncertainty is dominated by the scale variations. However, for the MSHT20nnlo\_nf3 prediction the PDF uncertainty grows rapidly with increasing $\sqrt{s}$ and exceeds the scale variation uncertainties at $\sqrt{s} \sim 13$ TeV. For ABMPtt\_3\_nnlo the PDF uncertainty is small and does not exceed 5\% in the entire kinematic range. The $\mu_f=\mu_{0c}, \mu_r=0.5\mu_{0c}$ curve (dashed blue curve in the lower panels of Fig. \ref{fig:thpred}), with all other parameters fixed to be central, gives an excellent description of the data (and pseudodata) for the ABMPtt\_3\_nnlo case, and also a reasonable description of the data (within about two standard deviations) for the MSHT20nnlo\_nf3 case. 

\begin{figure*}
	\centering
	\includegraphics[width=0.495\textwidth]{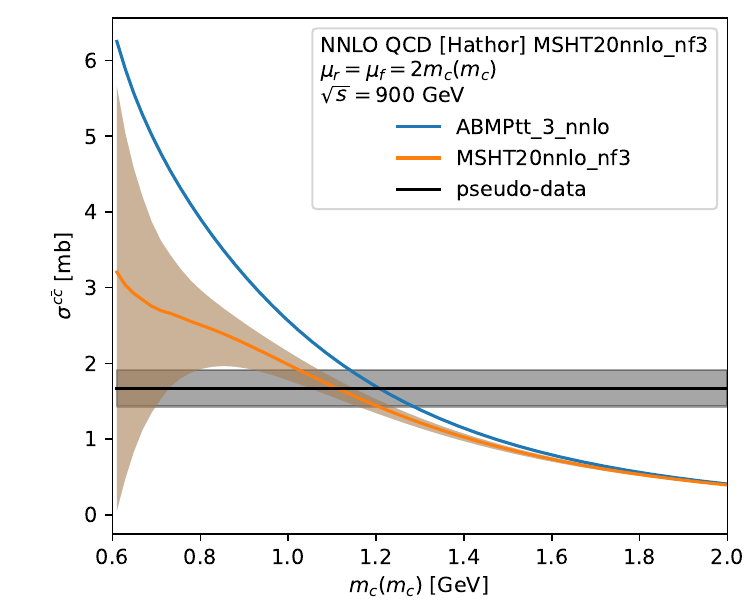}
	\includegraphics[width=0.495\textwidth]{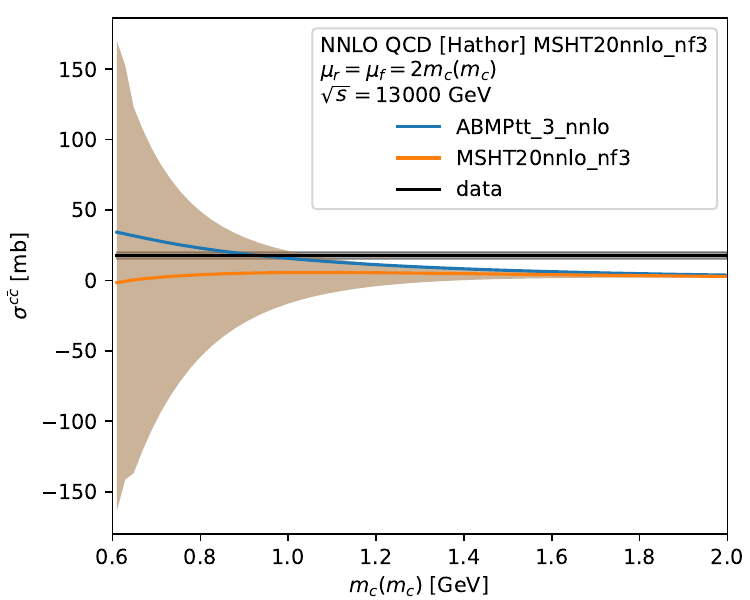}
	\caption{The total charm cross sections at NNLO QCD at $\sqrt{s}=0.9$ (left) and 13~TeV (right) computed using the ABMPtt\_3\_nnlo and MSHT20nnlo\_nf3 PDF sets. The corresponding data values with their uncertainties are shown as horizontal bands.}
	\label{fig:thpred_mcdep}
\end{figure*}

Overall, the measurements show good agreement with the QCD predictions up to the highest order known today.
In order to investigate the behavior of the predictions further, in Fig.~\ref{fig:thpred_mcdep} they are shown as a function of $m_c(m_c)$ for fixed $\sqrt{s}=0.9$ and $13$ TeV. At $\sqrt{s}=0.9$ TeV both predictions exhibit monotonic $m_c(m_c)$ dependence, as expected. On the contrary, at $\sqrt{s}=13$ TeV the central MSHT20nnlo\_nf3 predictions reaches its maximum at $m_c(m_c) \approx 1$~GeV, while at lower $m_c(m_c)$ it decreases and becomes even negative at $m_c(m_c) \approx 0.65$~GeV, accompanied by a large PDF uncertainty. Thus the present data on charm production can be used to constrain the proton PDFs, in particular the gluon distribution at small values of the partonic momentum fraction $x$, where it is not constrained by other data.

As a demonstration that the present charm data can pin down the uncertainty of some of the modern PDF sets, a profiling technique is employed~\cite{Paukkunen:2014zia,HERAFitterdevelopersTeam:2015cre}. It is based on minimizing the $\chi^2$ function constructed from theoretical predictions and data together with their uncertainties. In particular, for the present demonstration only the theoretical uncertainties arising from the PDFs are included in the $\chi^2$ function. They are included through nuisance parameters, and the values of these parameters at the minimum of $\chi^2$ are interpreted as optimized (profiled) PDFs, while their uncertainties determined using the tolerance criterion $\Delta\chi^2=1$ are interpreted as new PDF uncertainties.  Only the data point at $\sqrt{s}=13$ TeV is considered, since it corresponds to the lowest $x$ values probed in the process.
Also, since the scales are not varied, the ($\mu_f=2m_c(m_c), \mu_r=m_c(m_c)$) choice is used in the profiling, which already gives a good starting $\chi^2$ for the variation.
This profiling is of course not intended to replace a full PDF fit, but should
be suited to give an indication of the expected sensitivity. 

\begin{figure*}
	\centering
	\includegraphics[width=0.495\textwidth]{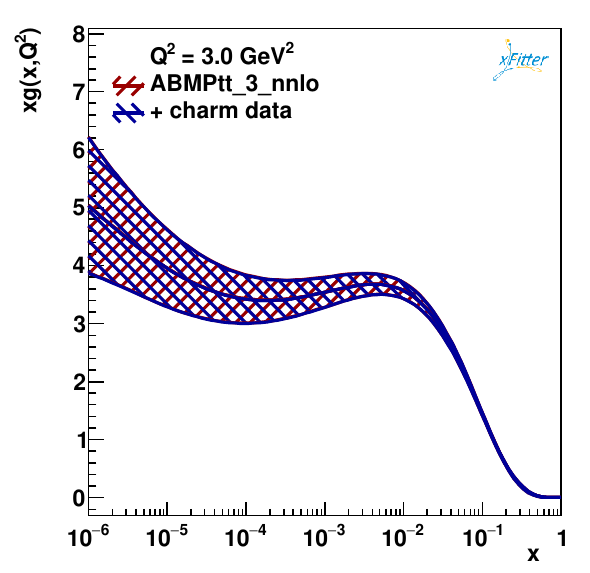}
	\includegraphics[width=0.495\textwidth]{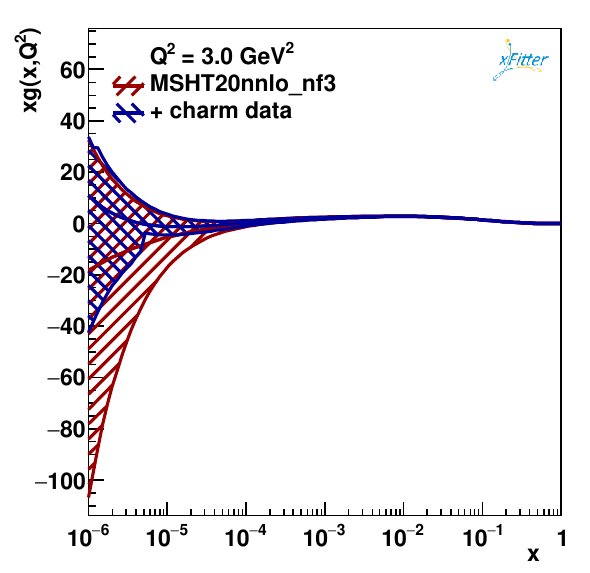}
	\caption{The gluon distribution of the original and profiled ABMPtt\_3\_nnlo (left) and MSHT20nnlo\_nf3 (right) PDF sets at the scale $\mu_f^2=3$~GeV$^2$.}
	\label{fig:pdfprof}
\end{figure*}

The original and profiled gluon PDF is shown in Fig.~\ref{fig:pdfprof}. For the ABMPtt\_3\_nnlo PDF set, the charm data do not provide any noticeable constraints. However, for the MSHT20nnlo\_nf3 set the profiled distribution is shifted towards larger values and has greatly reduced uncertainties at low $x$. This is due to the fact that the central MSHT20nnlo\_nf3 gluon distribution is negative at $x \lesssim 10^{-5}$ and does not describe the charm data well, as shown on Figs.~\ref{fig:thpred} and \ref{fig:thpred_mcdep}.

\begin{figure}
	\centering
	\includegraphics[width=0.5\textwidth]{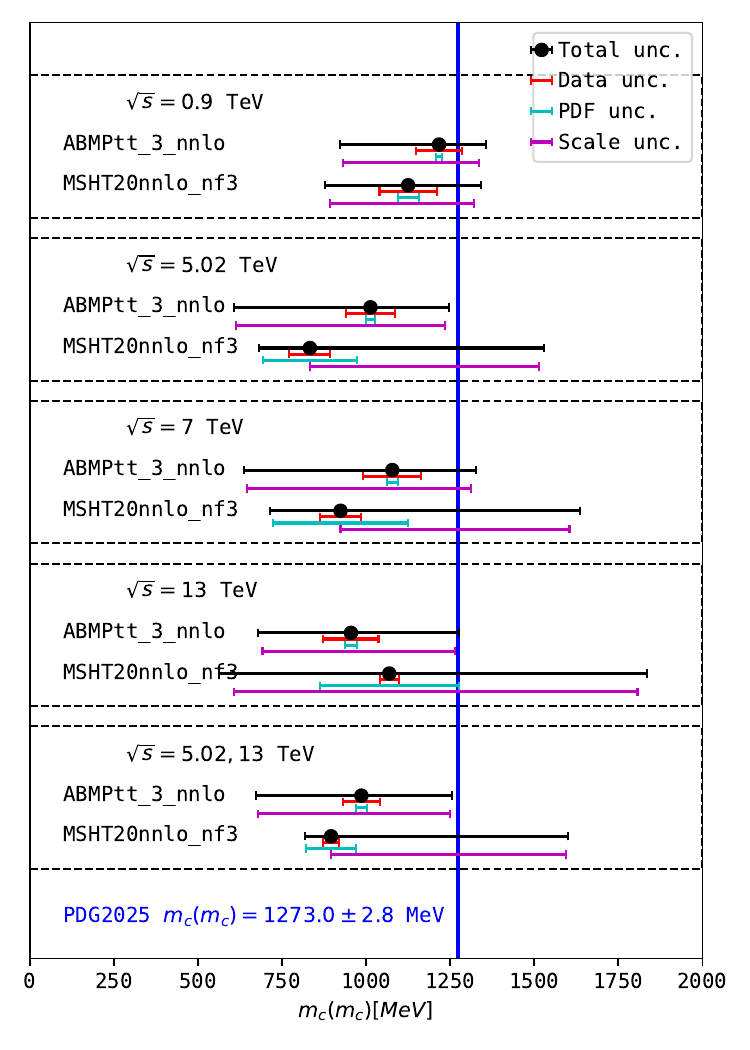}
	\caption{The $m_c(m_c)$ values extracted using various charm data points, together with their experimental, PDF and scale variation uncertainties. The world average value from Ref.~\cite{ParticleDataGroup:2024cfk} is shown as well.}
	\label{fig:mc}
\end{figure}

Furthermore, since the $m_c(m_c)$ variation of $0.2$~GeV shown on Fig.~\ref{fig:thpred} is comparable to the other theoretical uncertainties, it is possible to use the charm data to determine $m_c(m_c)$. This is done in the same way as the profiling procedure described above, by minimizing the $\chi^2$ between the data and theoretical predictions, but this time treating also $m_c(m_c)$ as a free parameter. Furthermore, to estimate the impact of scale variation uncertainties, the procedure is repeated by varying the scales according to the 7-point scale variation prescription. The resulting $m_c(m_c)$ values for each of the considered data or pseudodata points are shown in Fig.~\ref{fig:mc}. The uncertainties are dominated by those from the scale variations which are about $0.3$~GeV. The experimental uncertainty and PDF uncertainties are almost one order of magnitude smaller (except the extracted value using the data point at $\sqrt{s}=13$ TeV and MSHT20nnlo\_nf3 set, which is accompanied by the large PDF uncertainty as discussed above).
The values of $m_c(m_c)$ extracted using both the 5 and 13 TeV data points can be found in Tab.~\ref{tab:mc2} as well.
The result does not critically depend on which data point or combination of data points is being used.
The fact that the results are consistent across input cross sections with different respective ddFONLL extrapolation factors confirms that any potential extrapolation bias is covered by the uncertainties. 
Lower center-of-mass energy seems to slightly increase
the sensitivity to $m_c(m_c)$.  

The numerical results for the $m_c(m_c)$ extraction from the combination of the 5 and 13 TeV data points obtained in this work
are shown in Table \ref{tab:mc2}. 
The result for the ABMP case, which is less affected by the correlation with low-$x$ gluon PDF uncertainties, is
$$m_c(m_c) = 0.986^{+0.269}_{-0.313}  \ {\rm GeV}.$$
The uncertainties are dominated by the scale dependence. The specific
scale choice
$\mu_f=2m_c(m_c), \mu_r=m_c(m_c)$,
which describes the data well in Fig. \ref{fig:thpred},  gives the value
1.249~GeV.

\begin{table*}
 \begin{center}
  \caption{The $m_c(m_c)$ values extracted using the data points at $\sqrt{s}=5$ and 13 TeV obtained in this work, together with their uncertainties. Also the impact of each of the scale variations is given. The last row presents the central value extracted using the $(\mu_f/\mu_{0c}, \mu_r/\mu_{0c}) = (1, 0.5)$ choice.} \label{tab:mc2}
  \renewcommand{\arraystretch}{1.5}
  \begin{tabular}{|l|r|r|}
   \hline
                                             & ABMPtt\_3\_nnlo & 
MSHT20nnlo\_nf3 \\
   \hline
   $m_c(m_c)$, GeV                            & 0.986 & 0.896 \\
   \hline
   data uncertainties                        & $\pm 0.055$ & $\pm 0.023$ \\
   \hline
   PDF uncertainties                         & $\pm 0.016$ & $\pm 0.074$ \\
   \hline
   scale uncertainties                       & ${}^{+0.263}_{-0.308}$ & 
${}^{+0.699}_{-0.0}$ \\
   \hline
   \hline
   $(\mu_f/\mu_{0c}, \mu_r/\mu_{0c}) = (0.5, 0.5)$ & +0.188 & +0.232 \\
   \hline
   $(\mu_f/\mu_{0c}, \mu_r/\mu_{0c}) = (1, 0.5)$   & +0.263 & +0.140 \\
   \hline
   $(\mu_f/\mu_{0c}, \mu_r/\mu_{0c}) = (0.5, 1)$   & +0.154 & +0.699 \\
   \hline
   $(\mu_f/\mu_{0c}, \mu_r/\mu_{0c}) = (2, 1)$     & +0.007 & +0.039 \\
   \hline
   $(\mu_f/\mu_{0c}, \mu_r/\mu_{0c}) = (1, 2)$     & -0.308 & +0.288 \\
   \hline
   $(\mu_f/\mu_{0c}, \mu_r/\mu_{0c}) = (2, 2)$     & -0.068 & +0.060 \\
   \hline
   \hline
   $m_c(m_c)(\mu_f/\mu_{0c}, \mu_r/\mu_{0c}) = (1, 0.5)$, GeV & 1.249 & 1.036 \\
   \hline
  \end{tabular}
 \end{center}
\end{table*}

The uncertainties on $m_c(m_c)$ are almost two orders of magnitude larger than the one of the current PDG value \cite{ParticleDataGroup:2024cfk},
$$ m_c(m_c) = 1.2730 \pm 0.0028 \ {\rm GeV},$$
but the results are consistent and constitute, to the knowledge of the authors, the first such extraction
from purely hadronic collisions at LHC. This can be interpreted as a nontrivial consistency check of the validity
of the perturbative QCD approach, within its large uncertainties, down to the scale of the charm quark mass,
even at LHC energies.

\section{Conclusions and Outlook}
\label{sect:con}

Total charm-pair cross sections in $pp$ collisions are interesting because 
they can be calculated to NNLO in QCD without any reference to fragmentation
effects. On the other hand, the fiducial differential charm cross sections 
from which the total cross sections must be extrapolated are currently 
known to NLO+NLL at most (e.g. FONLL), and must be treated for known effects of 
non-universal charm fragmentation. A new procedure using the FONLL framework 
as input for an empirical parametrization of the data in both shape and 
normalization, with all its parameters actually fitted to data, is used to 
derive so-called data-driven FONLL (ddFONLL) 
parametrizations which can be used to extrapolate the differential 
cross sections to total cross sections with minimal bias. 
This includes an empirical 
treatment of all known non-universal charm fragmentation effects, 
in particular for the baryon-to-meson ratio as a function of 
transverse momentum.
These parameterizations are then no longer theory predictions, but
theory-inspired parametrizations of all relevant existing data. 

Such ddFONLL parametrizations have been obtained by fitting ALICE and LHCb
$D^0$ production data, with parameters constrained to be consistent with
all other existing charm final state measurements at LHC and in $e^+e^-$ and
$ep$ data, also adding uncertainty estimates on those parameters that have not
yet been measured. The parametrizations at 5 and 13 TeV $pp$ center of mass are
found to be consistent with each other, suggesting that the method will also
work for other intermediate or closeby center-of-mass energies, which should
however still be fitted independently whenever possible since a slight
$\sqrt{s}$ dependence can not be excluded. 

Using the measurements for all bins in which measurements are available,
and the ddFONLL data parametrization in all others,  
the resulting 5 TeV and 13 TeV total charm pair production cross sections 
are obtained to be 
%\begin{eqnarray}
\begin{align}
 \sigma^{\mathrm{tot}}_{c\bar{c}}(5\ \mathrm{TeV})
 = 8.43 &^{+0.25}_{-0.25}\mathrm{(data)} ^{+0.40}_{-0.42}(\tilde{f}) \\ \nonumber
 &^{+0.67}_{-0.56}\mathrm{(PDF)} ^{+0.13}_{-0.12}(\mu_f, \mu_r, m_c, \alpha_K) \\ \nonumber
 &^{+0.88}_{-0.65}(f^{pp}_{D^0}) \ {\rm mb} \\ 
 = 8.43 &^{+1.21}_{-0.99}(\mathrm{total})\ \mathrm{mb}, \hspace{8.2cm} \\
 \sigma^{\mathrm{tot}}_{c\bar{c}}(13\ \mathrm{TeV})
 = 17.43 &^{+0.56}_{-0.53}\mathrm{(data)} ^{+0.76}_{-0.78}(\tilde{f}) \\ \nonumber
 &^{+1.47}_{-1.22}\mathrm{(PDF)} ^{+0.24}_{-0.18}(\mu_f, \mu_r, m_c, \alpha_K) \\ \nonumber
 &^{+2.05}_{-1.19}(f^{pp}_{D^0}) \ {\rm mb} \\
 = 17.43 &^{+2.70}_{-1.96}(\mathrm{total})\ \mathrm{mb}. \hspace{8.2cm}
\end{align}
%\end{eqnarray}
in which $f^{pp}_{D^0}$ refers to the integrated $D^0$ fragmentation fraction
measured at 5 TeV or 13 TeV, respectively. The respective extrapolation 
factors for unmeasured phase space are about 1.8 and 1.9.
These results were obtained from $D^0$ final states in the specified fiducial range, as an example. 

One of the advantages of the ddFONLL method is that (within uncertainties) the total charm cross section result neither depends on the kinematic fiducial range nor on the type of charm hadron chosen as the starting point for the extrapolation.
The full treatment of charm fragmentation non-universality, which comes with increased baryon production and decreased meson production both on average and as a function of $p_T$ in $pp$ collisions compared to $e^+e^-$/$ep$ collisions, can substantially change the total charm cross sections compared to previous determinations assuming charm universality in both shape and normalization.

Specifically, with the $D^0$ example in this work, the central values for the total cross sections increase by factors of 1.5-1.6 with respect to \cite{nnloCharm1}. As detailed in Section \ref{sec:totXsec}, this increase is dominated by the change of the average $D^0$ fragmentation fraction in $pp$ measurements with respect to previous $e^+e^-$ measurements, while the increase of the final uncertainty has significant contributions from all the other considered
parameters. This result thus supersedes the previous determination.
The measurements are still consistent with the NNLO predictions, but now 
situated towards the upper edge of the NNLO theory uncertainty band.     

Since the total charm-pair cross sections obtained in this way are
consistent with NNLO predictions, they allow first studies 
of their sensitivity e.g. to the charm-quark mass and/or the NNLO gluon PDF at 
very low proton momentum fraction $x$.

A significant potential to constrain the gluon PDF at low $x$
($O(10^{-4}-10^{-6})$) is found e.g. for the MSHT20 parametrization, which might
not yet be significantly constrained in this region from other data.
The 13 TeV total cross section, which accesses the lowest $x$ values,
is found to be particularly constraining in this respect. 
Using either the ABPMtt or the MSHT20 PDF, the running charm mass $m_c(m_c)$
obtained from allowing its variation is found to be consistent with the PDG
value of $1.2370 \pm 0.0028$ GeV, within its large scale variation
uncertainty of about 0.3 GeV.
This is the first such extraction from LHC charm production data at NNLO, and,
although its uncertainty is two orders of magnitude larger than the PDG
uncertainty, the consistency is nontrivial and indicates that perturbative QCD
continues to work down to the charm mass scale even at LHC energies, within
the uncertainties evaluated from purely perturbative scale variations. 

So far the novel ddFONLL procedure outlined in this paper has only been applied
to charm production at a few center-of-mass energies in $pp$ collisions at LHC.
It however has significant potential for extensions.

The identical procedure could be applied to beauty production provided that
the corresponding relevant $\tilde f(p_T)$ and $f^{pp}$ functions and values are
available from data.
In the beauty case, also differential calculations are available at
NNLO(+NNLL)~\cite{Czakon:2024tjr}, so the empirical $\tilde f(p_T)$ modifier
could just as well directly be applied
to such calculations. The resulting total cross sections could then be used for an extraction of $m_b(m_b)$.

If nuclear modification effects can be controlled the procedure could also
be expanded at least to heavy ion $pA$ collisions ($AA$ collisions are less
obvious since the nuclear modification effects are large also in the
nonperturbative final state). 

The procedure can also be expanded to lower center-of-mass energies at RHIC
or in the fixed target regime, whenever enough data exist to constrain the
non-universal fragmentation effects. 
The $\sqrt{s}$ dependence of these effects can then be studied with a larger
lever arm, or, if under control, the larger lever arm can be used to better
constrain the QCD parameters.

%Through ddFONLL, and depending on whether the theoretical understanding of the
%so far purely empirical parametrization will make progress, it might become
%possible to interface even the measured differential charm and beauty cross
%sections into PDF fits such as those in \cite{prosa2019} using absolute cross
%sections instead of using ratios only. 

A method similar to ddFONLL might also be applied to MC predictions, replacing
the `FONLL with non-universal $e^+e^-$ fragmentation' reference by the
respective MC model, and deriving effective $\tilde f$ functions as corrections
relative to the MC model in question. 

The results of all these studies could, and presumably should, be investigated
concerning their impact on heavy flavour tagging in high $p_T$ jets at LHC
and elsewhere, which so far mostly still rely on charm and beauty hadron
compositions consistent with fragmentation universality. Due to the
consistency with asymptotic convergence to universality at high $p_T$
found in this work, this impact should be very small whenever the charm or
beauty hadron transverse momentum inside these jets exceeds 20 GeV or so.
Given that the hadron $p_T$ may only be a fraction of the jet $p_T$, some
noticeable effects at the lower end of the usual LHC jet $p_T$ spectrum
may however not be excluded. 

Ground state charm hadrons have semileptonic decay fractions varying from
about 4\% ($\Lambda_c$) to about 16\% ($D^+$). Since the $\tilde f$ modifiers
change the hadronic composition in a $p_T$-dependent way, they will also
change the average semileptonic charm decay rate as a function of $p_T$.
Apart from the effect on leptonic flavour
tagging, this will also have an effect on the shape and normalization
of neutrino spectra from charm semileptonic decays, be it for forward
neutrino detectors at LHC (e.g. FASER \cite{FASER:2024ref}), for current and
future fixed target neutrino studies (e.g. SHIP \cite{SHiP:2025ows}), or for
neutrino spectra from cosmic ray showers (e.g. for IceCube \cite{prosa2019}). 

So it is the author's hope and expectation that the results and procedures
described in this paper can
serve as the starting point for a whole series of further physics investigations
also by third parties, through the code and descriptions provided in this work 
and in the public repository \cite{ddfonll}.

\section*{Acknowledgements}

The work of O.Z. has received funding through the MSCA4Ukraine project, which is funded by the European Union.

%
%\newpage
%
% BibTeX users please use
\bibliographystyle{unsrt}
\bibliography{pheno_cTotXsecs}
\end{document}